\documentclass[fleqn,usenatbib,useAMS]{mnras}

\usepackage{graphicx}
\usepackage{amsmath}
\usepackage{amssymb}
\usepackage{multicol}
\usepackage{bm}	
\usepackage{nccmath}
\usepackage[T1]{fontenc}
\usepackage{ae,aecompl}
\usepackage{newtxtext,newtxmath}
\usepackage{threeparttable, tablefootnote}
\usepackage{xcolor}

\newcommand{\kms}{\,km\,s$^{-1}$}
\newcommand{\angstrom}{\mbox{\normalfont\AA}}

\usepackage{etoolbox}

\title[The cool CGM in SDSS galaxies]{Characterizing the abundance, properties, and kinematics of the cool circumgalactic medium of galaxies in absorption with SDSS DR16}

\author[Anand, Nelson \& Kauffmann]{Abhijeet Anand$^{1}$\thanks{E-mail: abhijeet@mpa-garching.mpg.de},
Dylan Nelson$^{1,2}$, 
Guinevere Kauffmann$^{1}$
\\\\
$^{1}$Max-Planck-Institut f\"{u}r Astrophysik, Karl-Schwarzschild-Str. 1, 85741 Garching, Germany\\
$^{2}$Universit\"{a}t Heidelberg, Zentrum f\"{u}r Astronomie, Institut f\"{u}r theoretische Astrophysik, Albert-Ueberle-Str. 2, 69120 Heidelberg, Germany\\
}

\pubyear{2021}

\begin{document}
\maketitle

\begin{abstract} 
In order to study the circumgalactic medium (CGM) of galaxies we develop an
automated pipeline to estimate the optical continuum of quasars and detect
intervening metal absorption line systems with a matched kernel convolution
technique and adaptive S/N criteria. We process $\sim$ one million quasars in the latest Data
Release 16 (DR16) of the Sloan Digital Sky Survey (SDSS) and compile a
large sample of $\sim$ 160,000 Mg\,\textsc{ii} absorbers, together with
$\sim$ 70,000 Fe\,\textsc{ii} systems, in the redshift range
$0.35<z_{abs}<2.3$. Combining these with the SDSS DR16 spectroscopy of
$\sim1.1$ million luminous red galaxies (LRGs) and $\sim 200,000$ emission
line galaxies (ELGs), we investigate the nature of cold gas absorption at
$0.5<z<1$. These large samples allow us to characterize the
scale dependence of Mg\,\textsc{ii} with greater accuracy than in previous
work.  We find that there is a strong enhancement of Mg\,\textsc{ii}
absorption within $\sim 50$ kpc of ELGs, and the covering fraction within
$0.5r_{\rm vir}$ of ELGs is 2-5 times higher than for LRGs. Beyond 50 kpc,
there is a sharp decline in Mg\,\textsc{ii} for both kinds of galaxies,
indicating a transition to the regime where the CGM is tightly linked
with the dark matter halo.  The Mg\,\textsc{ii} covering fraction
correlates strongly with stellar mass for LRGs, but weakly for ELGs, where 
covering fractions increase with star formation rate.  Our
analysis implies that cool circumgalactic gas has a different
physical origin for star forming versus quiescent galaxies.
\end{abstract}



\section{Introduction}

The gaseous halo surrounding galaxies, known as the circumgalactic medium (CGM), holds important clues to the process of galaxy formation and evolution. The CGM is the main venue for cosmological gas accretion and inflows, as well as galactic outflows produced by feedback processes. These flows pass through the CGM and play a pivotal role in several key processes regulating galaxy formation, such as determining the timescales of gas depletion and star formation \citep{whitaker12}, the origin of the observed bimodality and the way in which quenching occurs \citep{tumlinson17, schiminovich10}, and the distribution of the baryonic budget as well as the total metal content of the gas surrounding galaxies \citep{behroozi10, peeples14}.

The formation of galaxies and their evolution is driven by the gas flows from the CGM  \cite[see,][for a review]{tumlinson17}, implying that our understanding of galaxy formation is itself limited by our current understanding of the CGM. However, large sky surveys performed with ground and space-based telescopes such as SDSS, Keck, VLT, and HST have significantly deepened our understanding of the CGM over the past two decades.

In this regard, one of the most powerful tools has been transverse absorption line studies where the CGM is observed in absorption against a bright background source such as a quasar. Different metal absorbers detected at redshifts smaller than the redshift of the background source provide direct observational constraints on the gas flows around galaxies at different epochs \citep{zhu13b, turner14, huang21}. 

Galaxy-absorber pair studies have been at the centre of our understanding into the nature of the CGM since the first discovery of a galaxy-metal absorber pair by \citet{bergeron86}. More recently, there have been several galaxy-absorber pair studies investigating the physical relationships between galaxies and the CGM, covering  different types of galaxies over a broad range of redshifts \citep{steidel94, churchill05, chen10, nielsen13}. These studies have found that the amount of gas varies strongly with distance from galaxies, and can depend on galaxy stellar mass, star formation history, color and shape \citep{bordoloi14, borthakur16, lopez18, lan18, rubin18}. 

Among all the detected metal lines the most extensively examined is the Mg\,\textsc{ii} $\rm \lambda\lambda 2796, 2803$ doublet. It is one of the strongest absorption features that can be detected by ground based optical telescopes at modest redshifts ($ 0.3\lesssim z \lesssim 2.5$). Mg\,\textsc{ii} absorbers are tracers of low-ionization cold gas ($\lesssim 10^{4}$ K) in the CGM and in the intergalactic medium (IGM). Several Mg\,\textsc{ii} surveys have been performed to constrain the temperature and density profile of low-ionization gas in the CGM along with the statistical properties of weak ($\rm EW_{rest}^{2796}\leq 0.3$ \AA) and strong ($\rm EW_{rest}^{2796}>0.3$ \AA) Mg\,\textsc{ii} absorbers \citep{weymann79, tytler87, sargent88, caulet89, steidel92, churchill99, churchill00, york06, quider11, zhu13b}. In addition,  C\,\textsc{iv} $\lambda\lambda$1548, 1550 is also commonly probed, tracing $\sim 10^5$ K gas in the CGM around galaxies. \citet{cooksey13} compiled a large C\,\textsc{iv} absorber catalogue detected in SDSS DR7 quasars and quantified the number of C\,\textsc{iv} absorbers as a function of rest equivalent width. They found a monotonic and significant increase in comoving line density of C\,\textsc{iv} around galaxies over a large range of redshift, $0<z<6$. \citet{chen16} compiled a similar C\,\textsc{iv} absorber catalogue using SDSS DR9 quasars and compared C\,\textsc{iv} properties with Mg\,\textsc{ii} systems, finding generally larger velocity dispersion for C\,\textsc{iv}. 

\citet{pieri14} stacked several hundred thousand Ly $\alpha$ absorbers at $ 2.4<z<3.1$ to probe the CGM at high redshift. They estimated the average neutral hydrogen column densities in those systems and found evidence for metallicities near the solar value. On the other hand, most similar to our present work, \citet{zhu14} performed a detailed statistical galaxy-absorber pair study with DR7 Mg\,\textsc{ii} absorbers \citep{zhu13a} and Luminous Red Galaxies (LRGs) from DR11 of SDSS \citep{dawson13}. They estimated average cool surface densities as traced by Mg\,\textsc{ii} absorbers around massive galaxies out to projected radii of 10 Mpc. They observed a change of slope on scales of 1 Mpc, consistent with the expected gas distribution in the parent halo together with gas outside the halo. Extending this analysis, \citet{lan14} found that both LRGs and ELGs have high covering fractions of cold gas ($\sim$ 1 percent) even at impact parameters of 500 kpc. 

Recently, \citet{lan18} have proceeded with a larger dataset of several thousand galaxy-quasar pairs using the Extended baryon Oscillation Spectroscopic Survey (eBOSS; \citealt{dawson16}) in the Sloan Digital Sky Survey IV (SDSS-IV; \citealt{blanton17}) finding an anisotropic metal absorption distribution around emission line galaxies (ELGs). They also observed the amount of cool gas to be different around ELGs and LRGs. There have also been statistical studies investigating the nature of the CGM around star-forming versus quiescent galaxies \citep{bordoloi11, menard11, peek15, huang16, lan20, huang21}. It is observed that the mean covering fraction of Mg\,\textsc{ii} absorbers varies strongly with galaxy type. Similarly, \citet{zibetti05, zibetti07} performed image stacking to study the photometric properties of Mg\,\textsc{ii} systems in SDSS and found that weak and strong absorbers are originated in different types of galaxies. Apart from these galaxy centric CGM studies, large spectroscopic surveys have also provided the unprecedented opportunity to perform statistical studies and cross-correlations exclusively with absorbers \citep{quider11, zhu13b, zhu14}. 

The distribution of velocity separations between galaxies and absorbers allows the kinematics of cold gas around galaxies to be investigated \citep{tremonti07}. For example, the clustering of Mg\,\textsc{ii} systems around BOSS LRGs ($\rm M_{\star}\sim 10^{11.5}\, M_{\odot}$) shows an excess of Mg\,\textsc{ii} up to $R_{p}=20$ Mpc, as well as relative velocities of $\rm \Delta v\sim 10000$ \kms within a projected distance of $\lesssim 800$ kpc \citep{kauffmann17}. The implication is that cool circumgalactic gas can originate in either supernovae or supermassive black hole driven outflows, as well as due to infall and accretion.

Since the first light of Sloan Digital Sky Survey (SDSS; \citealt{york00}) more than a million quasar spectra have been observed, and these can be searched for intervening absorber systems. Several detection (both automated and visual) algorithms have been developed to find Mg\,\textsc{ii} absorption lines in SDSS quasars \citep{nestor05, york06, bouche06, prochter06, lundgren09, quider11, zhu13a}. With the continuous release of ever larger datasets, it is increasingly important to develop efficient automated pipelines to detect absorber systems in background sources, and thereby study gas absorption as a function of galaxy properties.

In this paper, we develop an automated continuum estimation and absorption detection pipeline using the approach of \citet{zhu13a} as our starting point. We run our pipeline on the full DR16 quasar sample and study the statistical variation of absorbers as well as galaxy-absorber pairs to measure the physical properties of CGM gas as a function of galaxy mass, star formation activity, impact parameter, and redshift.

The paper is divided into five sections: Section \ref{method} introduces the observational data and describes our methods for continuum estimation and automatic absorption detection. We explore the resulting Mg\,\textsc{ii} absorber catalogue and its statistical properties in Section \ref{catalogue}. In Section \ref{galaxy_quasar} we analyse the galaxy-centric CGM properties of ELGs and LRGs. Finally we discuss the implications of our findings in Section \ref{discussion} and summarize the results in Section \ref{conclude}.


\section{Methods}\label{method}

\subsection{Quasar catalogue} \label{dataset}

The latest Data Release 16 (DR16) quasar catalogue\footnote{\url{https://www.sdss.org/dr16/algorithms/qso_catalogue/}} compiled by \citet{lyke2020} was released in late July 2020 as part of Value Added catalogue (VAC) of latest SDSS DR16 \citep{ahumada20}. Each SDSS data release is cumulative and includes all the objects observed in any previous release. The latest DR16 quasar catalogue contains 750,414 quasars. However, for our analysis we download the spectra of all the objects that are classified as QSOs in SDSS database\footnote{\url{https://dr16.sdss.org/optical/spectrum/search}}. It includes 983,317 objects identified as quasars (QSOs) with $0<z<7$ ($\sim$ 3,000 QSOs with $z>4.8$). 

In order to create robust continua of quasars we use the previously available quasar catalogue, DR14Q \citep{paris18,abolfathi18}. It contains all quasars that were observed in SDSS-I/II/III \citep{york00,schneider10, eisenstein11, dawson13} and SDSS-IV/eBOSS \citep{dawson16, blanton17} and classified as quasars with SDSS pipeline \citep{bolton12}. In order to compile a complete and pure catalogue, pipeline classified quasars were inspected visually to remove failed or uncertain classifications. The completeness and purity within a given target selection are as high as 99.5 percent in the catalogue. This high fidelity sample enables us to construct eigenspectra, as described below.

\subsection{Continuum Estimation}\label{continuum}

The estimation of a robust continuum for a quasar is a crucial step in detecting absorbers in its spectrum. Among the empirical methods to model the quasar continuum the most common is principal component analysis (PCA). Though standard PCA is quite powerful, it does not use any information about the known uncertainties and missing data and assigns components to the variations that are purely due to errors. For more flexibility to work with known uncertainties and missing data, another powerful dimension reduction technique called Non-negative Matrix Factorization (NMF; \citealt{lee99}) is used. NMF factorizes a large non-negative matrix into two much smaller non-negative matrices, minimizing the error, one of which can be thought as the basis (W) and other as the coefficient (H). In contrast to PCA, NMF naturally handles the data uncertainties because of its unique iterative update rules that guarantee non-increasing approximate error. Even if the matrices are weighted by their data uncertainties it is guaranteed to converge to a local minimum \citep{blanton07}. Then each column of the original matrix can be approximately reconstructed as the linear combination of basis (or, eigenspectra, W), and corresponding coefficients (or, eigenvalues, H). We construct the NMF eigenspectra for DR14 quasars using the approach adopted by \citet{zhu13a}, dividing the continuum estimation into four main steps.

First, the spectral flux must be re-binned on a common rest-frame wavelength array. Each SDSS spectrum is an observed frame spectrum and has a common grid i.e. \mbox{$\rm log\lambda_{i+1}- log\lambda_{i} = 0.0001$\AA}\, implying $ \Delta v = 69\, \rm{km\,s}^{-1}$. We define a common rest-frame wavelength array that can incorporate all quasars in each redshift range defined in Table \ref{tab:nmf_table}. We shift each quasar into the rest-frame and compute the total coadded fluxes (provided in each spectrum) using linear interpolation. While doing so we take care of the boundary values by restricting the interpolation to the edges of the redshift range appropriate for each quasar. 

\begin{table}
    \centering
    \caption{Details of our normalization scheme used in the NMF fitting for DR16 QSOs, using eigenspectra constructed from DR14 QSOs.}
    \label{tab:nmf_table}
    \begin{tabular}{c|c|c}
    \hline
    \hline
     Normalization & Eigenspectra Construction & Continuum Fitting  \\
     Wavelength Range& Redshift Range ($ z_{\rm QSO}$) & Range ($ z_{\rm QSO}$)\\
      (QSO rest-frame)&&\\
     \hline
     4150-4250 \AA & $z < 1.0$  & $z < 0.97$ \\
     3020-3100 \AA & $0.4 < z < 1.8$ & $0.97 < z < 1.49$ \\
     2150-2250 \AA & $0.8 < z < 2.8$ & $1.49 < z < 2.10$ \\
     1420-1500 \AA & $2.0 < z < 4.8$ & $2.10 < z < 4.8$ \\
\end{tabular}
\end{table}

Second, we renormalize the flux for each quasar to a common scale at a fixed rest-frame wavelength, adopting the approach taken by \citet{zhu13a}. As we move through different rest-frame wavelengths as a function of $z_{\rm QSO}$, a uniform normalization can not be applied due to the wavelength coverage of SDSS. Thus we select four different rest wavelength ranges and divide the observed spectra by the mean flux within this range (see column 1 of Table \ref{tab:nmf_table}). The choices in the first and second columns of Table \ref{tab:nmf_table} are adapted from \citet{zhu13a} which are based on the median quasar flux distribution given in \citet{vberk01}. The aim of selecting these four specific wavelength regions is to avoid any emission lines from the quasar itself entering the spectral regions that are searched for absorption systems. We want to normalize the quasar spectrum by the mean of flux in the region where the spectrum is as featureless as possible, masking missing or bad pixels as necessary.

Third, we run the NMF algorithm on our dataset. Note that, for the four redshift ranges, NMF can be run independently and the coefficient matrix (H) and eigenspectra (W) can be computed. The presence of spectroscopic artefacts can strongly affect the NMF eigenspectra which are tackled with an iterative approach. To avoid the peculiar objects we adopt the following approach: after constructing the set of eigenspectra we remove those quasars for which the coefficients differ by $\geq 5\sigma$ from the mean eigenvalues and repeat the NMF procedure until no outliers are found. The code converges in $\rm \sim 15-20$ iterations. We always initialize the $\rm W$ and $\rm H$ matrices by the results obtained in the previous iteration to make convergence faster. 

Finally, we use the DR14 quasar eigenspectra to fit the continuum of DR16 quasars, using the eigenspectra set derived from the redshift range whose median redshift is closest to quasar (see the third column of Table \ref{tab:nmf_table}). This guarantees that the continuum of each quasar is modelled with a set of eigenspectra that is constructed from the maximal number of quasars covering the same wavelength range. Then we compute the residual, defined as the ratio of normalized flux to the NMF continuum. We smooth this residual with a median filter of kernel size $\rm  = 141$ pixels ($\sim$ 8 times the typical width of Mg\,\textsc{ii} lines) to remove intermediate-scale fluctuations. We then derive a second NMF continuum as the product of the first NMF continuum and median filtered residual. We again compute the residual w.r.t this second NMF continuum and smooth the residual with the median filter of kernel size $\rm  = 71$ ($\sim$ 4 times the typical width of Mg\,\textsc{ii} lines) to remove small-scale fluctuations. The final NMF continuum is then computed by multiplying the second NMF continuum with this median filtered residual. 

\subsection{Automatic Detection of Absorbers}\label{pipeline}

The doublet nature of Mg\,\textsc{ii} absorbers motivates us to develop an automated algorithm to detect its presence. Our approach is broken up into the following steps.

\begin{table}
\centering
  \caption{Parameters of threshold and weighting scheme. All parameters are set by atomic physics. Note that $\rm \beta = \alpha / \delta$ and $\delta = f_{\rm strong} / f_{\rm weak}$ and $f_{\rm strong}$ and $f_{\rm weak}$ denote the oscillator strengths of doublet components. For selecting an optimized $\alpha$, we show the detailed analysis in Appendix~\ref{appendix:roc}.} 
  \begin{tabular}{||cc||}
    \hline
    Parameters & Mg\,\textsc{ii}\\
    \hline
    $\alpha$ & 2.50\\
    $ f_{\rm strong}$ & 0.6123\\
    $f_{\rm weak}$ & 0.3054\\
    \hline
    strong & 2796.35 \AA\\
    weak & 2803.52 \AA\\
    \hline
    $\rm \lambda_{prim}$ & 2799.935 \AA\\
    $\rm \lambda_{s1}$ & 2792.765 \AA\\
    $\rm \lambda_{s2}$ & 2807.405 \AA\\
    $\rm \Delta \lambda_{rest}$ & 7.17 \AA\\
    \hline
    $\rm \lambda_{a}$ & 2795.65 \AA\\
    $\rm \lambda_{b}$ & 2797.05 \AA\\
    \hline
  \end{tabular}
  \label{tab:thres}
\end{table}

\subsubsection{Wavelength Search Window}

\begin{figure*}
    \centering
    \includegraphics[width=0.97\textwidth]{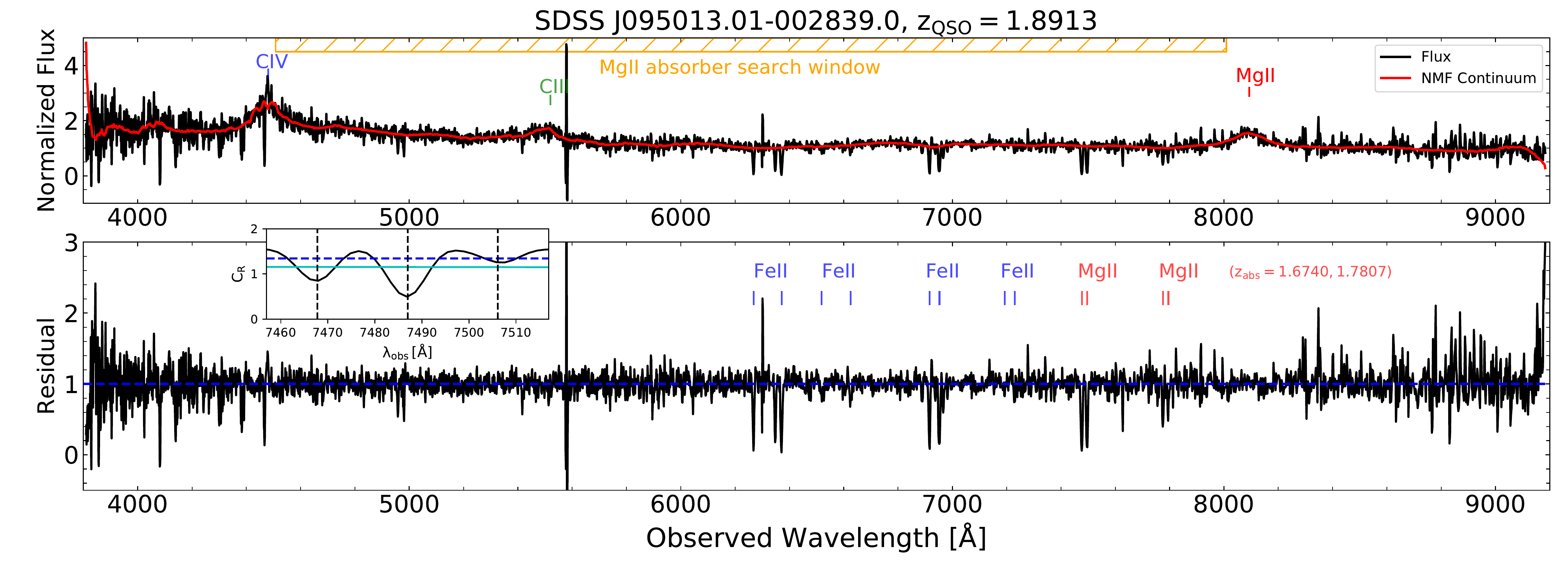}
    \caption{\textbf{Top:} the black line shows the normalized observed flux of quasar SDSS J095013.01-002839.0 ($\rm S/N = 7.46$). The red line is the best-fitting NMF continuum after median filtering. The hatched orange rectangle shows the wavelength search window where the spectrum is searched for intervening Mg\,\textsc{ii} absorbers. The intrinsic C\,\textsc{iv}, C\,\textsc{iii} and Mg\,\textsc{ii} emission lines of the quasar are shown in blue, green and red respectively. \textbf{Bottom:} the black line shows the residual spectrum defined as the ratio of the observed spectrum to the NMF continuum. The dashed blue line indicates unity. In this case the absorber detection pipeline identifies two intervening Mg\,\textsc{ii} absorbers at $\rm z_{abs} = 1.6740, 1.7807$ (shown in red). The corresponding Fe\,\textsc{ii} lines are shown in blue. Inset shows the convolved array with the thresholds (cyan: primary, blue: secondary) at the location of one of the absorbers. We clearly see three peaks as described in the text.}
    \label{fig:spectra_good}
\end{figure*}

First, we define a wavelength search window using the quasar intrinsic emission lines. We start from $\Delta z = 0.018$ redshifted from the quasar's C\,\textsc{iv} line, or the blue end of SDSS ($\rm \sim 3800$ \AA), and end at $\Delta z= 0.03$ blueshifted from quasar's Mg\,\textsc{ii} emission line, or the red end of SDSS ($\rm \sim 9200$ \AA). We select these $\Delta z$ values to detect intervening absorbers rather than quasar associated absorbers. This also avoids misidentification due to the cases where the continuum is not well fitted to the quasar's intrinsic C\,\textsc{iv} or Mg\,\textsc{ii} emission lines, which could produce spurious absorption dips. We also mask the possible Ca\,\textsc{ii} $\rm \lambda\lambda 3934, 3969$ lines (due to confusion with Mg\,\textsc{ii} absorbers at $z\sim 0.4$) and O\,\textsc{i} lines ($\rm \lambda 5577$ and $\lambda 6300$). In Figure~\ref{fig:spectra_good} we show an example quasar spectrum in normalized flux (top) and residual (bottom), with the wavelength search window indicated.

\subsubsection{Absorber Candidate Selection}

Next, we search for potential absorbers by defining a Gaussian kernel that mimics the Mg\,\textsc{ii} $\rm \lambda\lambda \, 2796, \,2803$ doublet. We convolve the residual with this kernel and apply a threshold on the convolved array ($\mathcal{C_{R}}$). To define the threshold we use the local noise estimate $\rm \sigma_{\mathcal{C_{R}}}$. To estimate this local noise for the convolved array, for each pixel we take the noise as the standard deviation of the convolved array within $\pm 100$ pixels around that pixel. This adaptive noise approach accounts for the noisy regions of the spectra, particularly at the edges. This convolution generically produces three peaks corresponding to the overlap of the kernel with the Mg\,\textsc{ii} doublet. The primary peak ($\rm \lambda_{prim}$) corresponds to the case when two Gaussians of the kernel fully overlap with Mg\,\textsc{ii} lines. The two secondary peaks ($\rm \lambda_{s1},\,\lambda_{s2}$) correspond to the case when kernel overlaps just one of the lines. We apply the following thresholds on the primary and secondary peaks to identify potential absorbers, selecting all the pixels that satisfy our threshold criterion. For the threshold parameters see Table \ref{tab:thres}.\\

\noindent \textbf{Rule 1: sigma criteria on primary peak:}

\begin{itemize}
    \item select all pixels $\{1+z_{\rm  abs_{i}} =\frac{\lambda_{\rm abs_{i}}}{\lambda_{\rm prim}}\}$ where $\rm \mathcal{C_{R}} \leq median(\mathcal{C_{R}}) - \alpha \cdot \sigma_{\mathcal{C_{R}}}$. This is a sigma criteria on the convolved residual and median($\mathcal{C_{R}}$) implies global median of the convolved array.
    \item for each pixel with its associated redshift $ z_{\rm abs}$ in $ \{z_{\rm abs_{i}}\}$, define the neighbourhood of secondary peaks:
    \begin{center}
        $ \lambda_{\rm sec_{1}} = \lambda_{\rm s1}\cdot(1+z_{\rm abs})\pm1.3$\\
        $\lambda_{\rm sec_{2}} = \lambda_{\rm s2}\cdot(1+z_{\rm abs})\pm 1.3$
    \end{center}
    \item measure the median residual amplitude for these pixels:
    \begin{center}
       $\mathcal{S_{R}}_{1}$ = median($\rm \mathcal{C_{R}}_{, \lambda_{sec_{1}}}$)\\
       $\mathcal{S_{R}}_{2}$ = median($\rm \mathcal{C_{R}}_{, \lambda_{sec_{2}}}$)\\
       $\rm \mathcal{T} =median(\mathcal{C_{R}}) - \beta \cdot \sigma_{\mathcal{C_{R}}}$ 
         where $\rm \beta=\frac{\alpha}{\delta}$
    \end{center}
\end{itemize}

\noindent \textbf{Rule 2: sigma criteria on secondary peaks:}

\begin{itemize}
    \item if $\mathcal{S_{R}}_{1}\leq \mathcal{T}$ or 
             $\mathcal{S_{R}}_{2}\leq \mathcal{T}$ then accept $ z_{\rm abs}$
    \item else, reject this $ z_{\rm abs}$ as a detected absorption pixel
\end{itemize}

For our chosen values of $\alpha$ (primary threshold), $\beta$ (secondary threshold), and $\delta = f_{\rm strong}/f_{\rm weak}$, where $f_{\rm strong}$ and $f_{\rm weak}$ are the oscillator strengths of Mg\,\textsc{ii} $2796$ and $2803$ lines, see Table \ref{tab:thres}. Note that the threshold on the secondary peaks is based on the theoretical strength of lines given the oscillator strengths. The threshold $\mathcal{T}$ is defined for the two smaller peaks centered at $\lambda_{\rm s1}$ and $\lambda_{\rm s2}$, which are weaker in the convolved residual. 

Finally, note that we apply our thresholds to all individual pixels in the spectrum. The resulting list of accepted pixels will in general contain several contiguous ranges which need to be grouped and considered as one detected absorption feature. To do so we combine contiguous pixels absorbers with $\Delta z<0.0026$, corresponding to $\rm \Delta\lambda_{rest, \, MgII}=7.17$ \AA. For each group we derive the absorber redshift as the mean of the pixel redshifts, weighting by the cube of the corresponding flux residual, such that the highest weight comes from the pixel closest to line center.

To provide the flexibility to detect absorbers of different strengths (rest equivalent widths) we separately run the kernel convolution varying the kernel FWHM from 3-8 pixels. We combine all the absorbers from each run, identifying and discarding duplicates by applying a similar weighted mean approach, weighting each redshift by the cube of the median of residual in the wavelength range $\lambda_{\rm a}\leq \frac{\lambda_{\rm obs}}{1+z_{\rm abs}}\leq \lambda_{\rm b}$ (see Table \ref{tab:thres}). Note that this grouping is applied only after running the detection pipeline for all widths to select the best location of the absorber. \\

\noindent \textbf{Rule 3: S/N criteria for final candidate selection:}\\ \\
The matched kernel convolution gives a list of potential Mg\,\textsc{ii} candidates. To select the final candidates we then use the S/N information of quasar spectra. Therefore, for each potential candidate from this list, we estimate the S/N ($\lambda$2796) and S/N ($\lambda$2803).\footnote{$\rm S/N = \displaystyle\sum_{i=p_{1}}^{p_{2}}\Big(1-\frac{F_{i}}{C_{i}}\Big)\, / \, \Big(\sum_{i=p_{1}}^{p_{2}}\sigma_{i}^{2}\Big)^{1/2}\, ;\,$ where F and C are flux and continuum respectively and $\sigma$ is the corresponding error. $p_{1}$ and $p_{2}$ are the starting and ending pixels within $\pm5$ pixels from line centre.} For the final candidate selection we apply the following criteria. We call this the \textit{Mg\,\textsc{ii} doublet criteria.} \\
\begin{center}
    $\rm S/N (\lambda2796) >3  \text{ and } S/N (\lambda2803) >2$
\end{center}

\subsubsection{Rejecting False Positives}

Given the absorber candidates that passed the above criteria, we next reject likely false positives. We fit each absorption profile with a double gaussian function and discard candidates which have peculiar doublet separation. In fitting we allow all six parameters (two amplitudes, two-line centres, and two-line widths) to vary. We keep only those absorbers for which the line separation is within 1.2 \,\AA \, of the fiducial value ($\rm \Delta\lambda_{rest}$, Table \ref{tab:thres}). At this stage we also estimate the mean redshift of the absorbers using the centroid of both lines. Finally, we check for cases where Fe\,\textsc{ii} lines are incorrectly identified as Mg\,\textsc{ii} lines. For spectra with more than one potential Mg\,\textsc{ii} absorber, we check whether its observed wavelength corresponds to any Fe\,\textsc{ii} $\rm \lambda\lambda 2586,\,2600$ lines, within $\pm 3$ \AA\,, corresponding to the redshifts of already identified Mg\,\textsc{ii} lines. If this is the case, we reject the Mg \textsc{ii} candidate having smaller S/N($\lambda$2796) than any of the two  Fe\,\textsc{ii} lines as Mg\,\textsc{ii} is a stronger transition than Fe\,\textsc{ii}. This selection might also reject a few true cases in which the Fe \textsc{ii} lines are also very strong, however, this is not very common. We show an example of a SDSS quasar spectrum with relatively high S/N ratio (= 7.5 $\rm pix^{-1}$) in Figure~\ref{fig:spectra_good}, highlighting two detected Mg\,\textsc{ii} absorbers and the locations of the  corresponding Fe\,\textsc{ii} lines.


\begin{figure*}
    \centering
    \includegraphics[width=0.9\linewidth]{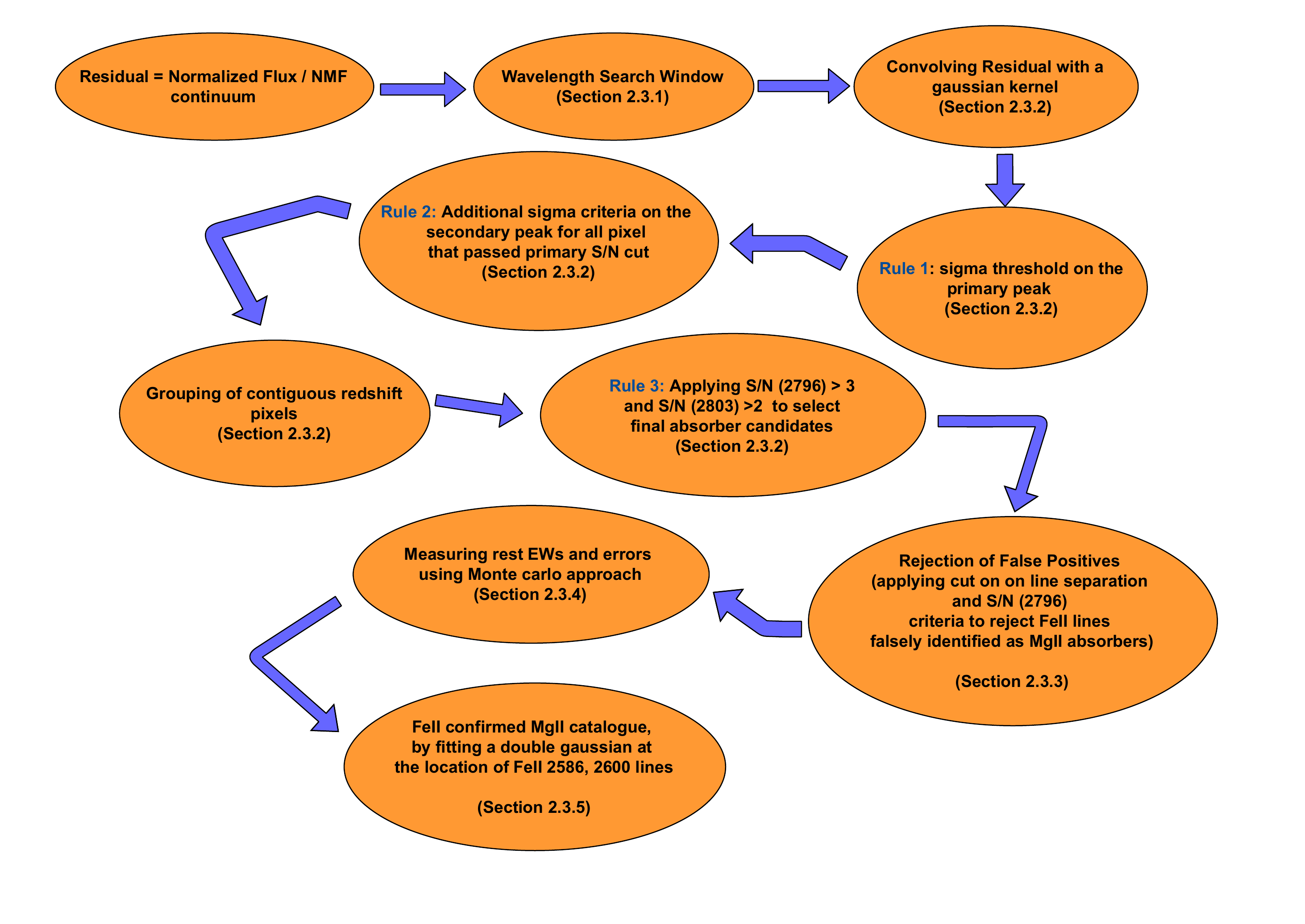}
    \caption{A short schematic of our automated detection pipeline.}
    \label{fig:pipe_flowchart}
\end{figure*}

\subsubsection{Rest equivalent widths of Absorbers}\label{ew_boot}

With the detected absorbers in hand, we apply a Monte Carlo approach to estimate errors on the fit parameters and ultimately the measured rest equivalent widths. We fit each doublet with a standard Levenberg-Marquardt minimization, first by adding randomly generated noise, using a normal distribution centred at 0 and with standard deviation equal to the mean of the errors on the residual. We also try the alternative approach of adding the true noise of the spectra multiplied by a standard normal distribution. In both cases we repeat this process 200 times. In case of failures, we record all parameters and rest equivalent widths as zeros. This guarantees that errors will be large in cases with many failures, if there is a false positive all 200 runs fail and the rest equivalent width is consistent with zero. Finally, we take the median of all 200 runs as the estimate of the rest equivalent width of lines. For errors we take the 16th (p16) and 84th (p84) percentiles and compute sigma as $\rm \sigma = (p84-p16) / 2$. Overall we find that these two error estimate methods agree well (shown in Figure~\ref{fig:error_plot}), also indicating that the majority of absorbers are genuine. For our purposes we use the uncertainties obtained from the true noise of the spectra. The typical error in the rest $\rm EW_{2796}$ ($\sigma_{EW_{2796}}$) is $\sim 0.2$ \AA.

\subsubsection{Fe - confirmed Mg\,\textsc{ii} catalogue}

With our Mg\,\textsc{ii} absorber catalogue we perform an additional step and attempt to confirm each Mg\,\textsc{ii} absorber with Fe\,\textsc{ii} $\rm \lambda\lambda \, 2586, \, 2600$ by fitting gaussian at the location of Fe\,\textsc{ii} lines. If the fitting succeeds (parameters are finite and within the boundaries) and the line separation of Fe\,\textsc{ii} lines is within $\rm 1.2$ \AA\, of the fiducial value ($\rm \Delta \lambda_{rest} = 13.48$ \AA) we flag this Mg\,\textsc{ii} absorber as `confirmed' with Fe\,\textsc{ii}. We also estimate the rest frame equivalent width and errors of the Fe\,\textsc{ii} $\rm \lambda\lambda 2586,\, 2600$ lines using a similar Monte Carlo approach as above. Note that Fe\,\textsc{ii} is a much weaker transition than Mg\,\textsc{ii} and the minimum Fe\,\textsc{ii} strength detectable with our algorithm is $\rm{EW}_{\rm 2600}\geq f_{\rm FeII,\, 2600}/f_{\rm MgII,\, 2796}\cdot \rm{EW}_{\rm 2796}$, assuming the same abundance for both Mg and Fe. With the minimum detected $\rm EW_{2796}\sim 0.2$ \AA\, in our catalogue, and $f_{\rm FeII,\, 2600}\sim 0.3$, we obtain a theoretical minimum $\rm{EW}_{\rm 2600}\sim 0.1$ \AA. In total we have $\sim 70,000$ Fe\,\textsc{ii} confirmed Mg\,\textsc{ii} absorbers ($\sim 44\%$ of the sample). Finally, we show a schematic in Figure~\ref{fig:pipe_flowchart} that summarizes the steps of our automated detection pipeline.

\section{Metal absorber catalogue}\label{catalogue}

\subsection{Properties of Individual Absorbers}

\begin{figure}
    \centering
    \includegraphics[width=0.95\columnwidth]{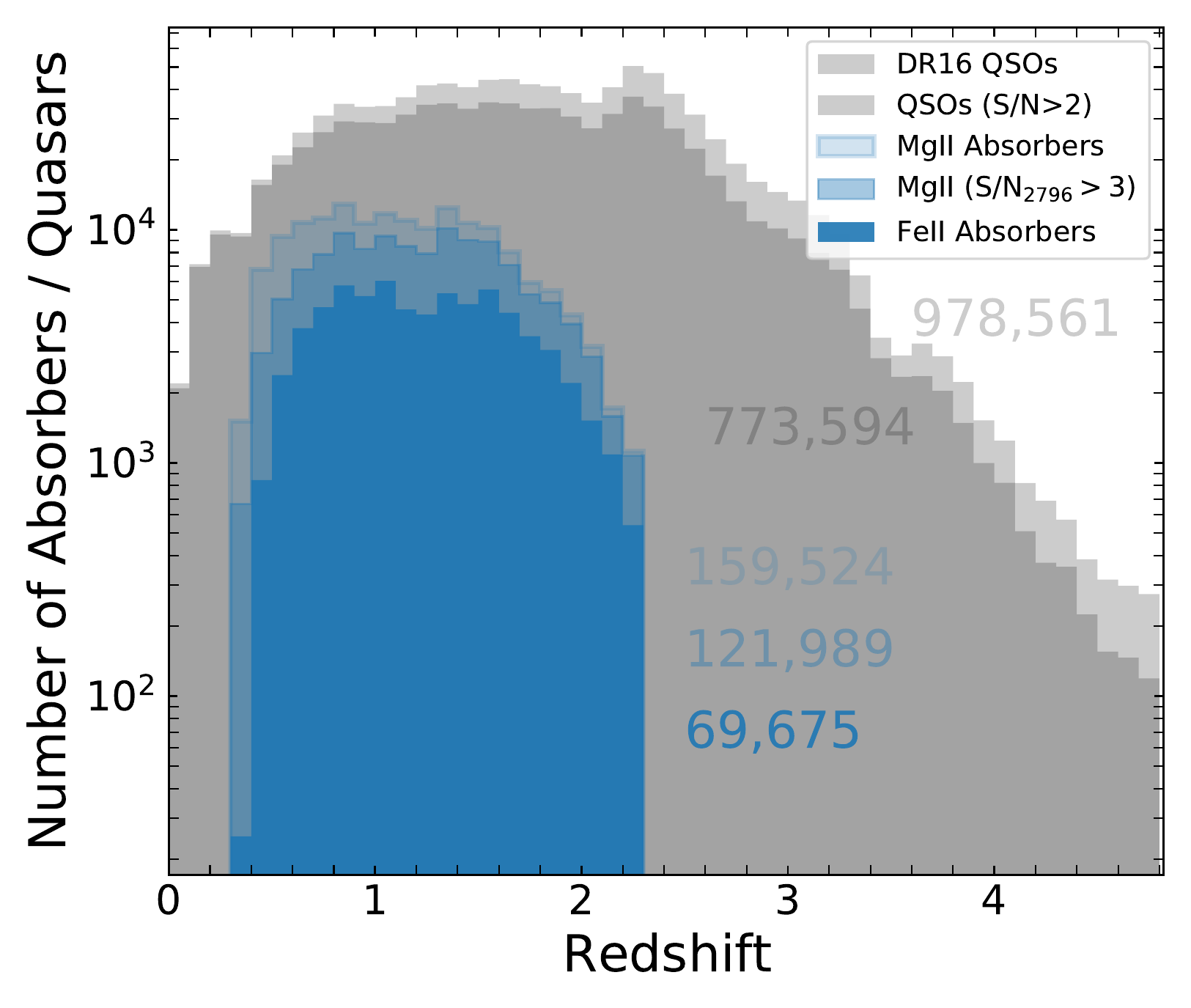}
    \caption{Redshift distribution of DR16 quasars (shown in light black, $\rm N =978,561$, dark black shows quasars with $\rm S/N_{QSO}>2$, $\rm N= 773,594$) and the Mg\,\textsc{ii} absorbers detected by our pipeline (shown in light blue, N = 159,524). The medium blue region shows the Mg\,\textsc{ii} absorbers with rest equivalent width greater than their three times the corresponding errors (N = 121,989). The dark blue region shows the Mg\,\textsc{ii} absorbers which are `confirmed' by Fe\,\textsc{ii} lines (N = 69,675). The redshift range spans from $z = 0.36$ to $z = 2.28$ with bin size of $\Delta z =0.1$.}
    \label{fig:z_histogram}
\end{figure}

\begin{table}
  \centering
  \caption{Sample statistics comparison versus SDSS data release (DR). References are: [a] \citet{paris17}; [b] \citet{paris18}; [c] \citet{ahumada20}; [$\star$] \citet{zhu13a} (trimmed case with the same masks as ours); \textbf{[$\dagger$] this work}. The increased quasar sample enables us to increase the statistics of Mg\,\textsc{ii} absorption systems significantly ($\sim 4$ times). Note that $\rm S/N_{2796}$ is defined as $\rm EW_{2796}/\sigma_{EW_{2796}}$ and $\rm S/N_{QSO}$ implies signal-to-noise of quasar spectra.}
  \begin{tabular}{||cccc||}
    \hline
    Objects & DR12 & DR14 & DR16\\
    \hline
    QSOs & 297,301$\rm ^{a}$ & 526,356$\rm ^{b}$ & 983,317$\rm ^{c}$ \\
    QSOs ($\rm S/N_{QSO}>1$) & -& - & 941,939 \\
    QSOs ($\rm S/N_{QSO}>2$) & -& - & 773,594 \\
    LRGs & - & $\lesssim 1$ million & 1,252,722$\rm^{c}$\\
    ELGs & - & 35,094$\rm ^{c}$ & 269,889$\rm ^{c}$\\ 
    Mg\,\textsc{ii} Absorbers & 39,219$^\star$ & - & \textbf{159,524}$^\dagger$\\
    Mg\,\textsc{ii} ($\rm S/N_{2796} > 1)$ & 38,327$^\star$ & - & \textbf{158,725}$^\dagger$\\
    Mg\,\textsc{ii} ($\rm S/N_{2796} > 2$) & 37,763$^\star$ & - & \textbf{150,236}$^\dagger$\\
    Mg\,\textsc{ii} ($\rm S/N_{2796} > 4$) & 33,376$^\star$ & - & \textbf{94,403}$^\dagger$\\
    Fe\,\textsc{ii} Absorbers & - & - & \textbf{69, 675}$^\dagger$\\
    Mg\,\textsc{ii} Absorbers  & - & - & \textbf{158, 494}$^\dagger$\\
    (with $\rm S/N_{QSO}>2$)&&&\\
    Fe\,\textsc{ii} Absorbers  & - & - & \textbf{69, 594}$^\dagger$\\
    (with $\rm S/N_{QSO}>2$)&&&\\
    \hline
  \end{tabular}
  \label{tab:sample}
\end{table}

Running the detection pipeline on the DR16 quasars we compile our final Mg\,\textsc{ii} absorber catalogue. In Figure \ref{fig:z_histogram} we show the redshift distributions of quasars (light gray shows all DR16 QSOs and dark gray shows QSOs with $\rm S/N_{QSO}>2$) and detected absorbers (blue). We construct three absorber catalogues: (i) all Mg\,\textsc{ii} detections (light blue), (ii) strong Mg\,\textsc{ii} absorbers defined as having $EW_{\rm 2796}>3\sigma_{2796}$ (medium blue), and (iii) `Fe\,\textsc{ii} confirmed' detections where we detect Fe\,\textsc{ii} $\rm \lambda \lambda 2586, \, 2600$ at the same redshift (dark blue). Within our wavelength search window we find 159,524 Mg\,\textsc{ii} absorbers which satisfy our detection pipeline. Out of these 69,675 absorbers have also passed our Fe\,\textsc{ii} confirmation test and 121,989 have $\rm EW_{\rm 2796}>3\sigma_{2796}$ (i.e. $\rm S/N_{2796}>3$).

\begin{figure}
    \centering
    \includegraphics[width=0.66\columnwidth, height=4.6cm]{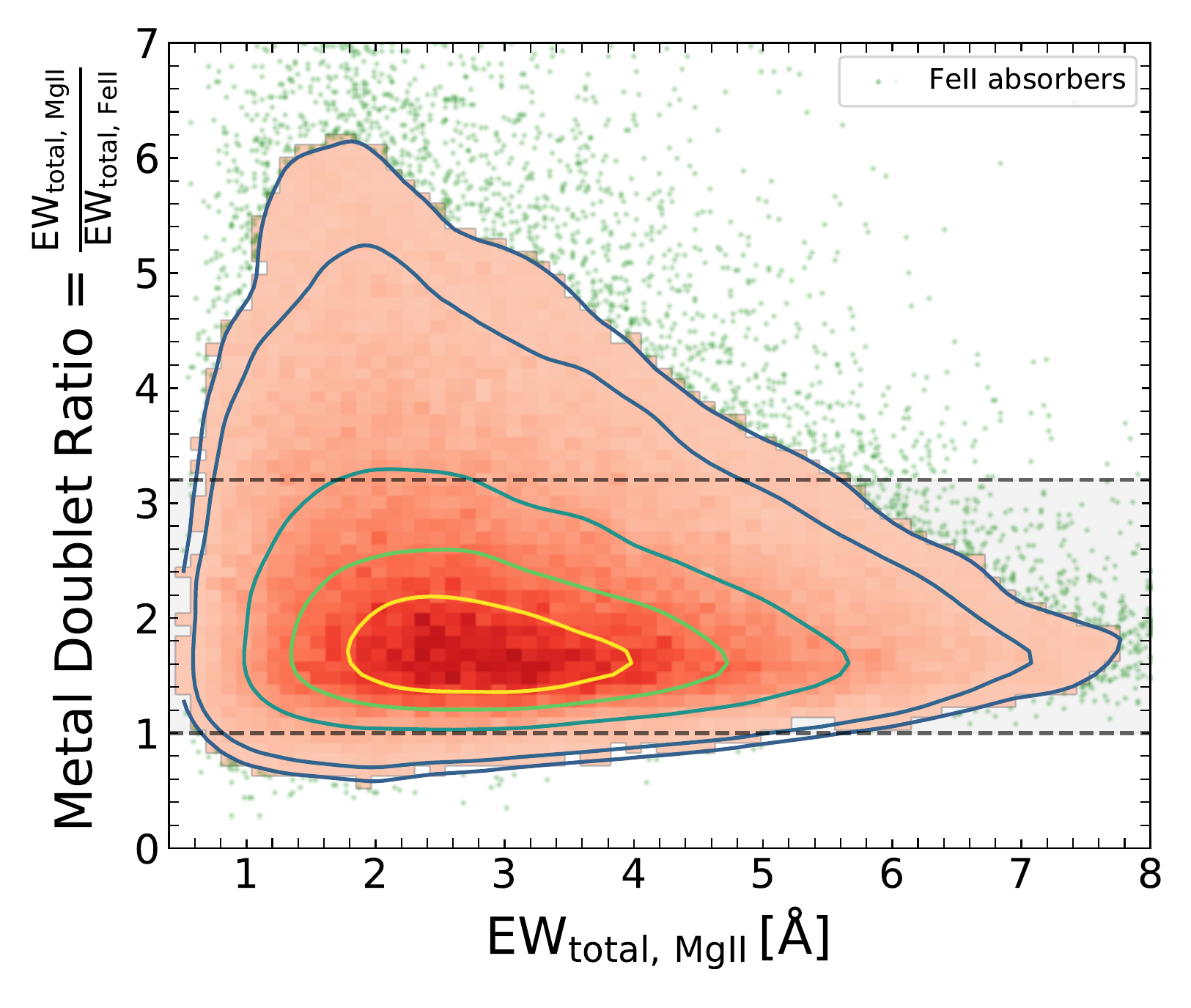}\\
    \includegraphics[width=0.66\columnwidth, height=4.56cm]{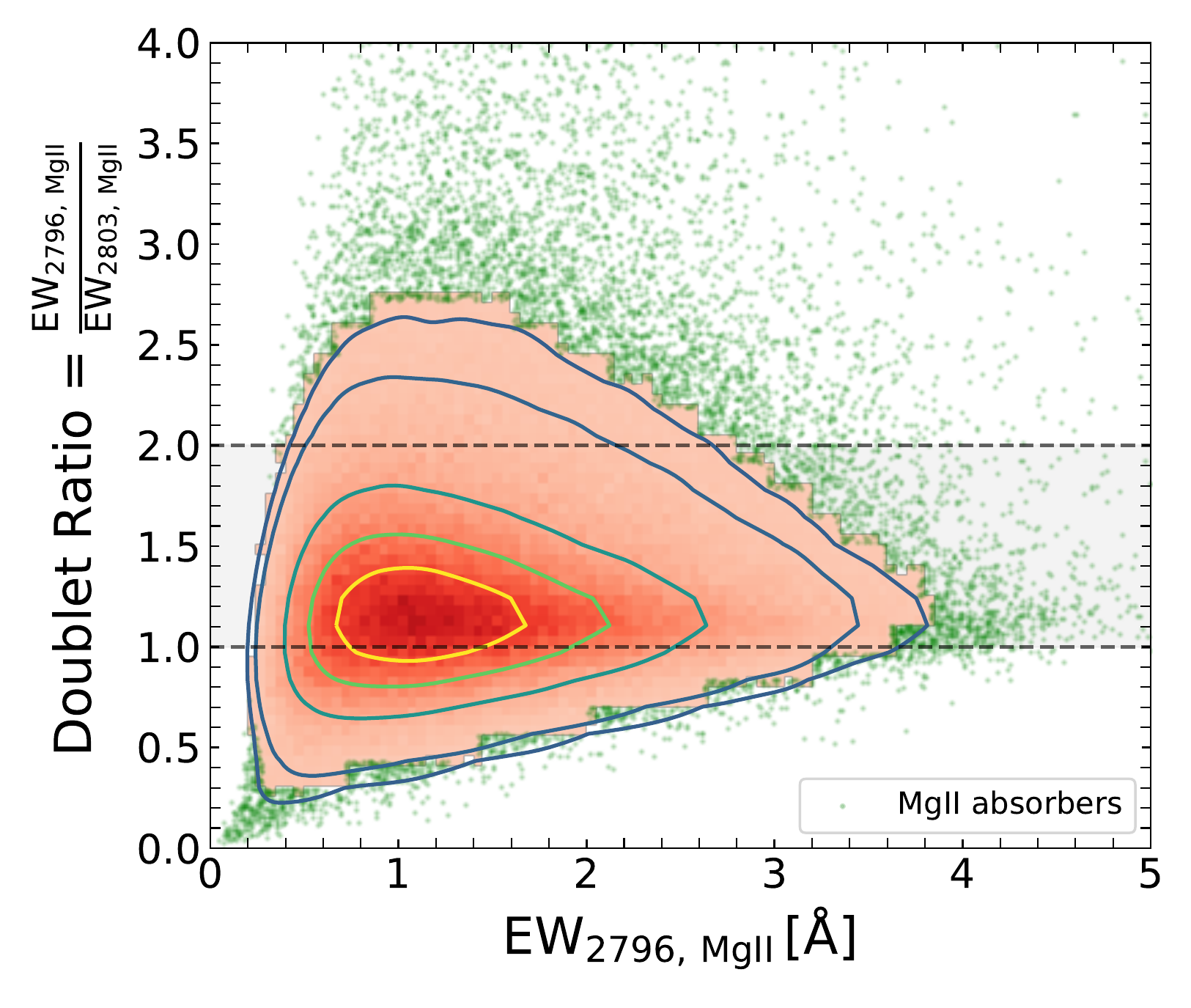}\\
    \includegraphics[width=0.66\columnwidth, height=4.56cm]{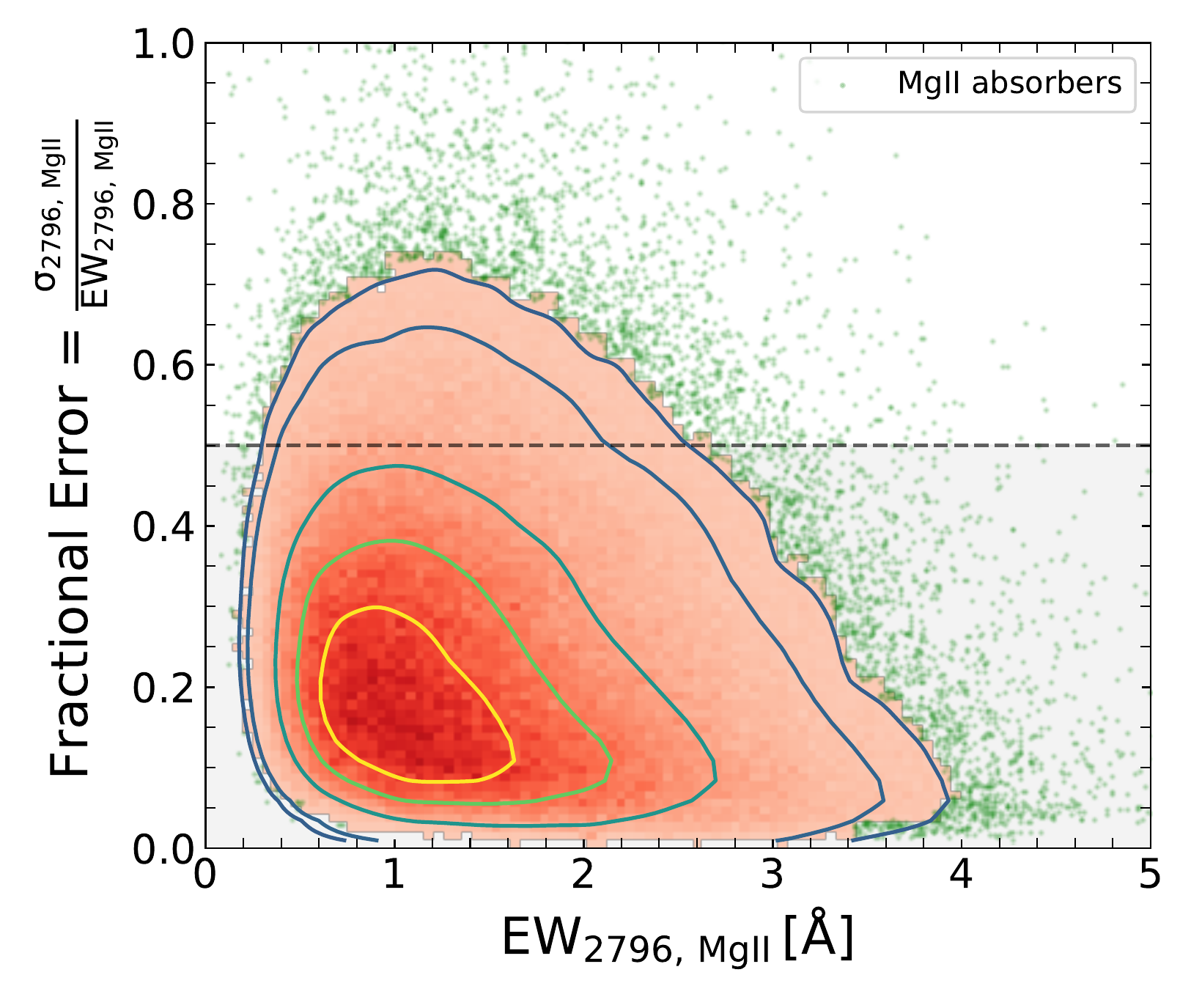}\\
    \includegraphics[width=0.66\columnwidth, height=4.56cm]{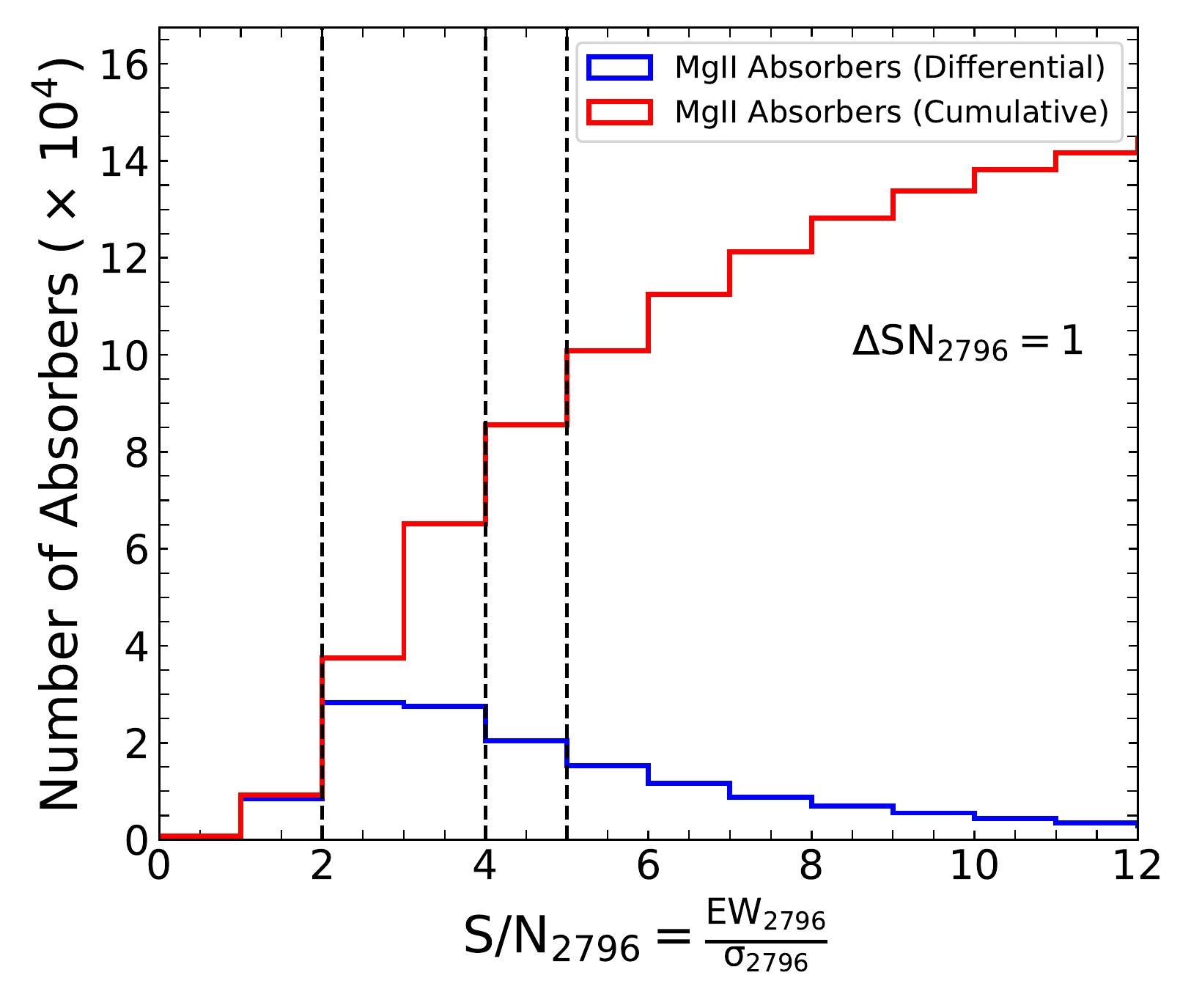}
    \caption{\textbf{Top:} Doublet ratio of total rest rest equivalent width of Mg\,\textsc{ii} $\rm \lambda\lambda 2796, \, 2803$ and Fe\,\textsc{ii} $\rm \lambda\lambda 2586, \, 2600$ as a function of total rest rest equivalent width of Mg\,\textsc{ii}. The shaded rectangle highlights the region when all lines are completely saturated (doublet ratio=1) and when lines are unsaturated (doublet ratio$\sim$ 3). Note that this theoretical limit is estimated using oscillator strengths and assuming abundances of Fe and Mg be same. Most Mg\,\textsc{ii} absorbers lie within this theoretical limit. \textbf{Second:} Mg\,\textsc{ii} $\rm \lambda\lambda 2796, \, 2803$ doublet ratio as a function of $EW_{\rm 2796}$, where the shaded rectangle highlights the region when both lines are saturated (i.e. doublet ratio=1) and when neither are saturated (i.e. doublet ratio=2). Most Mg\,\textsc{ii} absorbers lie within these two lines. \textbf{Third:} Fractional error in measurement of $EW_{\rm 2796}$ as a function of EW. Errors are estimated with bootstrap approach. In all panels contours enclose 25, 50, 75, 95 and 97.5 percentiles of the sample. \textbf{Bottom:} Cumulative (red) and differential (blue) distribution of $\rm S/N_{2796}$ of all detected Mg \textsc{ii} absorbers.} 
    \label{fig:metal_ews}
\end{figure}

In the four panels of Figure \ref{fig:metal_ews} we present the properties of individual Mg\,\textsc{ii} absorbers. The topmost panel shows the ratio of total rest equivalent width (EW) of Mg\,\textsc{ii} to Fe\,\textsc{ii}  (using $\rm \lambda\lambda 2586, \, 2600$ only) as a function of the total rest equivalent width of Mg\,\textsc{ii} absorption. This quantifies the relative strength of Fe\,\textsc{ii} absorbers, which also trace $\sim10^{4}$ K gas around galaxies. We see that most ($\gtrsim$ 80 percent) of the Fe\,\textsc{ii} confirmed Mg\,\textsc{ii} absorbers lie within the theoretical limit (assuming abundances to be same for Fe and Mg) for the line strength ratio of 1 (for the saturated case) and $\sim$3 (for the completely unsaturated case). The majority of absorbers regardless of EW have a line ratio of $\gtrsim1$, indicating they are intermediate between saturated and unsaturated. The median line ratio is $\sim 2$. We also note the large scatter and many outliers above $\sim 3$, likely indicating cases of non-solar abundance ratios.

The second (from top) panel of Figure~\ref{fig:metal_ews} shows the doublet ratio of $\rm{EW}_{\rm 2796}$ and $\rm \rm{EW}_{2803}$ as a function of $\rm{EW}_{\rm 2796}$ for all Mg\,\textsc{ii} absorbers. The doublet ratio indicates if the lines are saturated or unsaturated and is sensitive to the opacity of the medium. For Mg\,\textsc{ii} $\rm \lambda\lambda 2796, \, 2803$ the theoretical value of the doublet ratio varies between 1 (fully saturated) and 2 (completely unsaturated). We see that most strong absorbers ($\rm{EW}_{\rm 2796}>1$ \AA) have a doublet ratio close to 1, indicating saturation. For weak absorbers ($\rm{EW}_{\rm 2796}<1$ \AA) we see doublet ratio $\sim 2$ as expected due to low saturation.

The third (from top) panel of Figure~\ref{fig:metal_ews} shows the fractional errors on the $EW_{\rm 2796}$ measurement as a function of $ EW_{\rm 2796}$ for all deteced Mg\,\textsc{ii} absorbers. We see that, most ($\sim$ 94 percent) of the absorbers have low errors (fractional error < 0.5). Weak absorbers tend to have large errors, making it more problematic to measure their properties. The median $\rm{EW}_{\rm 2796}$ is $\sim 1.3$ \AA\, with typical $\sigma_{EW_{2796}}\sim 0.2$\, \AA\, for the sample. The bottom panel of Figure~\ref{fig:metal_ews} shows the   cumulative (red) and differential (blue) distributions of $\rm S/N_{2796}$. The majority of our absorbers have high $\rm S/N_{2796}$ (>2) and median $\rm S/N_{2796}$ is $\sim 4.7$.

\begin{figure*}
    \centering
    \includegraphics[width=6.9in]{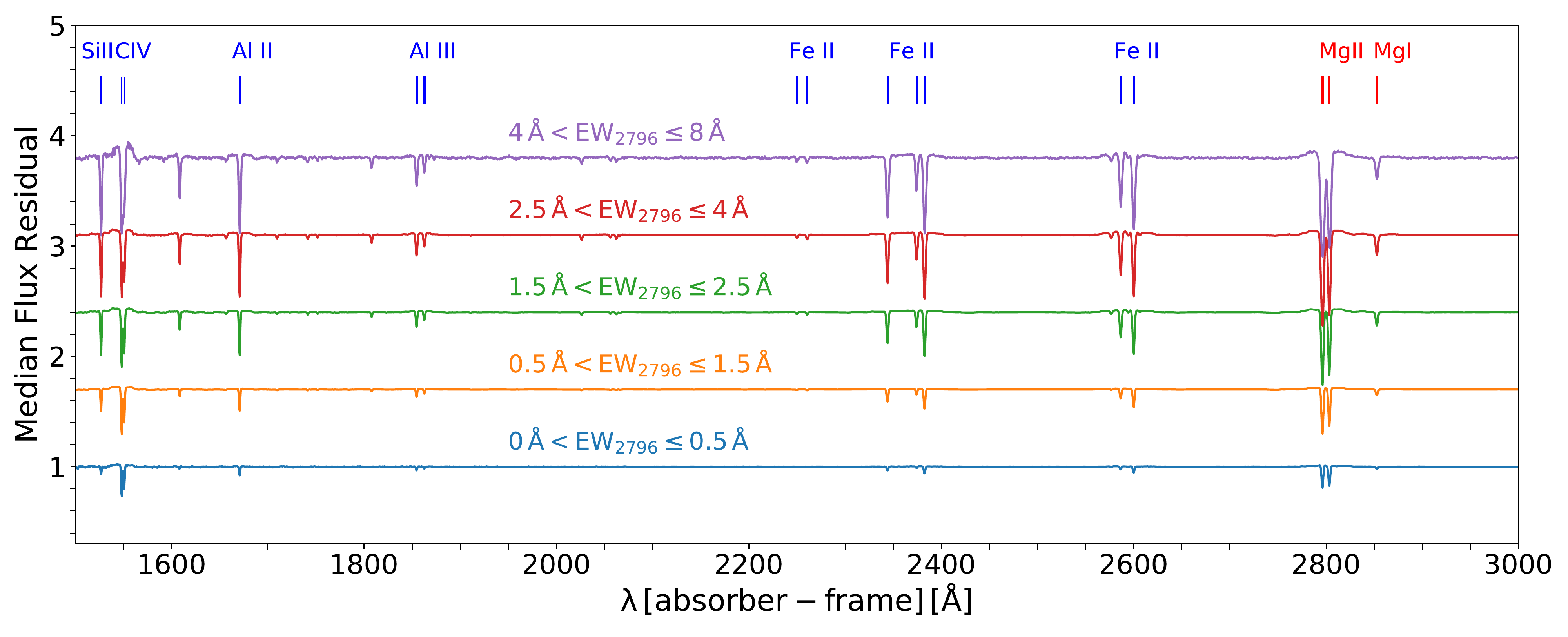}
    \includegraphics[width=6.8in]{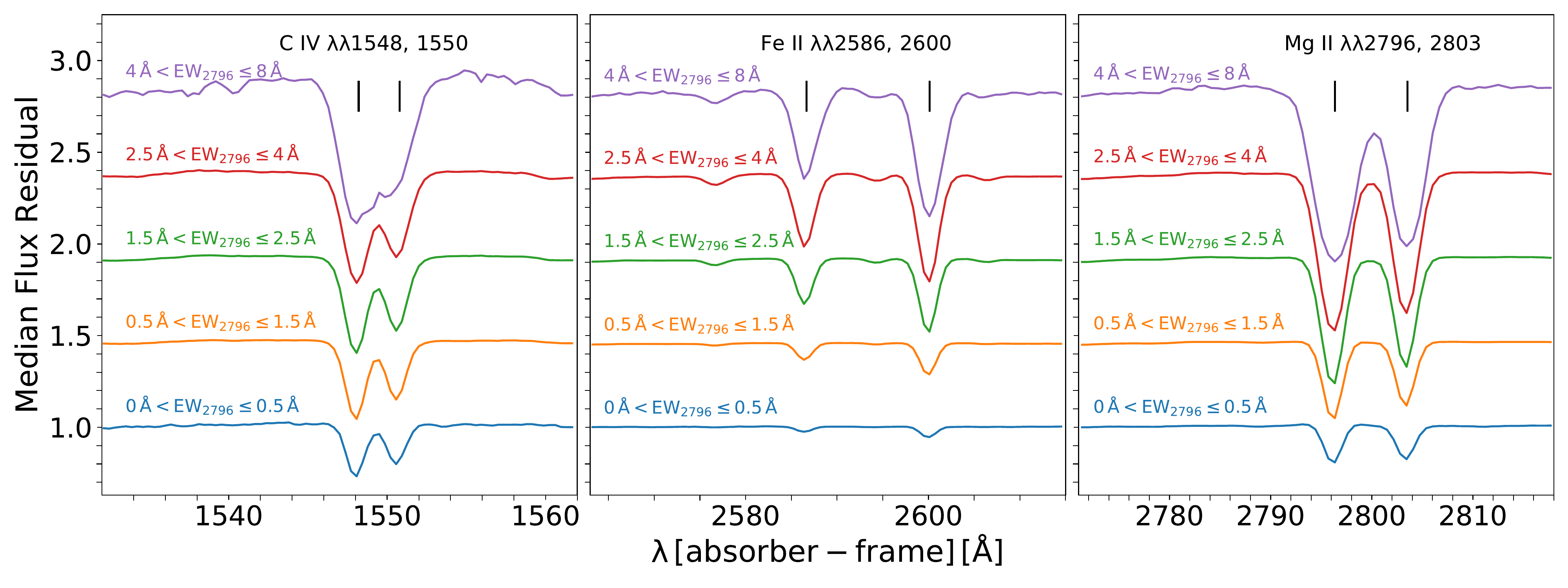}
    \caption{\textbf{Top:} Median composite spectra of quasars, stacked in the rest-frame of all DR16 Mg\,\textsc{ii} absorbers. We have divided the absorber sample into five sub-samples with $\rm 0\, \angstrom<EW_{2796}\leq0.5\,\angstrom\, \, (N_{abs} = 5718$), $\rm 0.5\, \angstrom<EW_{2796}\leq1.5\,\angstrom\, \, (N_{abs} = 90,317)$, $\rm 1.5 \,\angstrom<EW_{2796}\leq2.5\,\angstrom\, \, (N_{abs} = 50,083)$, $\rm 2.5 \,\angstrom<EW_{2796}\leq4\,\angstrom\, \, (N_{abs} = 12,869)$, $\rm 4\, \angstrom<EW_{2796}\leq8\,\angstrom\, \, (N_{abs} = 537)$. We have indicated the most prominent metal lines to guide the eye: in stacks we detect weaker transitions including Si\,\textsc{ii}, C\,\textsc{iv}, Al\,\textsc{ii}, Al\,\textsc{iii}, Mg\,\textsc{i} and other weak transitions of Fe\,\textsc{ii}.
    \textbf{Bottom:} Zoom into the median composite spectra to show the structure of three of the most prominent metal lines. For display purposes we have shifted the lines of different EWs by a small amount in the vertical direction in all panels.}
    \label{fig:fe_abs_median_spec}
\end{figure*}

\subsection{Properties of Absorbers in Stacked Spectra}

We stack the residual spectra of quasars in the rest frame of detected Mg\,\textsc{ii} absorbers to study features which become visible in these composite spectra. The top panel of Figure~\ref{fig:fe_abs_median_spec} shows a composite median spectrum, stacking on all Mg\,\textsc{ii} absorbers. We see several weak metal lines such as Si\,\textsc{ii}, C\,\textsc{iv}, Al\,\textsc{ii}, Al\,\textsc{iii}, Fe\,\textsc{ii} (several lines) and Mg\,\textsc{i}. The clear detection of these weak lines provides strong evidence that the majority of our absorbers are genuine. We have divided our sample into five $EW_{\rm 2796}$ bins to understand the corresponding variation of the strength of metal absorbers, which increases with increasing $EW_{\rm 2796}$. 

In spectral regions without absorption features the median flux residual is flat, indicating that our continuum estimation pipeline works well. In the bottom panel Figure~\ref{fig:fe_abs_median_spec} we zoom into the profiles of three prominent metal lines, C\,\textsc{iv} $\lambda\lambda$1548,1550 (left), Fe\,\textsc{ii} $\lambda\lambda$2586, 2600 (middle), Mg\,\textsc{ii} $\lambda\lambda$2796, 2803 (right). The absorption features are well represented by Gaussian profiles, as expected in a stack due to the central limit theorem. As we move to strong absorber systems the lines become saturated (doublet ratio $\sim$1) and doublet profiles overlap, particularly for C\,\textsc{iv}. For the strongest $\rm{EW}_{2796}$ bin noise begins to be visible in the residual due to the low statistics. As a check we also stack the spectra of all quasars with Mg\,\textsc{ii} absorbers without Fe\,\textsc{ii} confirmation (not shown). In this case we also detect the same ensemble of weaker metal transitions.

\begin{figure*}
    \centering
    \includegraphics[width=\linewidth, height=2.7in]{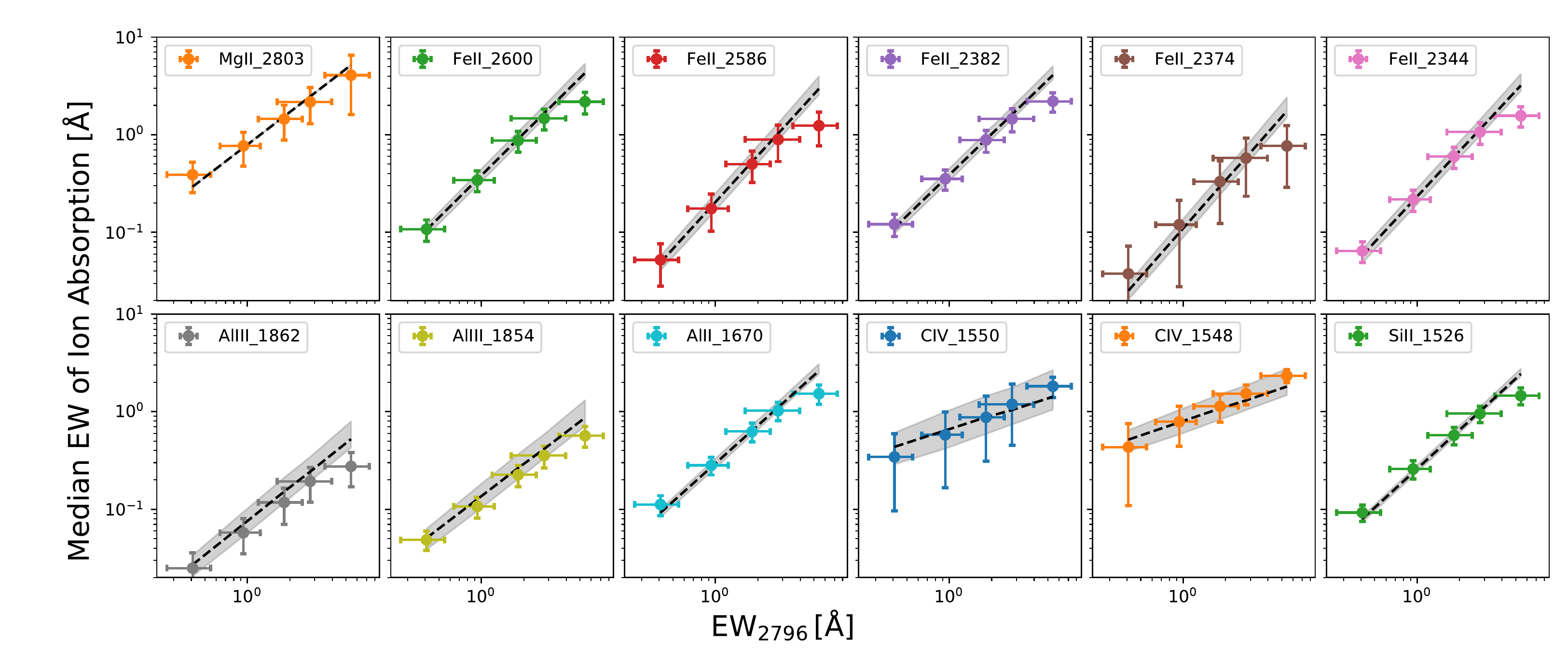}
    \caption{Rest equivalent width of different metal transitions as a function of rest equivalent width of Mg\,\textsc{ii} 2796 line for the median composite spectra. The solid circles show the measured rest EWs (rEWs) and dashed solid lines show the empirical rEWs scaling as a function of $\rm EW_{MgII,\, 2796}$ from \citet{lanfu17}. To compare we use the median redshift of absorbers in each rEW (2796) bin. The shaded region shows the corresponding 5th and 95th percentiles of redshifts in each rEW (2796) bin. We have estimated the rest equivalent widths by fitting suitable gaussian profiles and errors using the bootstrap method.}
    \label{fig:med_ew_stack}
\end{figure*}

In Figure~\ref{fig:med_ew_stack} we show the rest equivalent widths of visible metal lines in the composite spectra as a function of the total rest equivalent width of Mg\,\textsc{ii}. We see that Fe\,\textsc{ii} and C\,\textsc{iv} lines are the second most prominent features though the absorption strength of Fe\,\textsc{ii} is slightly higher compared to C\,\textsc{iv} lines. This is because Fe\,\textsc{ii} is a stronger transition and also at longer wavelength, Fe\,\textsc{ii} can be detected in SDSS quasar spectra with redshifts as low as $z\sim 0.5$, while C\,\textsc{iv} absorbers can only be detected in quasars with $z_{\rm QSO}\gtrsim1.5$ due to the wavelength coverage of SDSS. We note that one should limit the stacks to similar redshift path-lengths in order to properly quantify the relative scaling between the lines.

In order to investigate whether our pipeline estimates the consistent rest equivalent widths, we also make a comparison of metal equivalent widths against the empirical scaling relations (as a function of $\rm EW_{MgII,\, 2796}$ and redshift) derived in \citet{lanfu17}. We observe that our pipeline yields a consistent trend, though there is a slight discrepancy in the strongest $\rm EW_{MgII,\, 2796}$ bin, possibly due to low statistics. We also note that a direct comparison is not straightforward as the scaling also depends on the redshift of the absorbers. For the current comparison we take the median redshift of absorbers in each $\rm EW_{MgII,\, 2796}$ bin and this could also be a possible source of the slight discrepancy in some cases.

\subsection{Completeness of Detection Algorithm}\label{completeness}

\begin{figure*}
    \centering
    \includegraphics[width=5.35in]{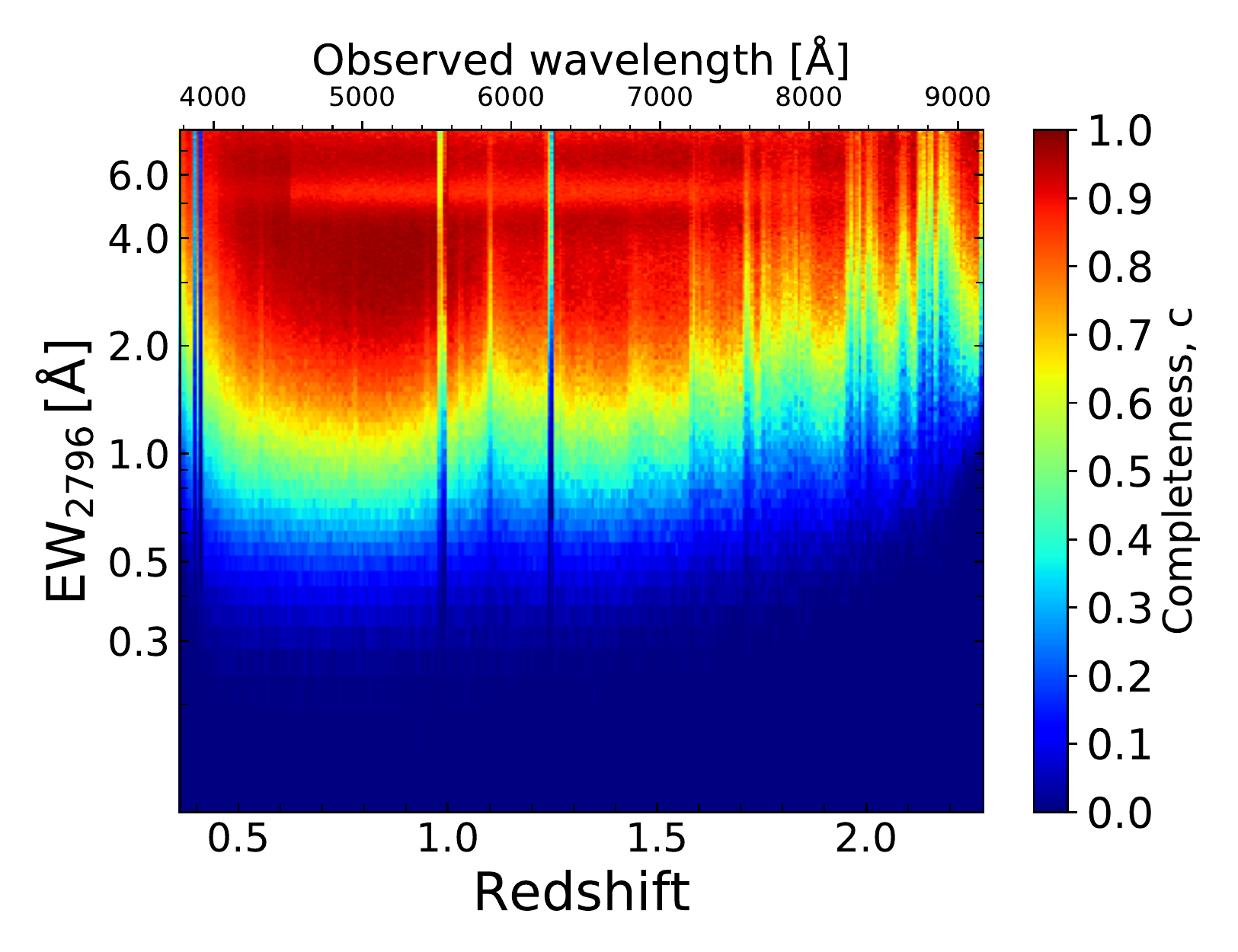} \\
    \includegraphics[width=3.4in]{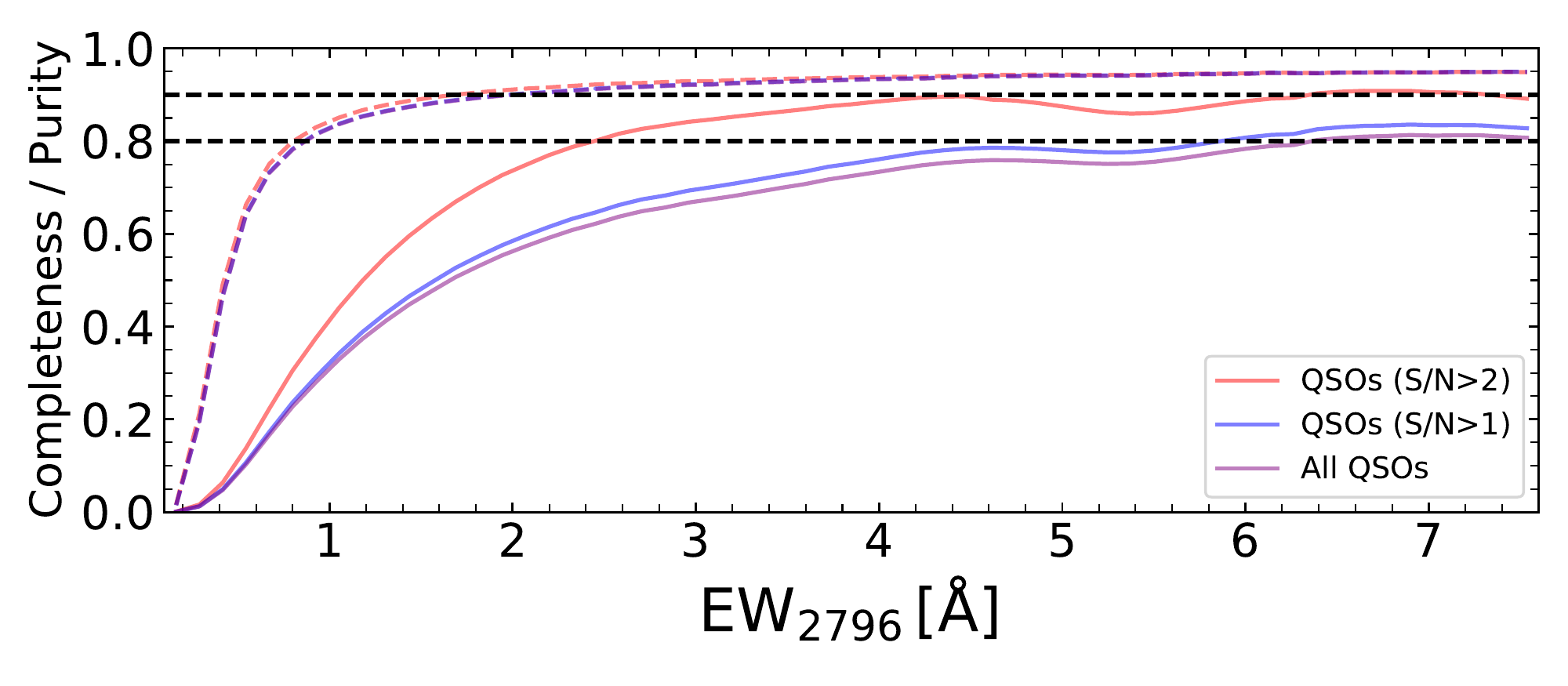}
    \includegraphics[width=3.4in]{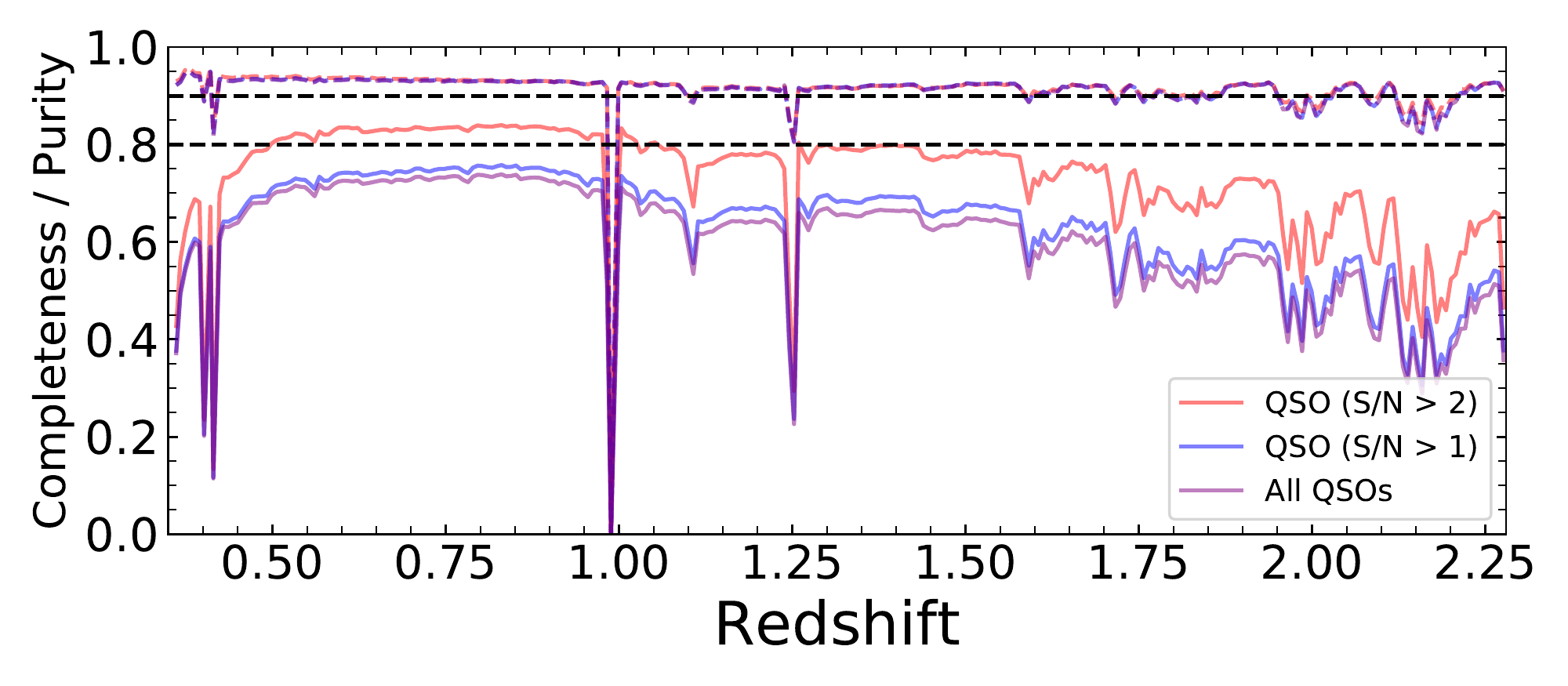}
    \caption{Completeness function $c({EW}_{\rm 2796},z)$. \textbf{Top:} 2D distribution as a function of rest equivalent width and redshift. We are showing the 2D completeness only for QSOs with $\rm S/N_{QSO}>2$. Completeness is low on both wavelength edges of SDSS spectra, as the edges are generally noisy. It also drops in regions with many sky lines such as O\textsc{I} $\lambda$ 5577, O\textsc{I} $\lambda$ 6300 and OH. The dip at $\sim$ 5900 \,\AA\, is due to sodium dichoric, while CaII $\lambda\lambda 3934,\, 3969$ is also present. \textbf{Bottom left}: as a function of $\rm{EW}_{\rm 2796}$ averaged over all redshifts. \textbf{Bottom right:} as a function of redshift smoothed over all $\rm{EW}_{\rm 2796}$. Weak and intermediate absorbers ($\rm{EW}_{\rm 2796}\leq 1$ \AA) have lower completeness, as expected. In both panels solid and dashed curves represent completeness and purity respectively. The dashed horizontal lines show 0.9 and 0.8 on vertical axis. The overall purity and completeness are very high for our detection pipeline for QSOs with $\rm S/N_{QSO}>2$.}
    \label{fig:completeness}
\end{figure*}

The search for absorption features in the relatively low S/N quasar spectra from SDSS is challenging, and an accurate characterization of our ability to recover absorbers of different strengths is required. To estimate the completeness of our detection pipeline we therefore implement a Monte Carlo simulation approach. We generate doublet profiles that mimic true absorber and insert these at a random location in a \textit{real} residual chosen randomly from the set of DR16 QSOs. While inserting we take care of all the masks that we have defined in our detection algorithm. To generate doublets we uniformly sample $\rm{EW}_{\rm 2796}$ from $\rm 0\leq EW_{2796}\leq 8$ \AA. We select both a doublet ratio ($\rm 0.25<\text{doublet ratio}<4.5$) and width ($0.34<\sigma<3.5$ \AA\,) from gaussian distributions. We then run our detection pipeline on the spectrum with a fake absorber and check for a successful detection. We define success as having satisfied the \textit{Mg\,\textsc{ii} doublet criteria} and recovered a sufficiently accurate rest equivalent width, $\rm |EW_{in}-EW_{out}|/EW_{in}<1$, where $\rm EW_{in}$ is the injected $\rm EW_{2796}$ and $\rm EW_{out}$ is the measured $\rm EW_{2796}$. In total, we have simulated over $\sim$ 33 million fake absorbers in $\sim 1$ million DR16 quasars. Since our completeness corrections are based on the final \textit{Mg \textsc{ii} doublet criteria}, this does not bias results in the galaxy section.

Finally, for absorbers in a given ${EW}_{\rm 2796}$ and redshift bin, we estimate the completeness, $c({EW}_{\rm 2796}, z)$ as the ratio of \textit{detected} absorbers to the \textit{injected} absorbers. Using this estimate we compute, for each absorber with given $\rm{EW}_{\rm 2796}$ and redshift, a completeness corrected effective `number of absorbers' $N_{\rm eff} = 1/c({EW}_{\rm 2796},z)$, the count of absorbers had our detection method been perfect. By definition, $c \leq 1$ and so $N_{\rm eff} \geq 1$. Similarly we estimate the purity as the ratio of detected absorbers to legitimate absorbers (i.e. all the absorbers that pass the Mg \textsc{ii} doublet criteria) found by our pipeline. We show the 2D distribution of completeness in ${EW}_{\rm 2796} - z$ space in the top panel of Figure~\ref{fig:completeness}.

In the bottom left panel of Figure~\ref{fig:completeness} we show the completeness (and purity) as a function of $\rm{EW}_{2796}$ averaged over all redshifts. The 2D distribution of purity is shown in Figure~\ref{fig:purity}. As expected, the completeness is higher for strong absorbers and lower for weak absorbers. The bottom right panel shows the completeness (and purity) as a function of $z$ averaged over all $EW_{2796}$, where dips at specific redshifts reflecting excluded wavelength search windows. Overall, the purity of our algorithm (shown in Figure~\ref{fig:purity}) is extremely high, at the cost of moderate completeness, however, completeness goes up significantly for QSOs with high S/N data as shown in the bottom left panel of Figure~\ref{fig:completeness}. Because the eBOSS spectra have relatively lower S/N \citep{dawson16}, we also include here completeness curves restricted to spectra above minimum S/N thresholds (i.e. $\rm S/N>1, 2$). As expected we find that it is difficult to detect absorbers primarily in noisy spectra, while our algorithm has excellent performance on high S/N data.

\begin{figure}
    \centering
    \includegraphics[width=0.95\columnwidth]{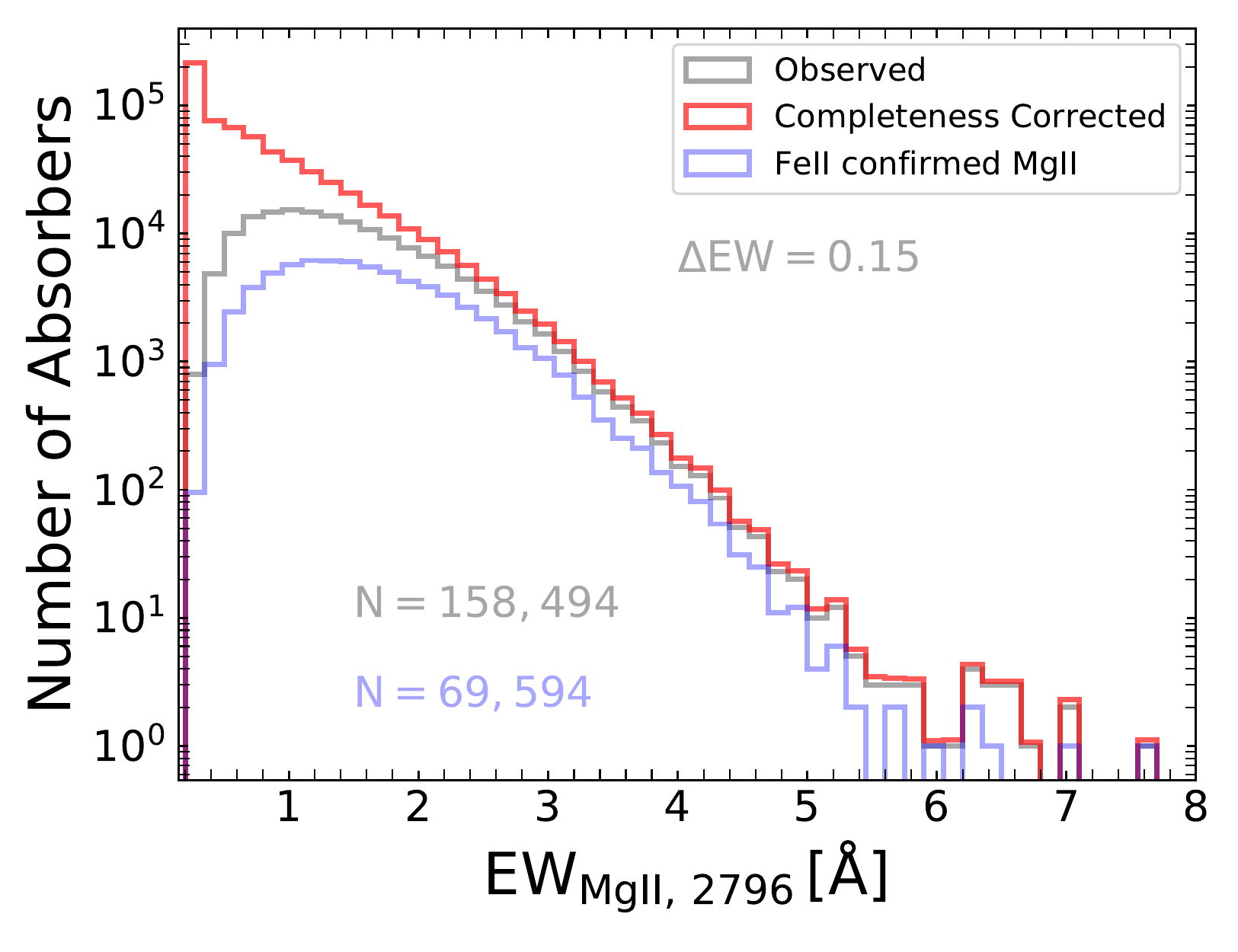}
    \caption{1D histogram of intrinsic observed (gray) and completeness corrected (red) of $\rm EW_{2796}$ for all Mg\,\textsc{ii} absorbers (associated with QSOs having $\rm S/N_{QSO}>2$), as well as the Fe\,\textsc{ii} confirmed subset (blue). The completeness corrected distribution follows an exponential profile in rest equivalent width, such that the weakest absorbers are also the most frequent. Our Mg\,\textsc{ii} catalogue spans 0.2 \AA $\lesssim \rm EW_{2796, MgII} \lesssim 8$ \AA.}
    \label{fig:ew_histogram}
\end{figure}

The observed and completeness corrected $\rm EW_{2796}$ distributions for all Mg\,\textsc{ii} absorbers (that are associated with QSOs having $\rm S/N_{QSO}>2$, i.e. our ``fiducial'' sample, defined explicitly in section \ref{fiducial}), as well as for the Fe\,\textsc{ii} confirmed subset, are shown in Figure~\ref{fig:ew_histogram}. The completeness corrected distribution follows an exponential profile, hence the mean value after completeness correction shifts to smaller values as weak absorbers dominate the distribution. The mean $\rm{EW}_{\rm 2796}$ for the observed distribution is $\sim$1.4 \AA\, while for completeness corrected distribution it decreases to $\sim 0.8$ \AA. We also see that the `completeness corrected distribution' is very high in the lowest bin, which may be due to the small number statistics of absorbers and, such a peak is possibly not significant. We observe that the fraction of Fe\,\textsc{ii} confirmed Mg\,\textsc{ii} absorbers for high rest equivalent widths does not reach unity even for very strong absorbers. This is because Fe\,\textsc{ii} is a lower wavelength transition than Mg\,\textsc{ii} and does not fall within the SDSS wavelength coverage for Mg\,\textsc{ii} absorbers at low redshifts. 

\section{Connecting to Galaxies}\label{galaxy_quasar}

With absorbers identified across a wide range of redshifts and over a significant fraction of the sky, we now need to connect them to nearby galaxies. Our main resource to obtain a large sample of galaxies with $0.4<z<1$ are the BOSS and eBOSS programs of SDSS. Namely, emission line galaxies (ELGs) and luminous red galaxies (LRGs), which we use to study the galaxy - Mg\,\textsc{ii} absorber connection. When necessary, we use a Planck-consistent cosmology for our analysis, i.e. $\Omega_{\rm m,\,0}=0.307,\, H_{0}=67.7$\kms\,$\rm Mpc^{-1}$.

\subsection{Fiducial Quasar Sample}\label{fiducial}
As discussed in Section \ref{completeness}, our detection pipeline has excellent performance on high S/N data. Therefore, for the galaxy-centric analysis we select only quasars having $\rm S/N_{QSO}>2$ and Mg \textsc{ii} absorbers associated with them. We make this cut to define the fiducial quasar sample and the corresponding Mg \textsc{ii} absorber catalogue because the completeness and purity both are very high for our detection algorithm and we miss intermediate or strong absorbers rarely. For example, at $EW_{\rm 2796}\sim 2$ \AA\, completeness and purity are $\gtrsim 75$ percent and $\gtrsim 90$ percent respectively as shown in bottom left panel of Figure~\ref{fig:completeness}. The high completeness catalogue also brings more confidence in the results as we are not applying big corrections. As presented in Table \ref{tab:sample}, out of $978,561$ quasars $773,594$ have $\rm S/N_{QSO}>2$, which is $\sim 80\%$ of the total sample size. We show the histogram of redshifts of quasars with $\rm S/N_{QSO}>2$ in Figure~\ref{fig:z_histogram}. We lose quasars uniformly in redshift space. The median $\rm S/N_{QSO}$ of QSOs is $\sim 7.5$. In our Mg \textsc{ii} catalogue we have $\sim1,030$ absorbers associated with quasars having $\rm S/N_{QSO}<2$ and we remove them for consistency. From now onwards we perform all galaxy-centric analysis with this fiducial quasar sample and Mg \textsc{ii} absorber catalogue.

\subsection{Galaxy Samples}

\subsubsection{Emission Line Galaxies}\label{elg}

The main selection criteria for ELGs in SDSS is a cut in the $g-r$ vs $r-z$ colour-colour diagram and $g$ band magnitude \citep{raichoor17}. Their typical stellar mass is $\rm M_{\star}\sim10^{10.5}\,M_{\odot}$ \citep{raichoor17} and they reside in dark matter haloes with $\rm M_{halo}\sim 10^{12.2}\, M_{\odot}$ \citep{favole16}.

The characteristic features of ELG spectra are strong gas emission lines such as [O\,\textsc{ii}] $\lambda3727$ [O\,\textsc{iii}] $\lambda5007$ and H $\beta$, due to high star formation activity. Based on [O\,\textsc{ii}] $\lambda3727$ luminosity and models described in \cite{kennicutt98} the star formation rate in ELGs is estimated to be from 1 to 20 $\rm M_{\odot}\, yr^{-1}$. The presence of hot and young stars make ELGs appear blue.

We select the latest DR16 ELG catalogue compiled by \citet{raichoor21}\footnote{VAC available at: \url{https://data.sdss.org/sas/dr16/eboss/lss/catalogues/DR16/eBOSS_ELG_full_ALLdata-vDR16.fits}}, which contains 269,178 objects. The catalogue includes stellar masses estimated by the FAST spectral fitting code \citep{kriek09}, with the \citet{bruzal03} stellar model and \citet{chabrier03} initial mass function (IMF). We include only ELGs with reliable redshifts (selected by \textsc{IMATCH==1} condition). We also apply cuts on stellar mass and redshift to connect with Mg\,\textsc{ii} absorbers. The final sample contains $188,323$ ELGs at $z>0.4$ and with $\rm 9 < log M_{\star}\, [M_{\odot}]< 12$. The typical measurement error in redshift is $\sim 20$ \kms. The median stellar mass, halo mass and redshift of the ELGs in the final sample are $\rm M_{\star} \sim 10^{10.4}\, M_{\odot}$, $\rm M_{halo} \sim 10^{12.1}\, M_{\odot}$ and $z \sim 0.84$, respectively. The ELG catalogue does not include a stellar mass uncertainty for each galaxy, however the typical uncertainty in stellar mass for ELGs is 0.05 dex and the majority are within 0.25 dex \cite[see section 6.3]{raichoor17}\footnote{Hong Guo, private communication.}. The stellar mass-redshift distribution is shown in the left panel of Figure ~\ref{fig:mass_z_hist}. For this sample we estimate the star-formation rate (SFR) using the scaling relation from \cite{kewley04}.

\begin{equation}
\centering
    \rm SFR\, [M_{\odot}yr^{-1}] = \frac{L_{[O\,\textsc{ii}]}\, [ergs\,s^{-1}]}{1.52 \times 10^{-41}}
    \label{eqn:sfr}
\end{equation}

\noindent where $\rm L_{[O\,\textsc{ii}]}$ is luminosity measured as $\rm L_{[O\,\textsc{ii}]} = F\cdot 4\pi D_{L}^{2}$; $F$ is the measured [O\,\textsc{ii}] $\lambda3727$ flux and $D_{\rm L}$ is the luminosity distance. We note that the measured [O\,\textsc{ii}] $\lambda3727$ flux (F) is not corrected for reddening and the SFR may be underestimated by a factor of a few on average and up to factor of 10 or more for the most massive, star-forming systems. Possible contamination by AGN is also discussed below. Our measured SFRs vary from 1 to 25 $\rm M_{\odot}yr^{-1}$ with a median value of $\sim 7$ $\rm M_{\odot}yr^{-1}$.

\subsubsection{Luminous Red Galaxies}\label{lrg}

The luminous red galaxies in SDSS are observed as described in \citet{padmanabhan12}. The selection of LRGs is based on cuts on SDSS $g, r$, and $i$ band magnitudes. Due to low star formation activity, a large fraction of these galaxies have no emission lines in their spectra. The presence of old stellar populations makes them look optically red. They are typically more massive than ELGs and have $\rm M_{\star}\sim10^{11.2}\, M_{\odot}$ and reside in dark matter haloes with $\rm M_{halo}\sim10^{13.5}\, M_{\odot}$ on average \citep{white11}.

\begin{figure*}
    \centering
    \includegraphics[width=0.46\textwidth]{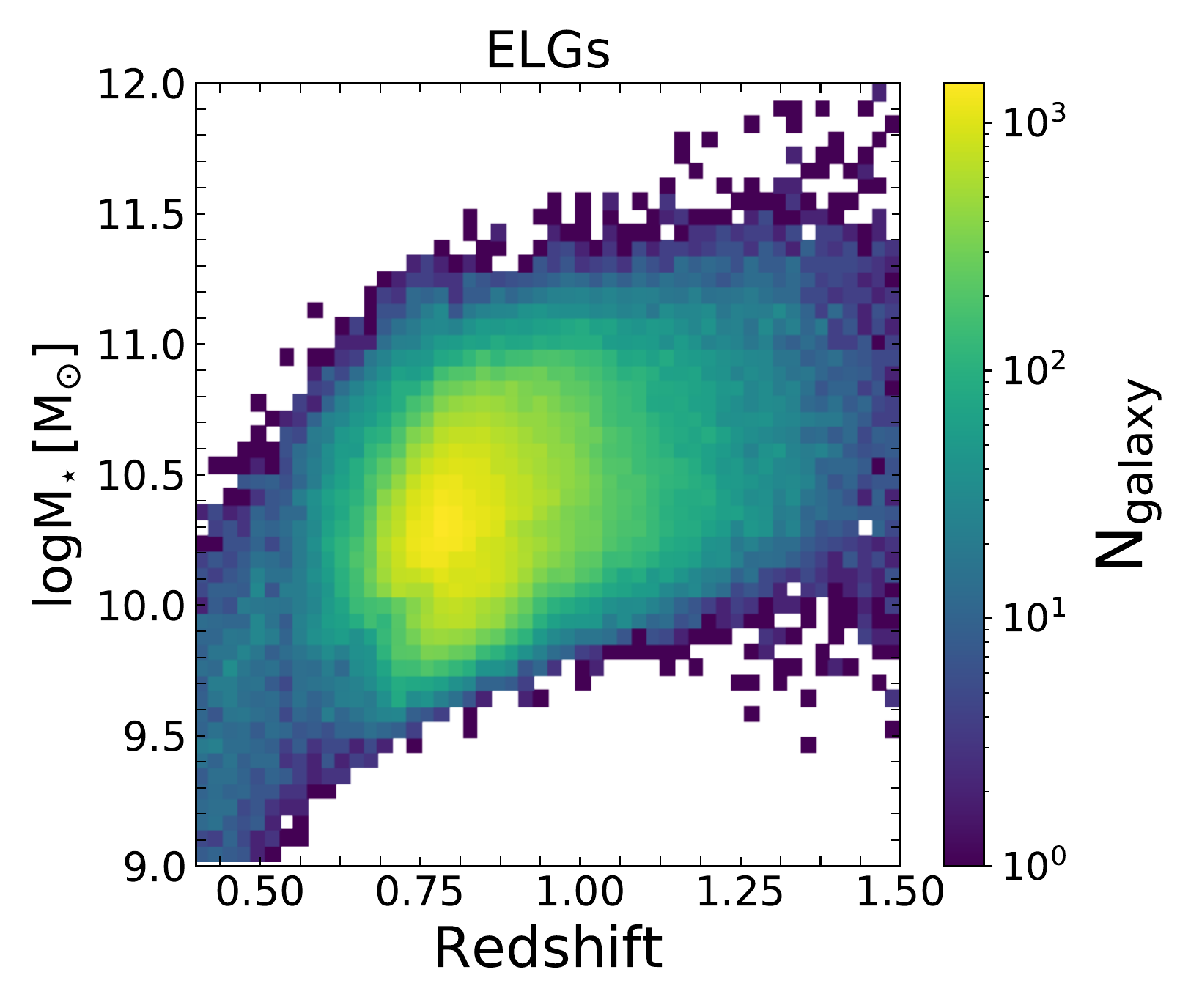}
    \includegraphics[width=0.46\textwidth]{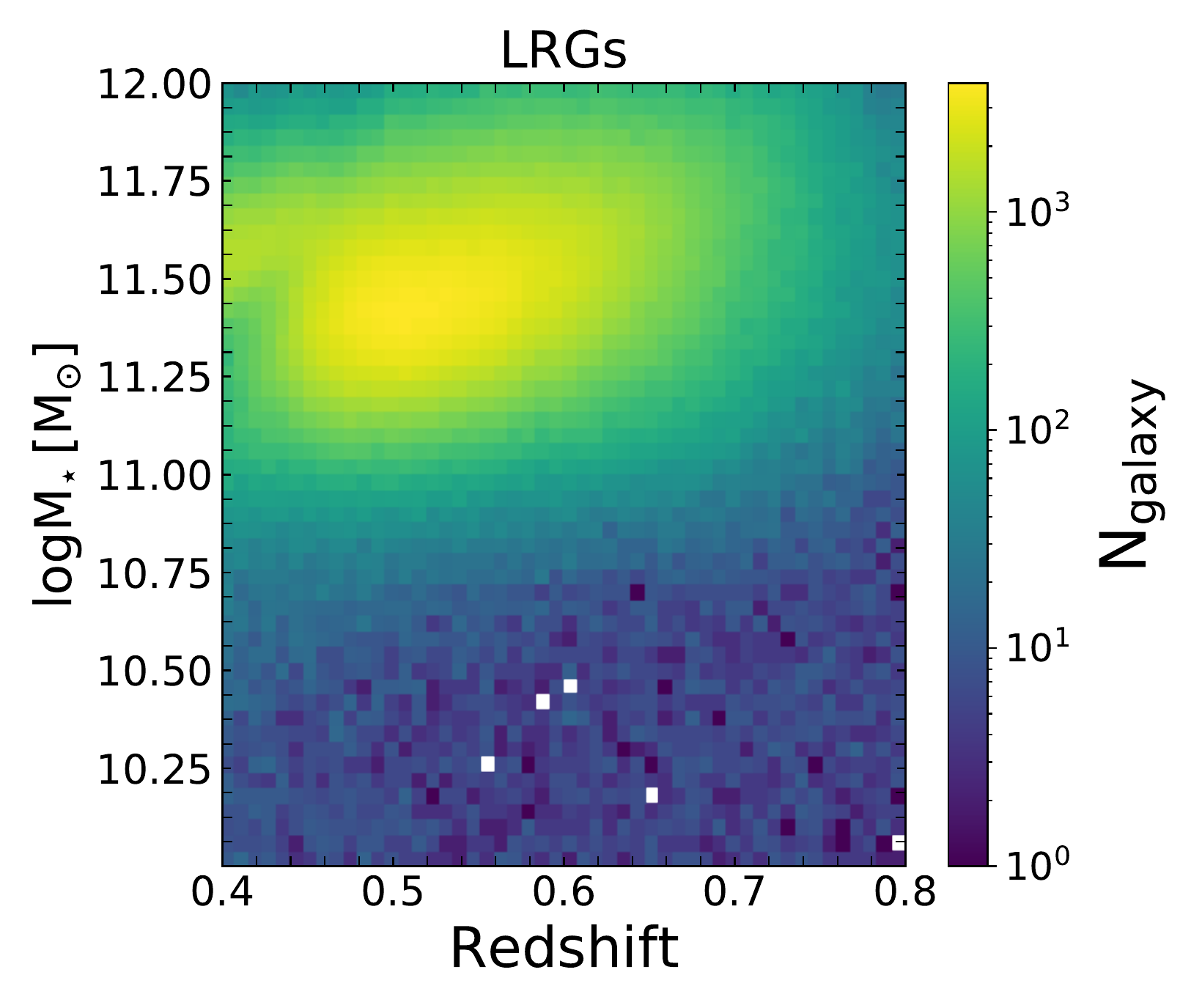}
    \caption{Joint distribution of stellar mass and redshift of all the galaxies used in this work. \textbf{Left:} Distribution of ELGs, where we have selected galaxies with $z>0.4$ and $\rm 9< logM_{\star}\, [M_{\odot}]< 12$. The mean redshift and stellar mass of ELGs are $<z>\sim0.86$ and $\rm <M_{\star}>\sim10^{10.4}\, M_{\odot}$ respectively. \textbf{Right:} Distribution of LRGs, where we have selected galaxies with $z>0.4$ and $\rm 10< logM_{\star}\, [M_{\odot}]< 12$. The mean redshift and stellar mass of LRGs are $<z>\sim0.54$ and $\rm <M_{\star}>\sim10^{11.5}\, M_{\odot}$ respectively.}
    \label{fig:mass_z_hist}
\end{figure*}

For LRGs we take the Wisconsin PCA-based catalogue\footnote{\url{https://www.sdss.org/dr16/spectro/galaxy_wisconsin/}} which contains a total of $\sim 1,489,670$ objects. The stellar masses were estimated using a principal component analysis (PCA) method described in \citet{chenY12}, assuming \citet{kroupa01} IMF. The authors also added 0.057 dex to adjust to a \citet{chabrier03} IMF \citep[see][]{herrmann16}. The stellar libraries and single stellar population models were taken from \citet{maraston11}. The typical measurement error in redshift is $\sim 30$ \kms. For our galaxy-absorber correlation analysis we select LRGs having $z>0.4$ with $\rm 10< log M_{\star}\, [M_{\odot}]< 12$. Our final LRG catalogue contains $\sim 1,081,329$ galaxies and the median stellar mass, median halo mass, and redshift  $\rm M_{\star} \sim 10^{11.5}\, M_{\odot}$,  $\rm M_{halo} \sim 10^{14}\, M_{\odot}$ and $z \sim 0.54$, respectively. The typical uncertainty in stellar mass is 0.16 dex. The stellar mass-redshift distribution is shown in the right panel of Figure ~\ref{fig:mass_z_hist}.

\subsection{Methods}

\subsubsection{Covering Fraction of Mg\,\textsc{ii} absorbers}\label{covering_fac}

We define $\rm N_{gal,\, j}^{abs}|_{\Delta R}^{\Delta z}$ the number of detected absorbers around the j$^{\rm th}$ SDSS galaxy within an annulus $\rm \Delta R$ satisfying a maximum $\Delta z$ separation. Analogously, $\rm N_{gal,\, j}^{QSO}|_{\Delta R}$ is the corresponding number of quasars, i.e. the number of sightlines. The subscript $\Delta R$ denotes an annulus with projected inner and outer radii $\rm R_{1}$ and $\rm R_{2}$, and $|\Delta z| = |z_{\rm galaxy} - z_{\rm abs}|$ is the redshift separation; we adopt $|\Delta z|\leq\rm 0.01$.

Given a galaxy with a position on the sky and redshift $(\alpha,\delta,z)$ we derive the apparent covering fraction of absorbers, defined as the fraction of quasar sightlines, in a given radial bin $\Delta R$, which have one or more absorbers satisfying a chosen $\rm EW_{2796}$ threshold

\begin{ceqn}
\begin{equation} \label{eqn:fc_def}
  f'_{\rm c}|_{\Delta R} = \rm \frac{N(sightlines\, with\, absorbers)}{N(QSO\, sightlines)}
  = \frac{ \sum_{j} N_{gal,\,j}^{abs}|_{\Delta R}^{\Delta z} }{ \sum_{j} N_{gal,\,j}^{QSO}|_{\Delta R} } .
\end{equation}
\end{ceqn}

\noindent where $j$ varies over all galaxies in given galaxy sample. Using our estimated completeness as a function of absorber rest equivalent width and redshift we then derive the true covering fraction $f_c$ by correcting the apparent covering fraction $f'_c$ as $f_c = f'_c / c({\rm{EW}_{\rm 2796}},z)$. 

In addition to this differential covering fraction, we also estimate cumulative covering fractions. For a given projected distance $D < D_{\rm proj}$ we count all absorber-galaxy pairs and QSO-galaxy pairs up to that distance, applying the same definition as above.

We measure the differential (cumulative) mean absorption strength per absorber around a given galaxy sample, in (up to) a given radial bin, as the sum of rest equivalent widths of all absorbers (weighted by their corresponding completeness corrected number), normalized by the sum of Mg\,\textsc{ii} absorbers (completeness corrected) in (up to) that bin. This quantity is the average total (both lines of the doublet) rest equivalent width per absorber. 

\subsubsection{Random Galaxy Samples}

To better understand the bias and clustering of Mg\,\textsc{ii} absorbers around galaxies, we compare to the expected average absorption signals for random sky sightlines. To do so, we define 40 random galaxy samples, each equal in size to our fiducial galaxy sample. Starting from true galaxies, we shuffle sky positions as well as redshifts. This shuffling de-correlates the positions and redshifts of the galaxies while preserving the original redshift distribution and sky coverage. We compute all observational measurements for both the true and random galaxy samples.

\subsubsection{Bootstrap Error Estimates}

To estimate errors we use the bootstrap approach, repeating the absorber/QSO-galaxy pair identification procedure 100 times. In each iteration we select galaxies from the original sample at random, with repetition, to account for Poisson statistics. We take the mean over these 100 iterations to estimate each quantity and take the error as the standard deviation of these samples.

\subsubsection{Halo Mass and Virial Radius}

We perform analysis requiring two unobservable quantities: halo mass and virial radius. To estimate halo mass for a given stellar mass and redshift we make use of the Stellar Mass-Halo Mass (SMHM) model developed in \citet{behroozi10}. Given halo mass, the virial radius $\rm r_{200}$ of the galactic halo is found using the $\Delta_{c}\sim200$ definition of \citet{brayan98}

\begin{equation}
r_{200} = r_{vir} = 211.83 \, M_{12}^{1/3}\,[\Omega_{m,0}(1+z)^{3}\,+\,\Omega_{\Lambda, 0}]^{-1/3}\, [\rm kpc]
\label{eqn:vir_radius}
\end{equation}

\noindent where $\rm M_{12} = M_{halo}/10^{12}\, [M_{\odot}]$. Note that adopting other SMHM relations would lead to different halo mass and radii estimates.

\begin{figure}
    \centering
    \includegraphics[width=\columnwidth]{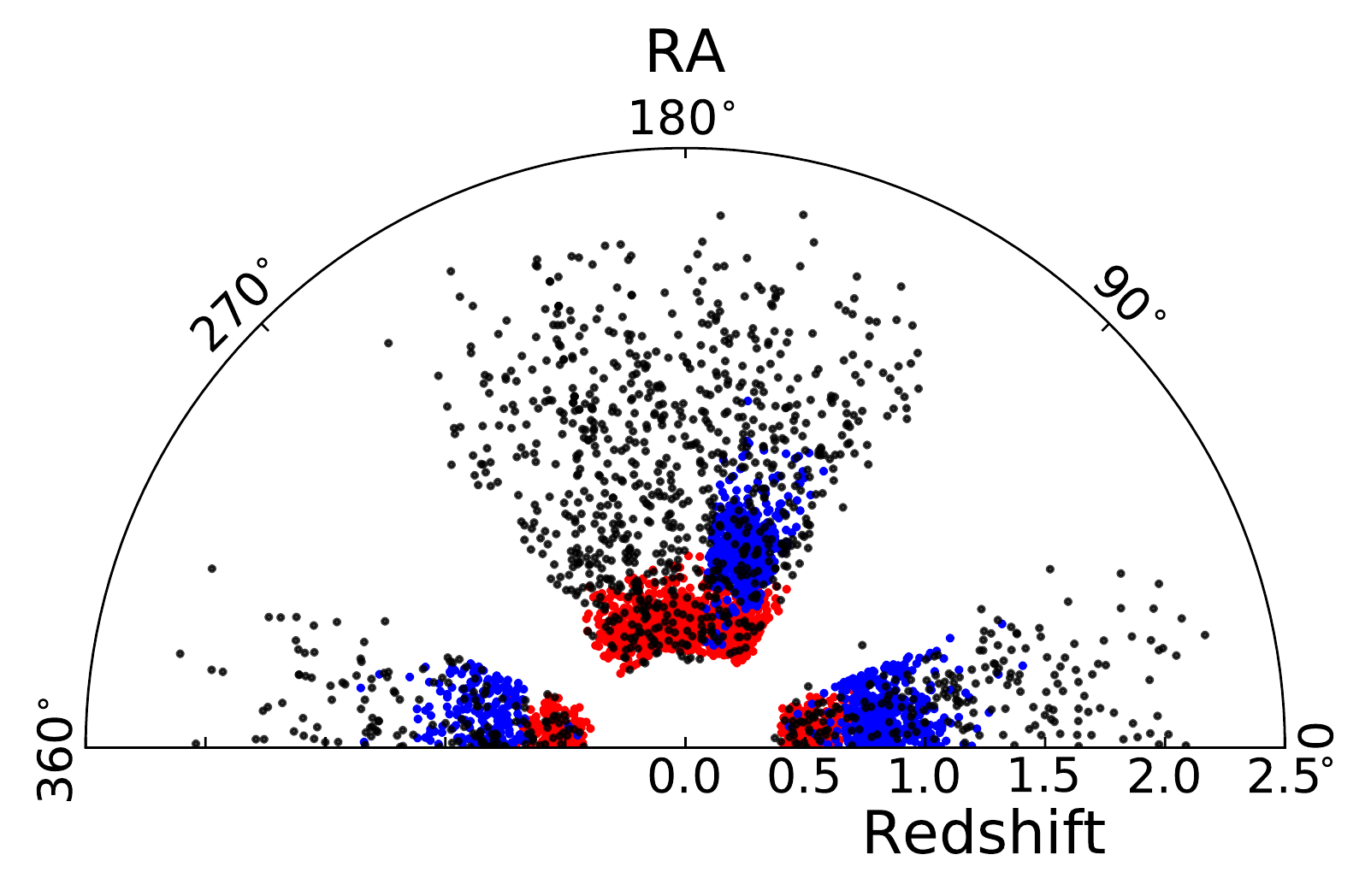}
    \caption{Wedge map of DR16 Mg\,\textsc{ii} absorbers, and the two galaxy samples in RA-$z$ space. A subsample of $\rm N = 1000$ objects is selected randomly to show the distribution of galaxies and absorbers on the sky. The black circles indicate the Mg\,\textsc{ii} absorbers, blue circles indicate the ELGs and red circles indicate the positions and redshifts of LRGs.}
    \label{fig:sky_map}
\end{figure}

\subsection{Galaxy-Absorber Correlation}\label{gal_abs_corr}

Figure~\ref{fig:sky_map} shows the positions and redshifts of 1000 objects chosen randomly from the galaxy samples and our DR16 Mg\,\textsc{ii} absorber catalogue. The black dots indicate the positions of detected Mg\,\textsc{ii} absorbers and blue and red dots denote ELGs and LRGs, respectively. ELGs clearly extend out to $z\sim1$ compared to LRGs that extend to $z\sim 0.7$. The map also shows that the Mg\,\textsc{ii} absorber distribution is quite uniform in RA-$z$ space and that many absorbers fall within the volume of the galaxy samples.

\begin{figure}
    \centering
    \includegraphics[width=0.9\columnwidth]{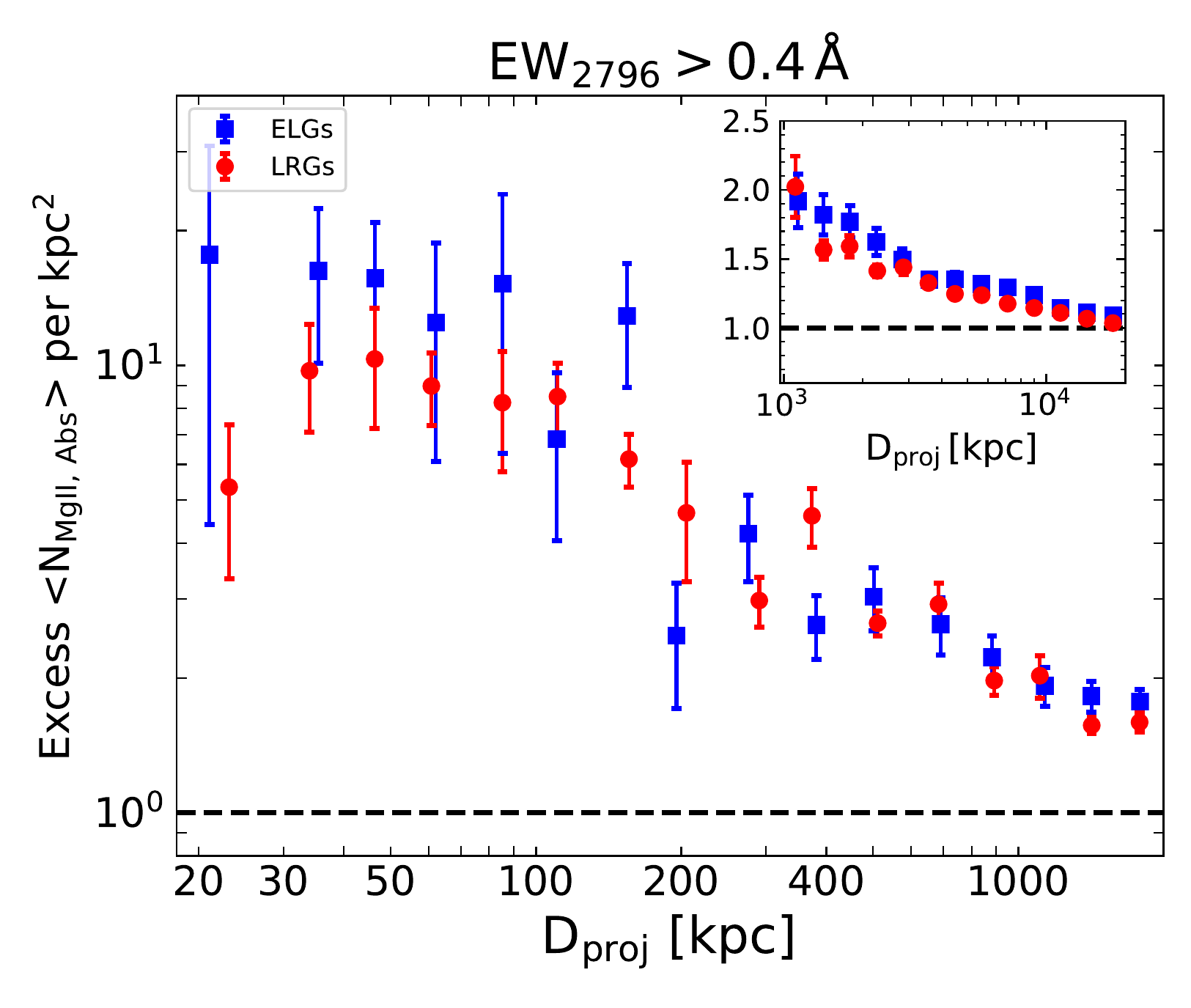}
    \includegraphics[width=0.9\columnwidth]{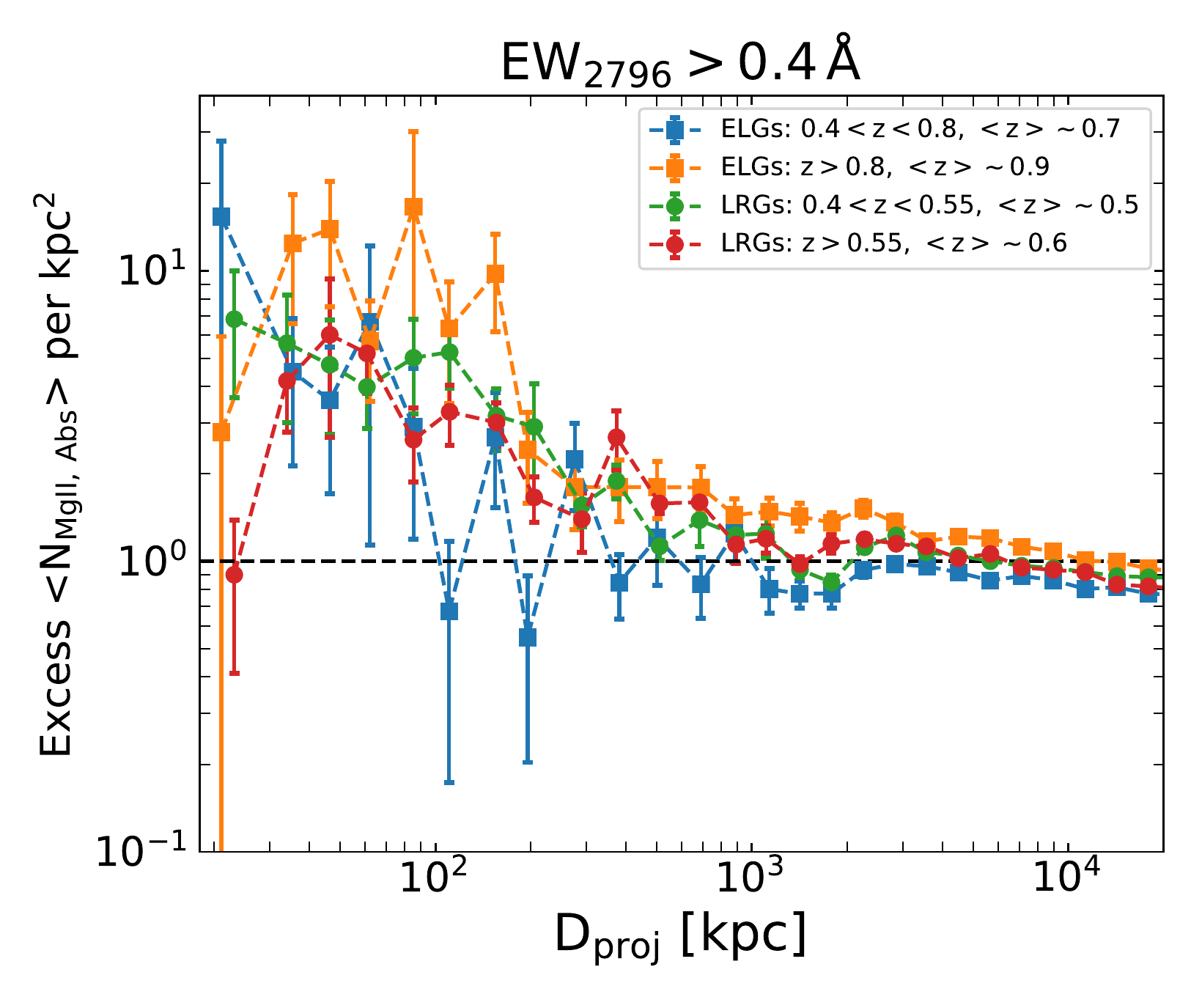}
    \caption{Excess mean surface density of absorbers ($d_{1}<D_{\rm proj}<d_{2}$): the mean number of absorbers per galaxy per $\rm kpc^{2}$ for Mg\,\textsc{ii} absorbers with $\rm{EW}_{\rm 2796}>0.4$ \AA\ divided by the corresponding value around the random galaxy sample. \textbf{Top:} The blue and red squares denote ELGs and LRGs, respectively. The corresponding dashed black line denotes unity, i.e. the expectation for random sightlines. The inset shows that the values converge to their random expectation only at $D_{\rm proj} \gtrsim 10\, \rm Mpc$. \textbf{Bottom:} Excess mean surface density of absorbers around ELGs and LRGs in two redshift bins. The surface density of absorbers increases towards higher redshift at larger distances.}    
    \label{fig:norm_surf_dens}
\end{figure}

We begin with the \textit{excess mean number of Mg\,\textsc{ii} absorbers per $kpc^{2}$} (hereafter, excess mean surface density), i.e. the surface density around true galaxies divided by the surface density around the random galaxy sample, as a function of projected distance ($D_{\rm proj}$). Figure~\ref{fig:norm_surf_dens} shows the result for Mg\,\textsc{ii} absorbers with $\rm{EW}_{\rm 2796}>0.4$ \AA. For other EW bins the trends are similar, with mean surface density decreasing for stronger absorbers, due to their relative scarcity. 

For ELGs the excess surface density rises to a maximum at $\sim 50-80$ kpc and then declines with projected distance, however, for LRGs, the trend is consistent with decreasing as a function of distance. The decreasing trend of Mg\,\textsc{ii} surface density with projected distance around LRGs was also identified with a different stacking based methodology \citep{zhu14}, as well as with individual galaxy-absorber pairs in the COS-LRG survey \citep{zahedy19}. We also see an enhancement of Mg\,\textsc{ii} absorbers around ELGs relative to LRGs below $ D_{\rm proj}\sim 100$ kpc, by a factor of $2-4$. For these two samples the mean surface density is consistent within the error bars beyond $D_{\rm proj}\sim 100$ kpc.

The excess mean surface density of absorbers converges to the random expectation (unity) only for $D_{\rm proj}\gtrsim 15\, \rm Mpc$ for both LRGs, and ELGs, as shown in the inset. To understand how excess surface density depends on redshift of galaxies we further divided ELGs and LRGs into redshift bins and estimate the excess surface density around each. As shown in the bottom panel of Figure~\ref{fig:norm_surf_dens} we find that within a given sample the excess surface density is slightly higher for low-$z$ galaxies below $D_{\rm proj}\sim 100$ kpc. They all converge to random values at $D_{\rm proj}\gtrsim 10$ Mpc.

\begin{figure}
    \centering
    \includegraphics[width=0.9\columnwidth]{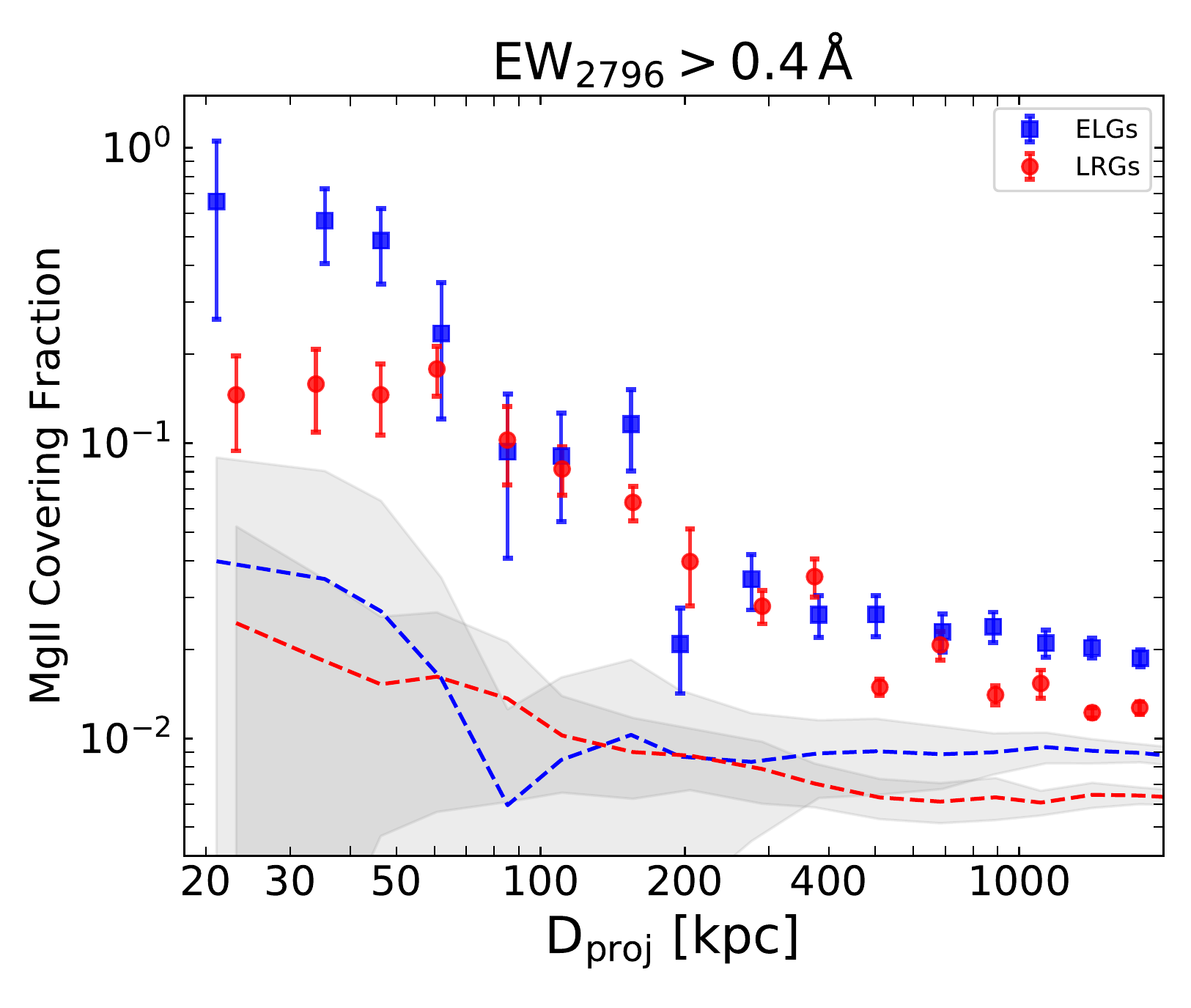}
    \includegraphics[width=0.9\columnwidth]{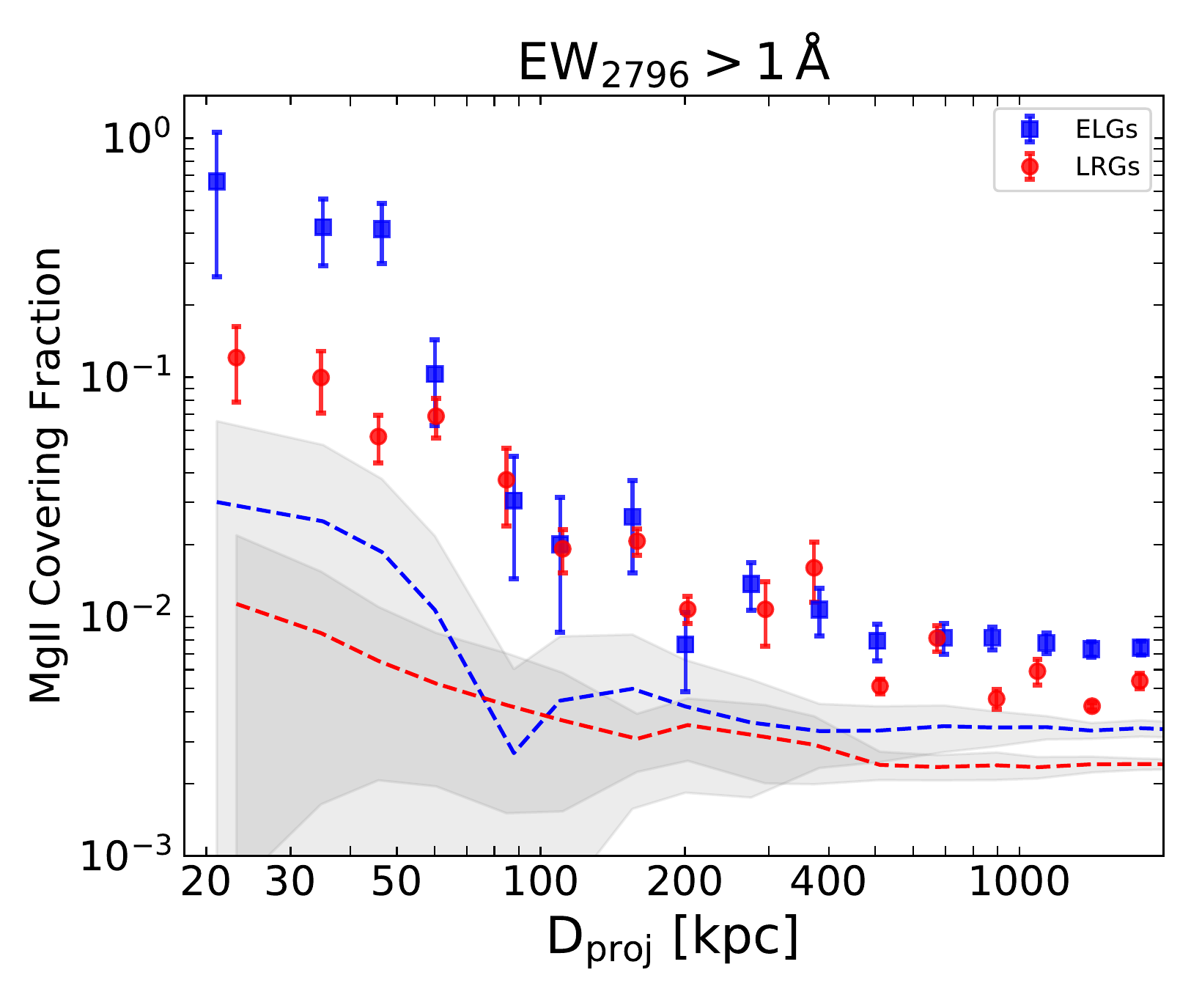}
    \includegraphics[width=0.9\columnwidth]{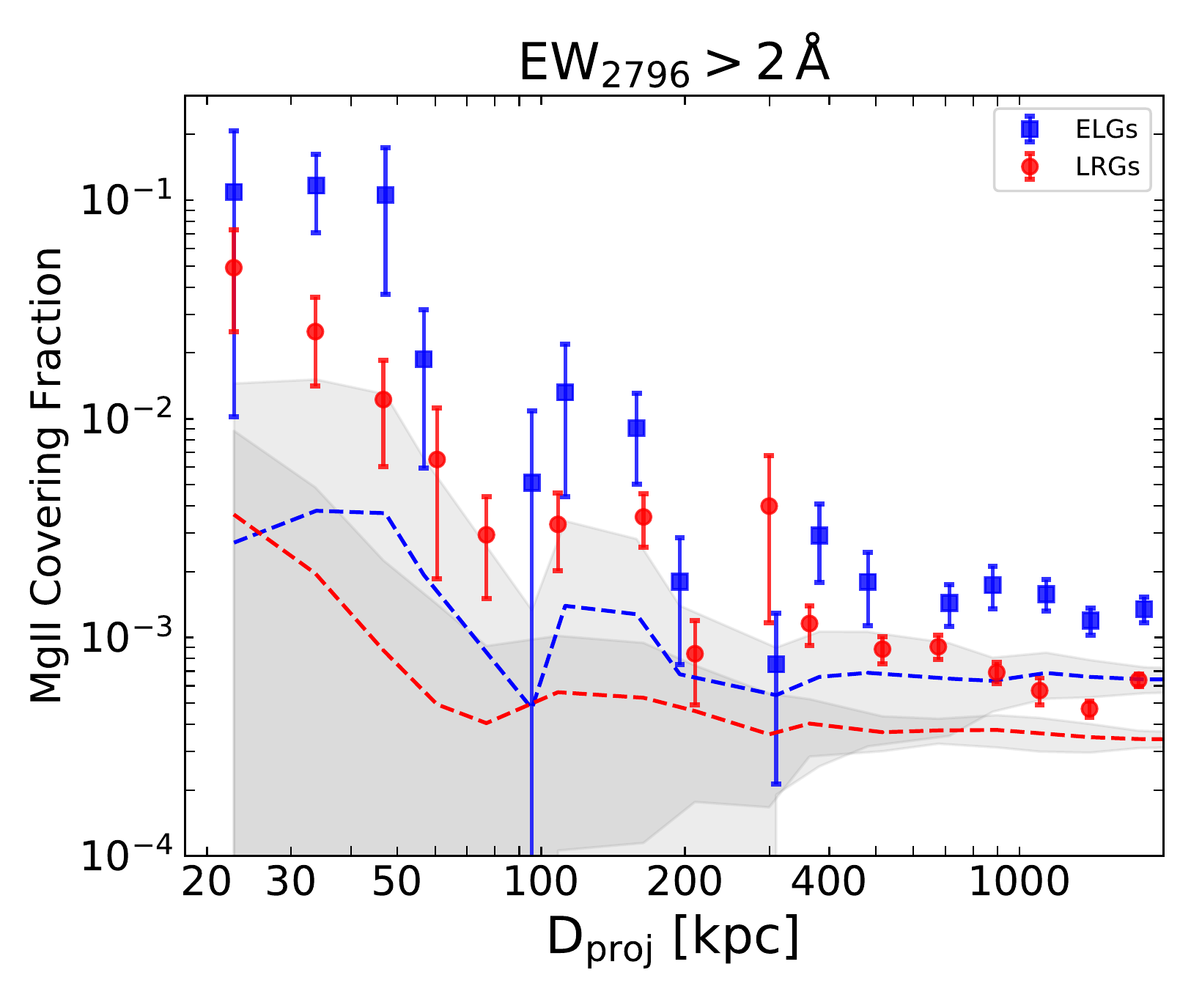}
    \caption{Differential covering fraction of Mg\,\textsc{ii} absorbers in three $\rm{EW}_{\rm 2796}$ bins (shown at the top of each panel). The blue and red squares denote ELGs and LRGs, respectively. The corresponding dashed color lines show the covering fractions expected around the random samples. The ELG sample has a pronounced excess relative to the LRGs within $D_{\rm proj} \lesssim 50$ kpc.}
    \label{fig:fc_mgii_gals}
\end{figure}

Next, we estimate the covering fraction, $f_{c}(d_{1}<D_{\rm proj}<d_{2})$, of Mg\,\textsc{ii} absorbers around galaxies. We show the variation of $f_{c}$ as a function of projected distance in Figure~\ref{fig:fc_mgii_gals}. The covering fraction decreases with projected distance and varies strongly with galaxy type. In each $\rm{EW}_{\rm 2796}$ bin ELGs have 2-5 times higher covering fraction than LRGs below $D_{\rm proj}<50$ kpc. For example, in the $\rm{EW}_{\rm 2796}>0.4$ \AA\, bin, ELGs have a Mg\,\textsc{ii} covering fraction of $f_{c}\gtrsim 50-70$ percent compared to $f_{c} \lesssim 15$ percent for LRGs. At large distances ($D_{\rm proj}>100$ kpc) there is no significant variation with galaxy type. Several previous studies have reported similar trends using different samples and analysis \citep{lan14, nielsen13,lovegrove11}. The covering fraction converges to the random expectation for $D_{\rm proj}\gtrsim 10$ Mpc. 

In addition to the trend with galaxy type, the covering fraction also depends strongly with the strength of absorbers (note the vertical limits vary with each panel). There is a clear anti correlation between covering fraction and absorber strength. This can be attributed to the rarity of strong absorbers around galaxies, such that absorption with $EW_{\rm 2796}>2$ \AA\, is roughly an order of magnitude less frequent in the inner halo. Particularly for LRGs, however, there are low number statistics and the measurement errors are high. Nonetheless, the strong enhancement of $f_c$ seen within 50 kpc is clearly present for all rest equivalent widths. As we discuss below, this small distance component implies an association to the ongoing star formation activity of ELGs and the resultant galactic-scale outflows.

\begin{figure}
    \centering
    \includegraphics[width=0.9\columnwidth]{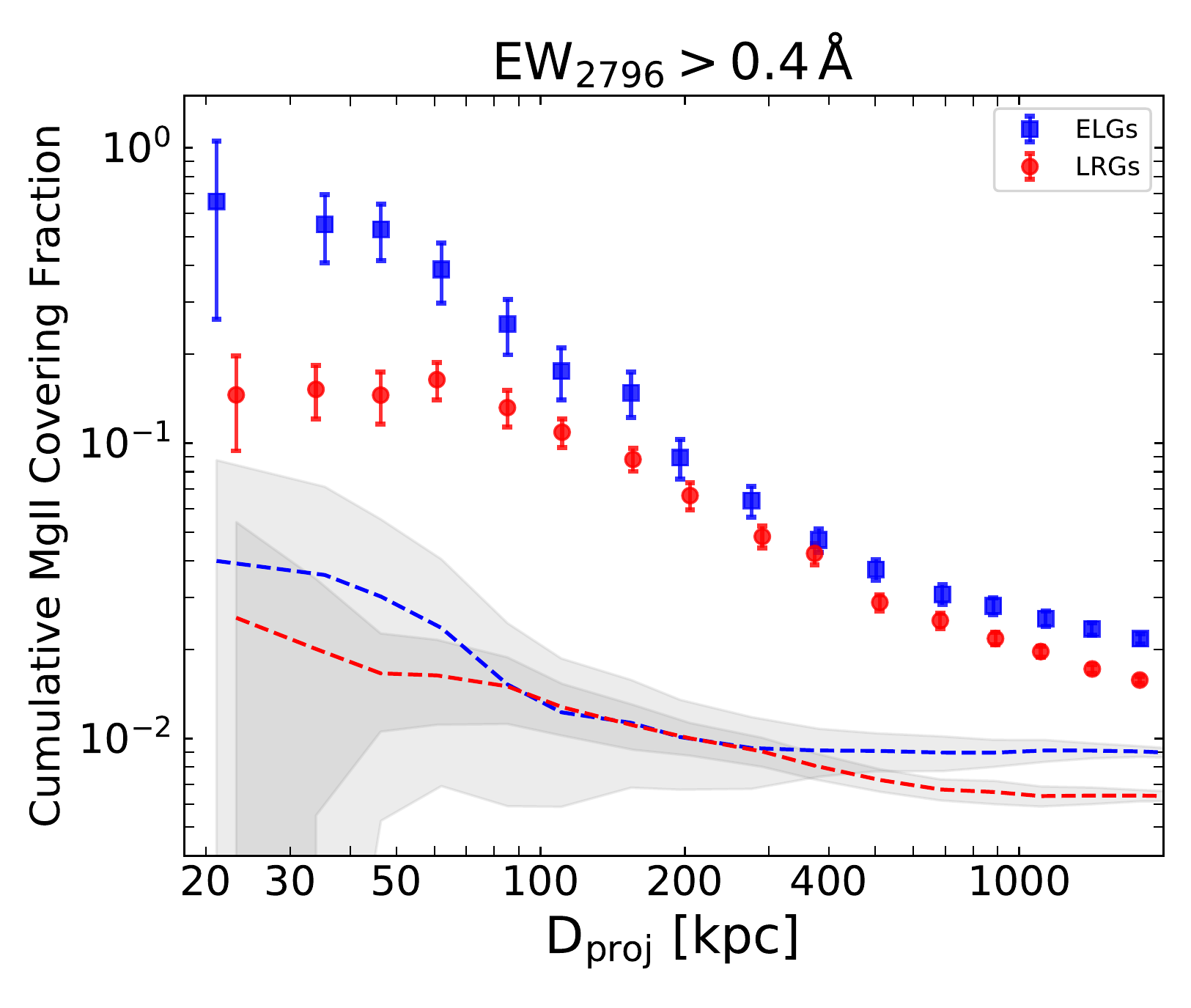}
    \caption{Cumulative covering fraction of Mg\,\textsc{ii} absorbers ($d < D_{\rm proj}$) $\rm EW_{2796}>0.4$ \AA\, bin around star-forming ELGs (blue) and passive LRGs (red) as a function of projected distance. The corresponding dashed color lines show cumulative covering fractions estimated using the random samples. Excess absorption around ELGs is visible out to $D_{\rm proj}\sim 200$ kpc.}
    \label{fig:cumu_fc_mgii}
\end{figure}

\begin{figure}
    \centering
    \includegraphics[width=0.9\columnwidth]{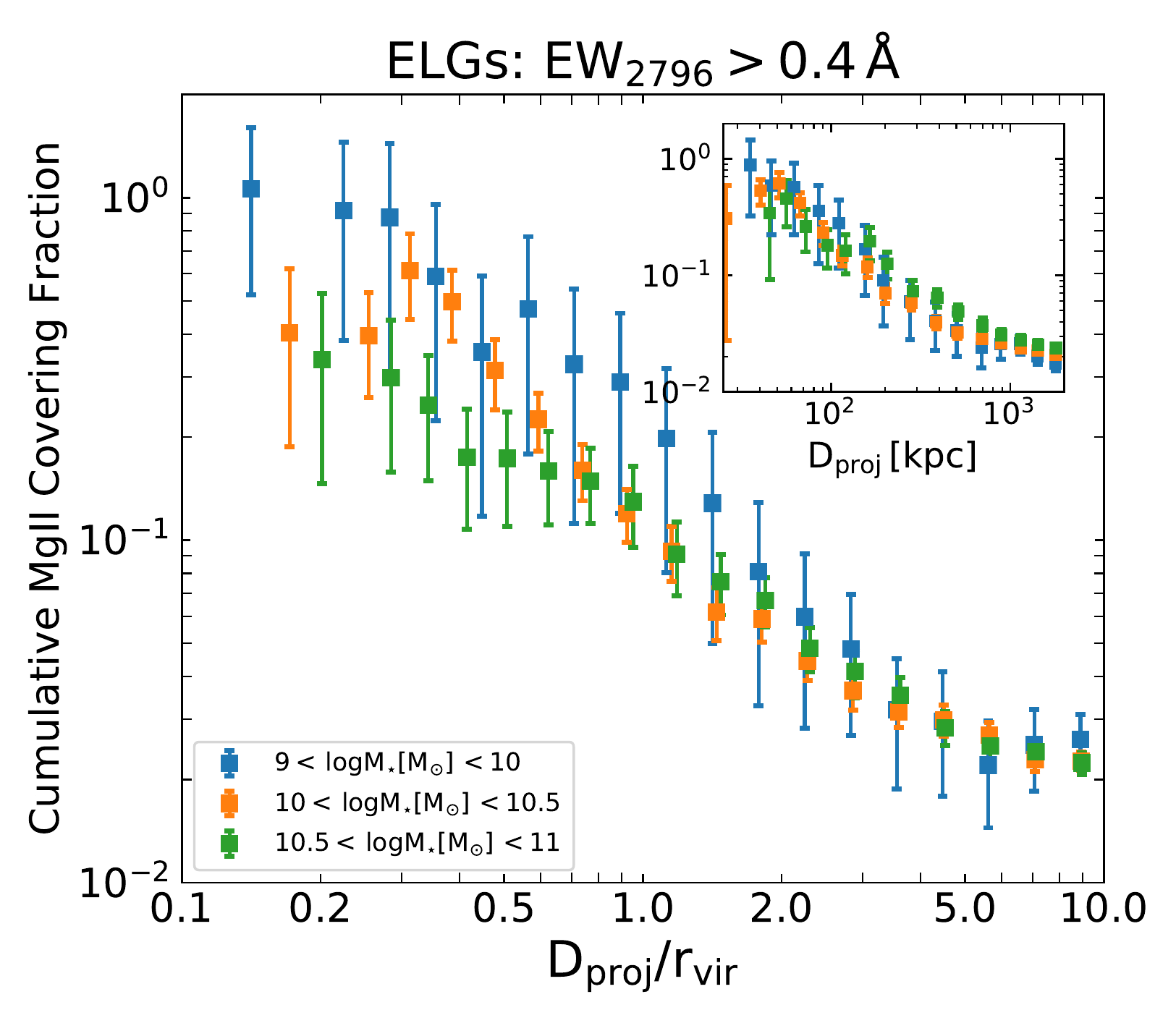}
    \includegraphics[width=0.9\columnwidth]{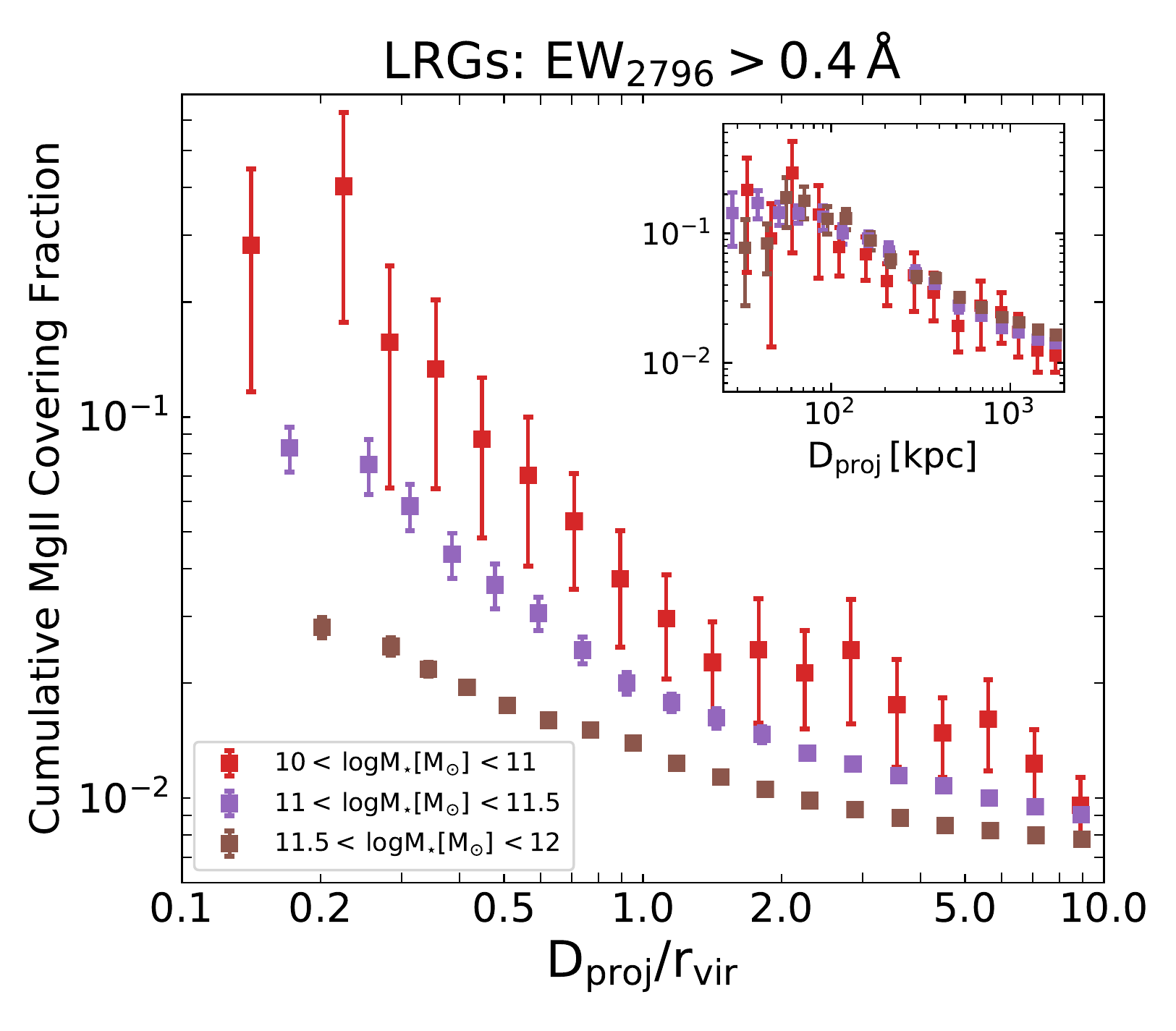}
    \caption{Dependence of the cumulative Mg\,\textsc{ii} covering fraction on stellar mass, around ELGs (top panel) and LRGs (bottom panel), as a function of distance normalized by the virial radius. The insets show the covering fraction as a function of physical kpc. In each panel, the galaxies are divided into stellar mass bins as shown. We focus here on absorbers with $\rm{EW}_{\rm 2796}>0.4$ \AA\, for clarity, since the other two EW bins show similar behavior. We see a very weak stellar mass trend for ELGs, however, there is a strong negative correlation for LRGs. Symbols are shifted horizontally for visual clarity.}
    \label{fig:mass_rvir_fc_mgii}
\end{figure}

We also investigate the cumulative covering fraction $f_{c}(d<D_{\rm proj})$, and the result as a function of projected distance is shown in Figure~\ref{fig:cumu_fc_mgii}. The trend reflects what we have previously seen with the differential covering fraction: star-forming galaxies (ELGs) have 2-4 times higher cumulative values than passive galaxies (LRGs) up to $D_{\rm proj}<100$ kpc. The measurement errors are also smaller as the cumulative estimation significantly enhances the statistics. We now clearly see a difference in $f_c$ between the two samples which extends up to $\sim$ 200 kpc. This partially reflects the different physical sizes of the gaseous (and dark) haloes hosting LRGs versus ELGs. For LRGs, $100$ kpc corresponds to $\lesssim0.2 r_{\rm vir}$ (near the central galaxy), and gas accreted from halo or IGM may not reach into these central regions \citep{huang16}. However, for ELGs, this corresponds to $\sim 0.3 r_{\rm vir}$ where metal rich gas ejected by powerful galactic outflows or winds can be deposited \citep{muratov17, nelson19, mitchell20}.

\begin{figure}
    \centering
    \includegraphics[width=0.9\columnwidth]{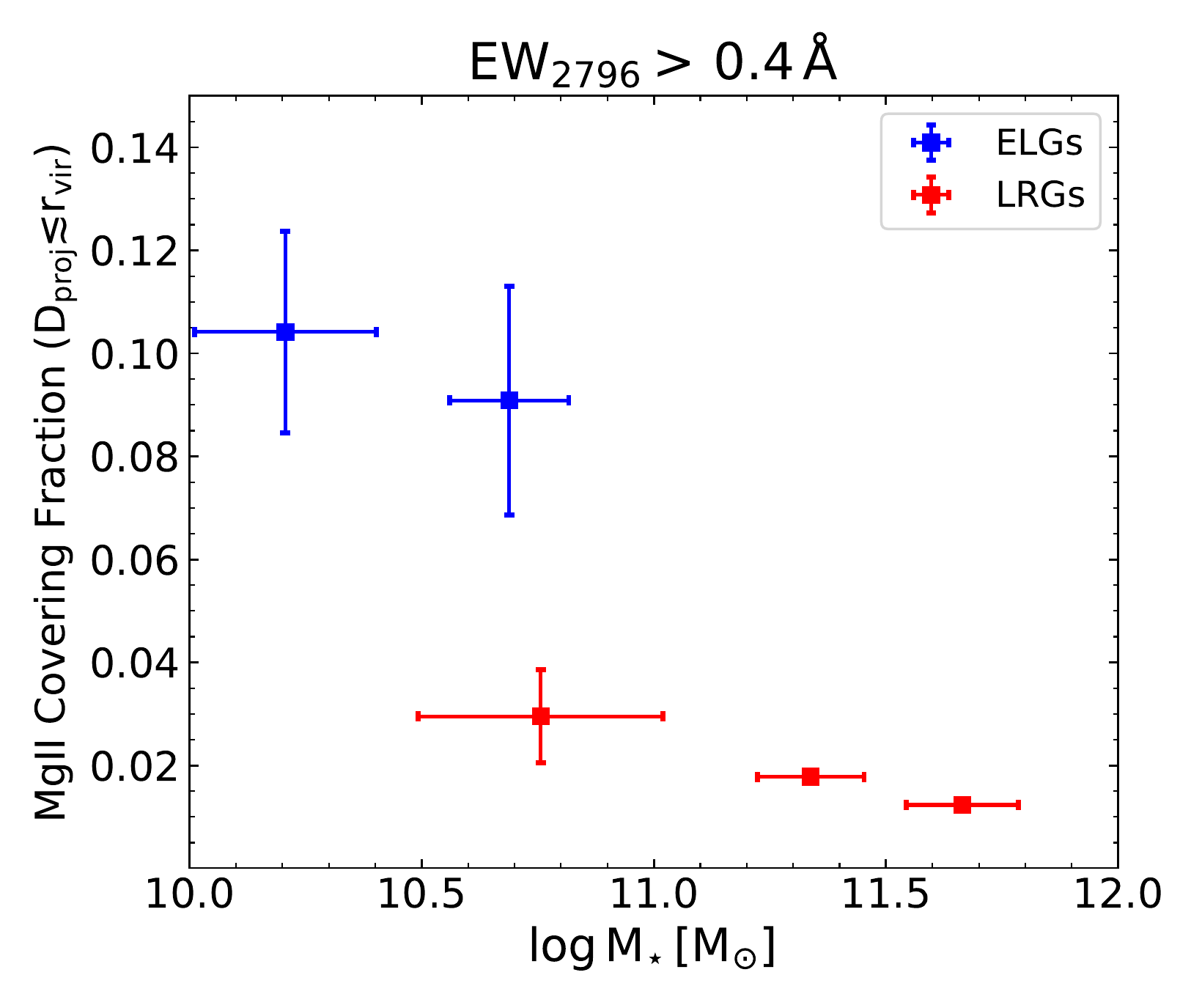}
    \includegraphics[width=0.9\columnwidth]{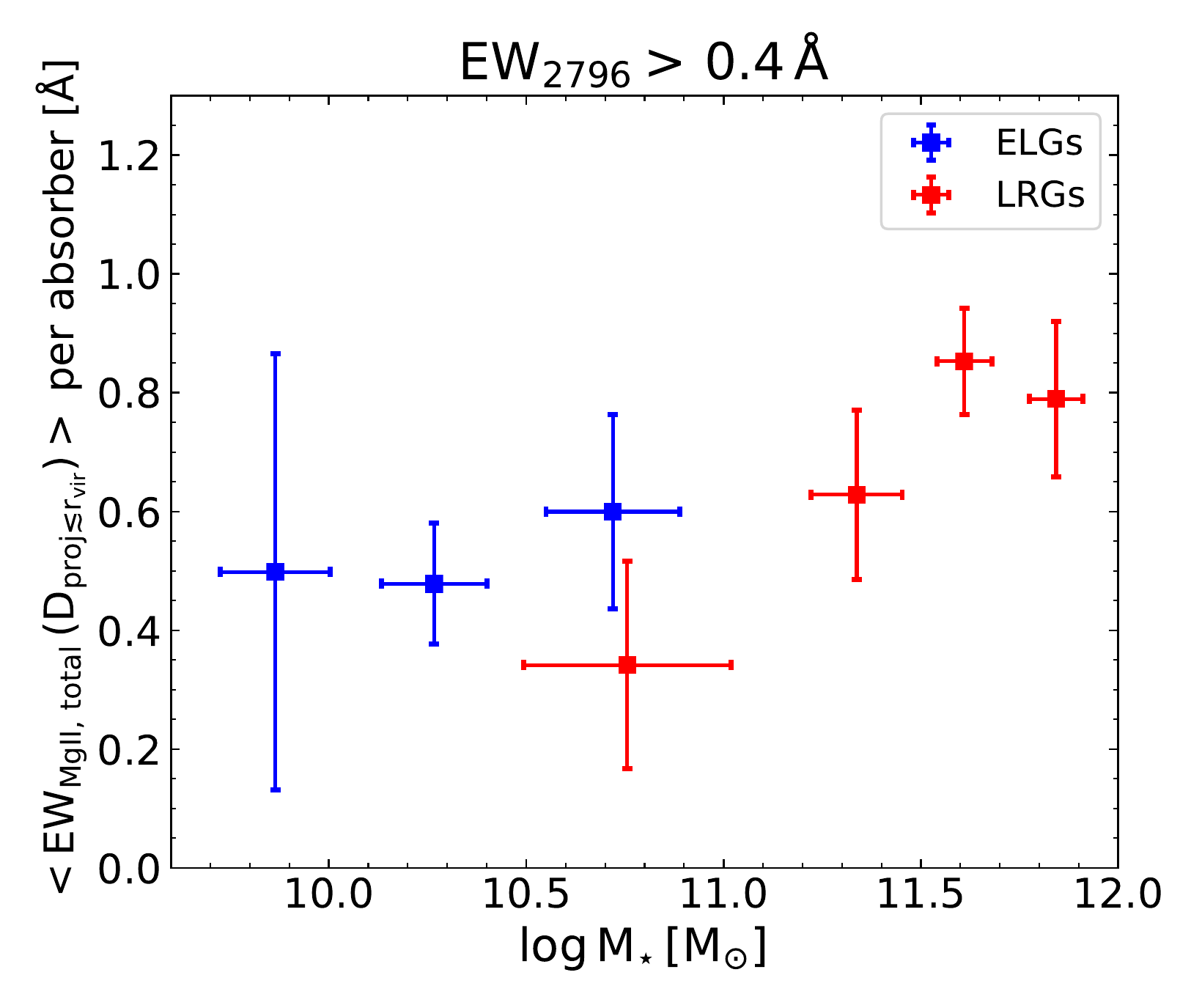}
    \caption{\textbf{Top:} Mg\,\textsc{ii} covering fraction at $D_{\rm proj} \lesssim r_{\rm vir}$ around ELGs (blue) and LRGs (red) as function of stellar mass. Both samples show the same signal: more massive galaxies have lower covering fractions of cool gas in their CGM. \textbf{Bottom:} Total Mg\,\textsc{ii} rest equivalent width per absorber within $D_{\rm proj} \lesssim r_{\rm vir}$ as a function of stellar mass. The absorption strength increases with the stellar mass of galaxies for both galaxy types. In both panels we show absorbers with $\rm{EW}_{\rm 2796}> 0.4$ \AA\, only as the other two bins show qualitatively similar results.}
    \label{fig:rvir_fc_mgii_mass}
\end{figure}

\subsection{Dependence on Stellar and Halo Mass}

To understand how the covering fraction evolves with the size of the dark matter halo we normalize the projected distances by the virial radii of galaxies. In Figure~\ref{fig:mass_rvir_fc_mgii} we show the cumulative Mg\,\textsc{ii} covering fraction in stellar mass bins (different for ELGs and LRGs). We see a weak anti-correlation between the stellar mass of ELGs and the covering fraction of Mg\,\textsc{ii} absorbers when normalizing by $r_{\rm vir}$. In contrast, LRGs show a strong stellar mass dependence in $f_c$. Higher mass passive galaxies have systematically lower cumulative Mg\,\textsc{ii} covering fractions at all distances. For example, LRGs with $\rm M_{\star}< 10^{11}\, M_{\odot}$ have cumulative Mg\,\textsc{ii} covering fractions up to $\sim 25\, $ percent compared to $2-3$ percent around more massive LRGs ($\rm M_{\star}>10^{11.5}\, M_{\odot}$) below $ D_{\rm proj}\sim0.3r_{\rm vir}$.

The reason for this strong stellar mass dependence may be two-fold. First, the tight correlation between stellar and halo mass implies that the size, and total mass, of the circumgalactic medium increases rapidly for more massive galaxies. Second, gas is also thermalized to higher temperatures in more massive haloes, which would naturally inhibit the formation of cooler gas phases. In this regime, heating by the virial shock produces long-term `hot-mode' growth, which is the dominant accretion mode for LRG-type galaxies \citep{birnboim03, keres05, nelson13}.

We show the stellar mass dependence of covering fractions versus physical kpc in the insets of Figure~\ref{fig:mass_rvir_fc_mgii}. Here we see only a weak stellar mass dependence of $f_c$ on  physical 
separation in kpc, and predominantly at large distances, where massive galaxies have higher covering fractions. The lack of a strong trend at fixed physical distance suggests that the signature above is largely driven by the $r_{\rm vir}$ normalization, i.e. the increasing size of more massive haloes.

To better explore these mass trends, we derive the cumulative Mg\,\textsc{ii} covering fraction at $D_{\rm proj} \lesssim r_{\rm vir}$, a rough outer boundary of the CGM, as a function of stellar mass. The result is shown in the top panel of Figure~\ref{fig:rvir_fc_mgii_mass}, where red and blue markers represent LRGs and ELGs, respectively. For ELGs we see a decreasing trend of Mg\,\textsc{ii} covering fraction from $\sim11$ percent for $\rm M_{\star}\sim 10^{10}\, M_{\odot}$ to $\sim9$ percent for $\rm M_{\star}\sim 10^{11}\, M_{\odot}$. For LRGs we see a similar decreasing trend of covering fraction with stellar mass from $\sim 3$ percent for low mass galaxies to just $\sim 1$ percent for massive galaxies. Qualitatively this shows a clear picture: more massive galaxies host less cool gas, on average, in their circumgalactic media.

In the bottom panel of Figure~\ref{fig:rvir_fc_mgii_mass} we show the cumulative total Mg\,\textsc{ii} absorption EW per absorber within $D_{\rm proj} \lesssim r_{\rm vir}$ (see section \ref{covering_fac}) as a function of stellar mass. We see an increasing trend in absorption strength with stellar mass for ELGs, albeit with large error bars. The cumulative absorption EW varies from $\sim 0.5$ \AA\, below $\rm M_{\star}\sim 10^{10}\, M_{\odot}$ to $\sim 0.6$ \AA\, around $\rm M_{\star}\sim 10^{11}\, M_{\odot}$. For LRGs, the trend is much stronger and values vary from $0.3$ \AA\,to $0.8$ \AA\, from the least to most massive galaxies. Broadly, this result implies that either Mg\,\textsc{ii} absorbers have larger EWs around more massive galaxies, or that sightlines intersect a larger number of individual absorbers in more massive haloes. A similar positive correlation between equivalent width of Mg\,\textsc{ii} absorbers and stellar mass was observed in star-forming galaxies \citep{bradshaw13}.

\begin{figure}
    \centering
    \includegraphics[width=0.9\columnwidth]{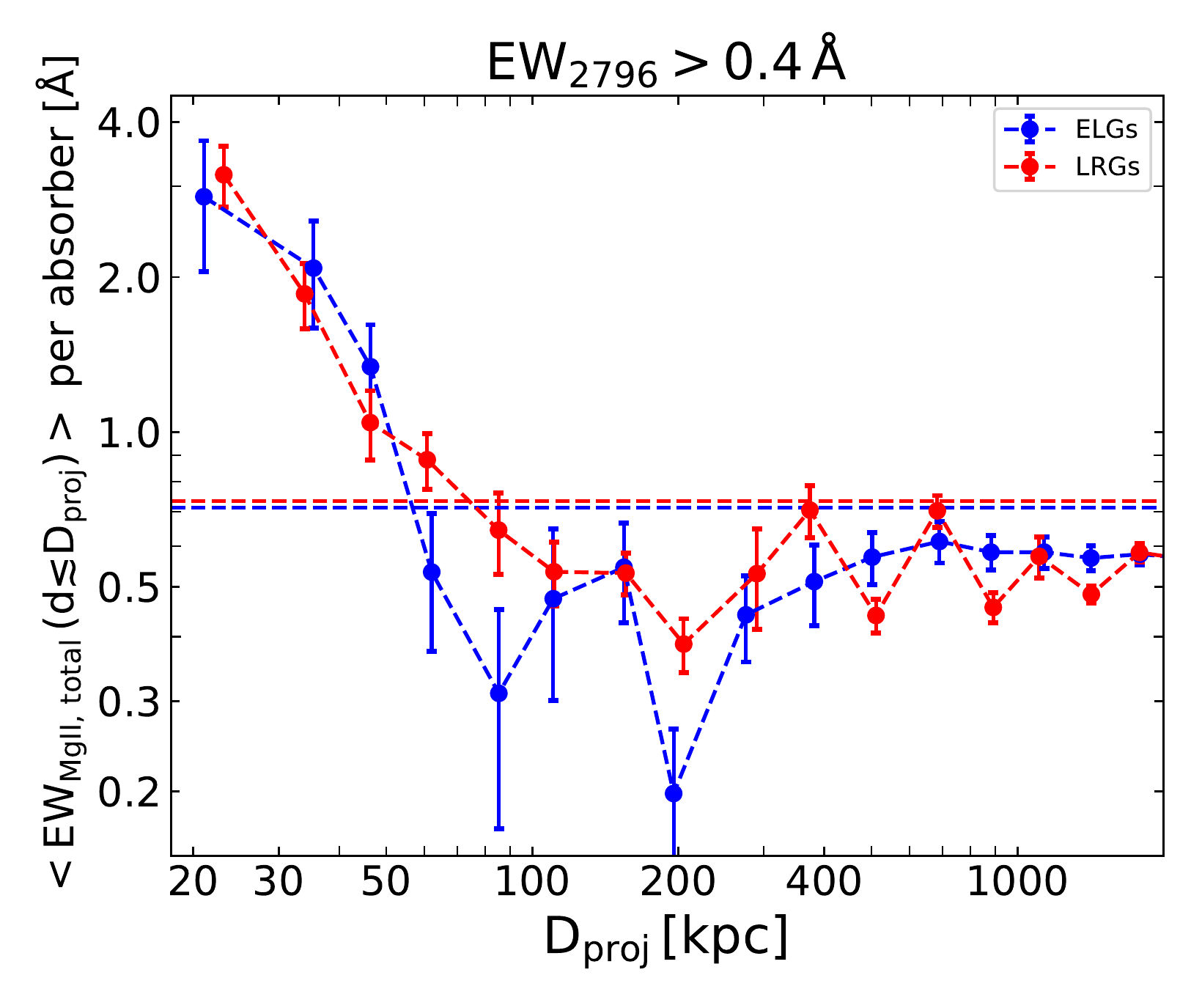}
    \includegraphics[width=0.9\columnwidth]{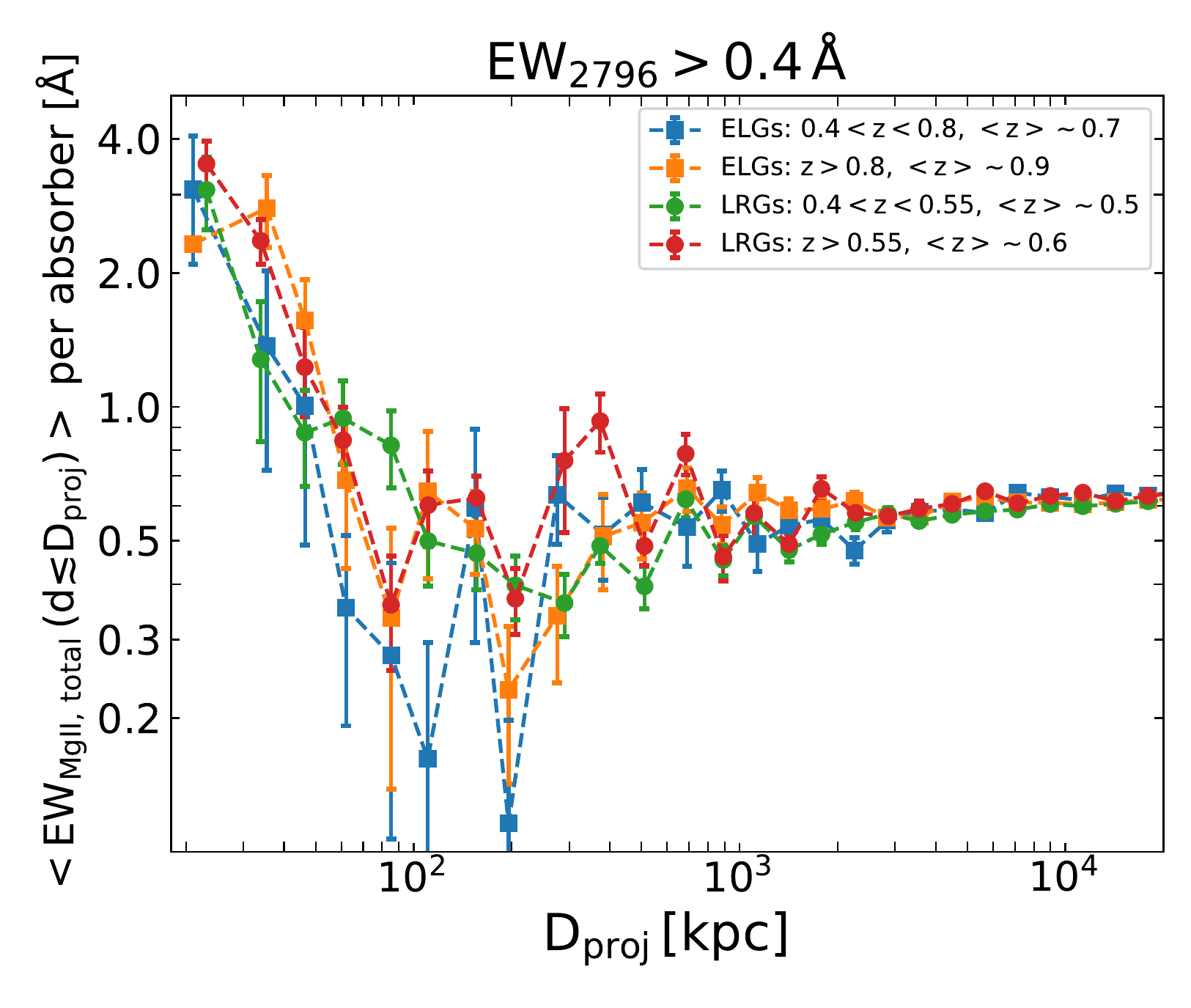}
    \caption{\textbf{Top:} Total Mg\,\textsc{ii} rest equivalent width ($\rm{EW}_{\rm 2796}+\rm{EW}_{2803}$) per absorber within $d \lesssim D_{\rm proj}$ around ELGs (blue) and LRGs (red) as a function of projected distance. The dashed lines show the mean values. There is a clear decreasing trend below $D_{\rm proj}<50$ kpc for both ELGs and LRGs. \textbf{Bottom:} Redshift evolution of total Mg\,\textsc{ii} rest equivalent width around ELGs and LRGs. We see a weak positive correlation between average redshift and total EW.}
    \label{fig:ew_tot_mgii}
\end{figure}

In the top panel of Figure~\ref{fig:ew_tot_mgii} we show this same mean total rest equivalent width per absorber, i.e. $\rm EW_{2796}+EW_{2803}$, cumulative as a function of $d \lesssim D_{\rm proj}$. At small distances (less than 100 kpc) the absorption strength declines with projected distance. For LRGs at $D_{\rm proj}<50$ kpc, we see an enhancement up to$3$ \AA,\, compared to $\lesssim$ 0.8\, \AA,\, at larger radii. ELGs show a similar behavior, but have systematically lower rest equivalent widths, up to $D_{\rm proj}\lesssim 400$ kpc, remaining slightly above the LRGs at larger distances. This is principally due to the significant difference in the average redshift between the two galaxy samples and the reason could be possibly related to the different gas properties at different epochs. To understand the possible redshift contribution to this difference we divide the ELGs and LRGs into low and high-$z$ bins (splitting at the medians). We clearly see a slightly higher EWs around high-$z$ galaxies at distances $D_{\rm proj}\gtrsim 1$ Mpc. We show the redshift evolution of EW in the bottom panel of Figure~\ref{fig:ew_tot_mgii} for both ELGs and LRGs. The weak positive correlation between total EW and redshift is clearly visible for a given galaxy sample.

Nevertheless, we can still see in each redshift interval the same characteristic scale dependence of the total rest equivalent width per absorber: a strong decrease
to a tentative minimum value at projected radii of $\sim 150$ kpc, followed by a slight rise to value that remains constant beyond 1 Mpc. This feature is more clear for ELGs than LRGs, and may be related to gas inflow processes onto dark matter haloes, which act to shock-heat infalling gas at the boundary of the halo.

\subsection{Dependence on Star Formation Rate}

\begin{figure}
    \centering
    \includegraphics[width=0.9\columnwidth]{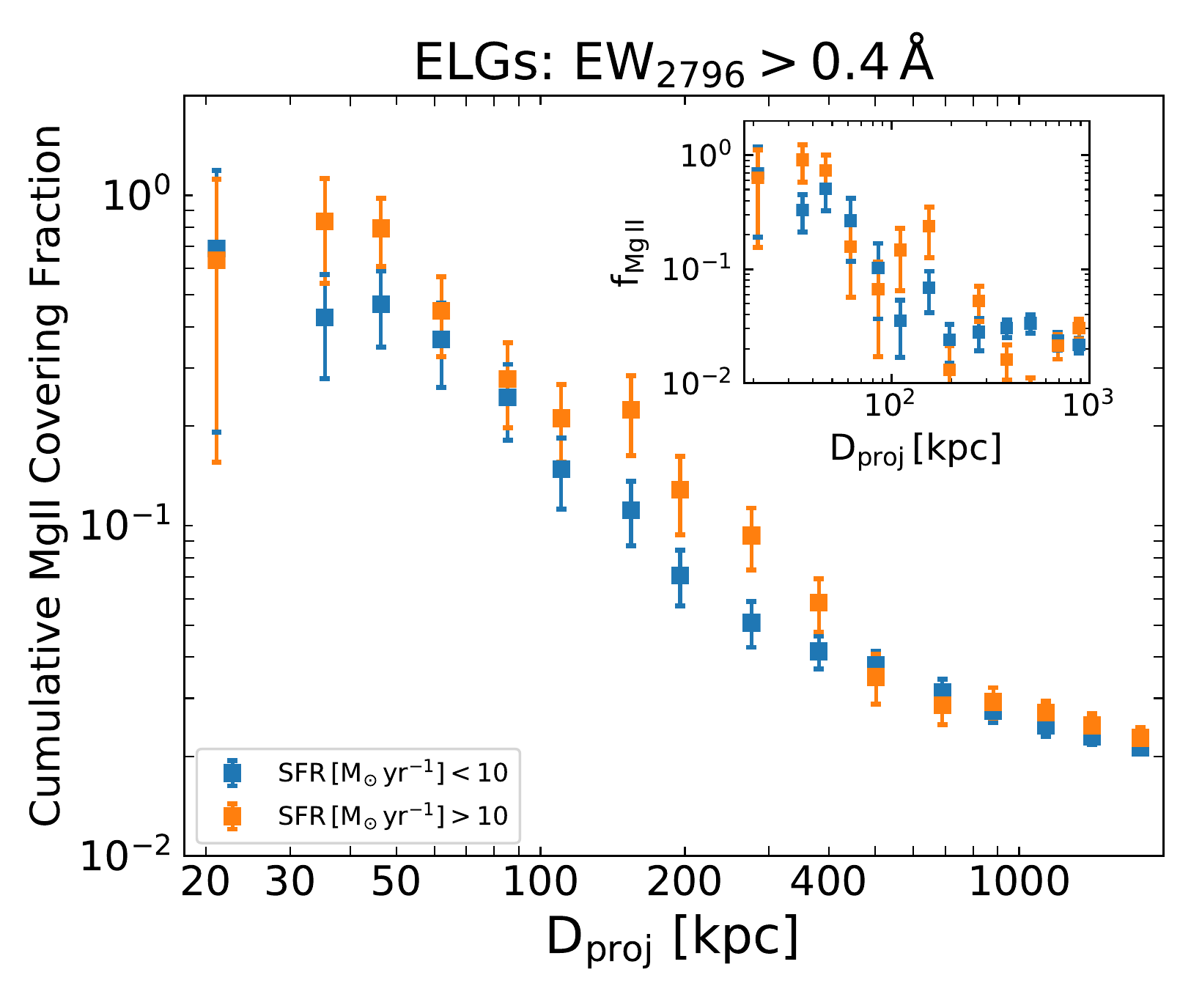}
    \includegraphics[width=0.9\columnwidth]{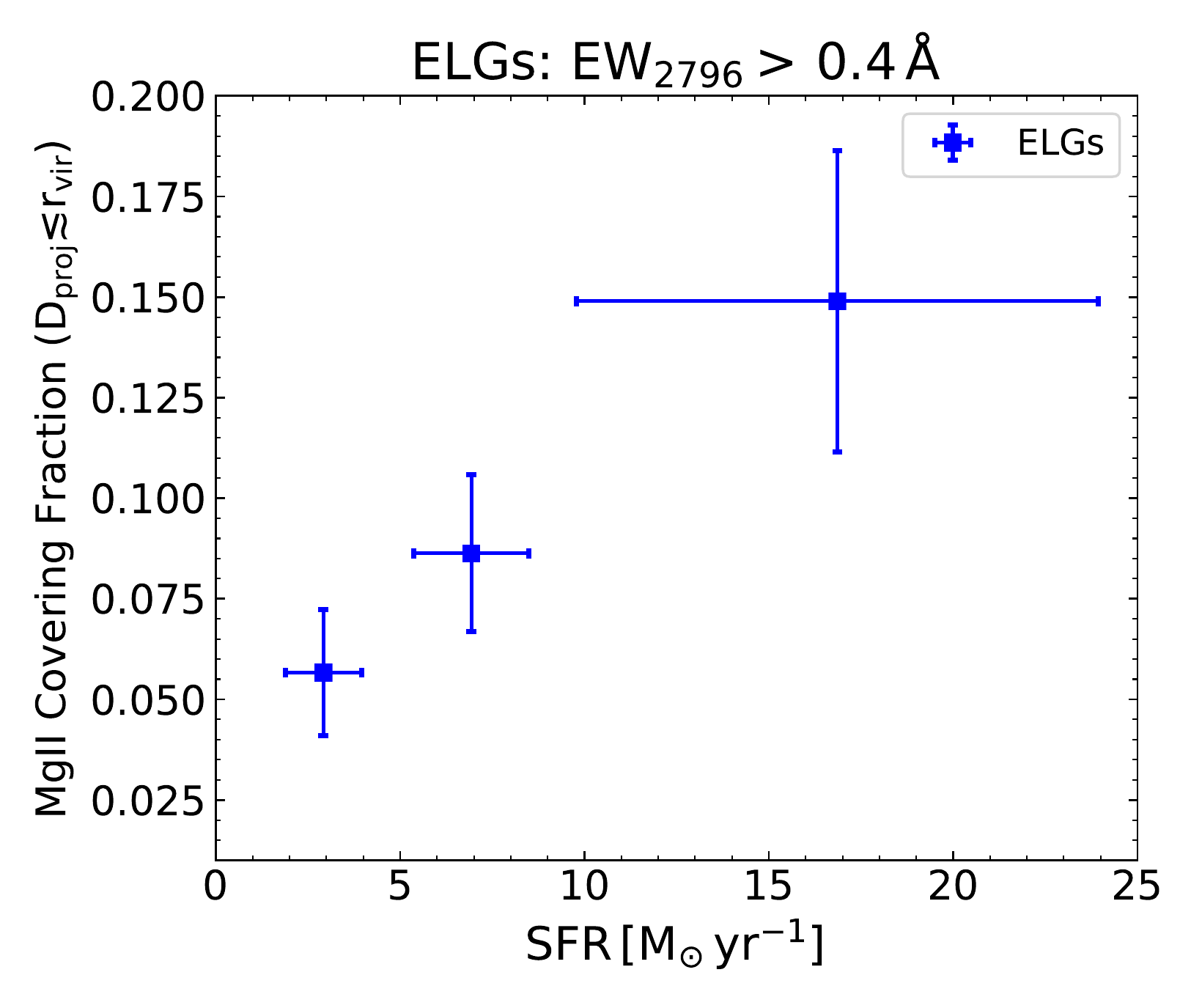}
    \caption{\textbf{Top:} Star formation rate dependence of cumulative Mg\,\textsc{ii} covering fraction, $f_{c}(d< D_{\rm proj})$ around ELGs as a function of projected distance from galaxy. Inset shows the differential covering fraction as a function of projected distance from galaxy in two SFR bins. \textbf{Bottom:} Mg\,\textsc{ii} covering fraction at $ D_{\rm proj}\lesssim r_{\rm vir}$ as a function of SFR of ELGs. A strong positive correlation between covering fraction and SFR of ELGs is visible.}
    \label{fig:fc_sfr}
\end{figure}

A key physical property of galaxies that may affect the distribution of cold gas traced by Mg\,\textsc{ii} absorbers is the star formation rate \citep[SFR;][]{lan18, rubin18}. The ELG catalogue includes measurements for [O\,\textsc{ii}] $ \lambda3727$ flux which traces the star formation activity in ELGs for $z>0.4$ \citep{kennicutt98}. Despite its simplicity and neglect of dust attenuation effects, this SFR value can still be used to perform qualitative studies. 

In the top panel of Figure~\ref{fig:fc_sfr} we show the dependence of cumulative and differential Mg\,\textsc{ii} covering fraction on star formation rate for ELGs. We divide the galaxies into two SFR bins: low ($\rm SFR<10\, M_{\odot}yr^{-1}$) and high ($\rm SFR>10\, M_{\odot}yr^{-1}$). We observe a strong variation of covering fraction with star-formation activity. Galaxies with high star-formation rate have 2-3 times higher cumulative covering fraction than their low SFR counterparts, all the way up to $D_{\rm proj} \sim 400$ kpc, close to the $r_{\rm vir}$ of ELGs. One the other hand the differential covering fraction (shown in the inset) is 2-5 times higher below $\sim 50$ kpc in galaxies with high star-formation rate.

In the bottom panel we show the cumulative covering fraction at $D_{\rm proj} \lesssim r_{\rm vir}$ as a function of SFR. We again see a strong positive correlation with SFR, though the highest SFR bin has large uncertainties. This increasing trend of covering fraction with SFR indicates that SF activity in ELGs plays an important role in enriching the cold gas in their CGM. We also point out that for massive star-forming galaxies a possible source of influence is the central active galactic nucleus (AGN). We have not tried to exclude AGN from our samples. The galactic outflows powered by AGNs can significantly enrich the metal abundance in the CGM of massive star-forming galaxies \citep{veilleux05}, and future work will focus on disentangling the roles of star formation versus AGN activity.

\begin{figure*}
    \centering
    \includegraphics[width=0.46\textwidth]{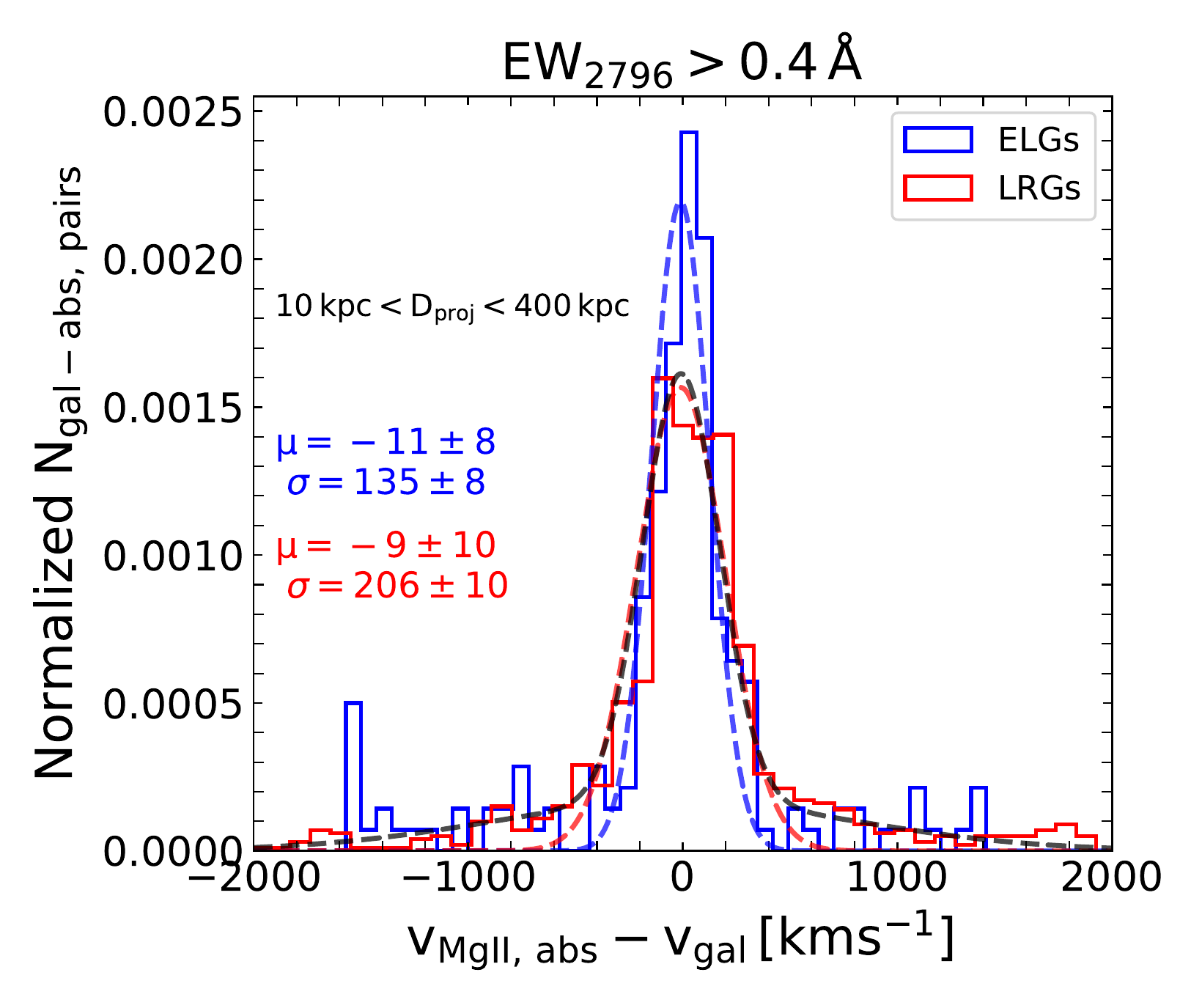}
    \includegraphics[width=0.46\textwidth]{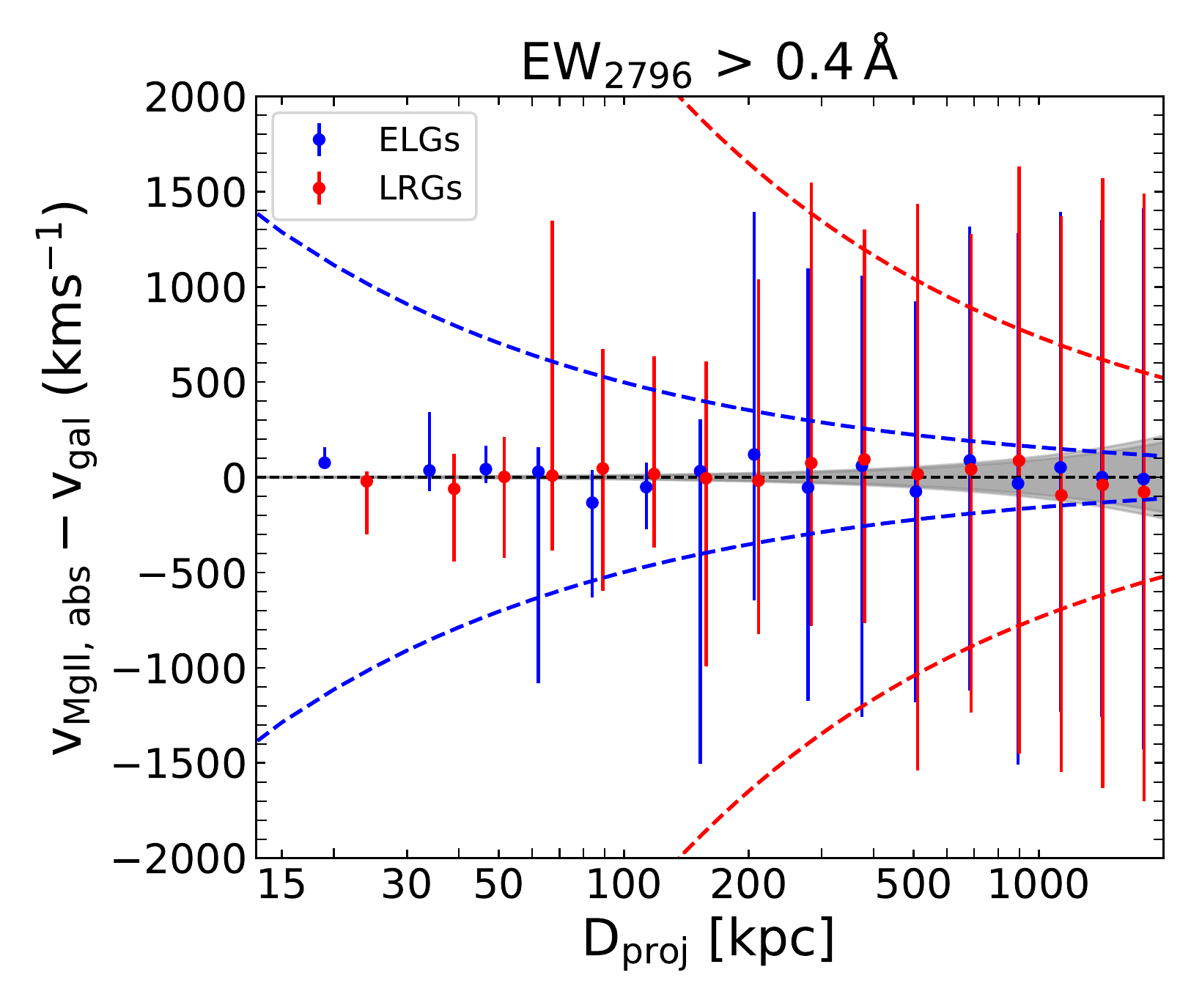}
    \caption{Line-of-sight (LOS) velocity difference between galaxies and detected cool gas absorption. \textbf{Left:} Distribution of relative velocities for galaxy-absorber pairs within $10\, \rm kpc < D_{\rm proj}<400\, \rm kpc$, blue and red curves showing ELGs and LRGs, respectively (solid lines). The distributions are characterized by single gaussian profiles shown in dashed color lines. A two-component gaussian fit to the LRG distribution is also shown in the dashed black line, which better characterizes the high-velocity tails. Best fit parameters (only for single gaussian) are shown in the panel. \textbf{Right:} Line-of-sight velocity separation between galaxy-absorber pairs as a function of projected distance. The colored circles denote the median value in each radial bin and vertical bars show the 5th to 95th percentile ranges in each bin. The blue and red dashed line shows the escape velocity corresponding to the typical halo masses of ELGs and LRGs, respectively. The Hubble flow is indicated by the shaded grey region. Note that we have slightly offset horizontally the points and lines for visual clarity.}
    \label{fig:dv_vs_dist}
\end{figure*}

\begin{table*}
\centering
  \caption{Line-of-sight velocity separation between galaxy-absorber pairs (within $\rm 10\, kpc< D_{proj}<400\, kpc$) as a function of galaxy stellar mass. $\langle\Delta v\rangle$ and $\sigma$ are the mean and standard deviation of each distribution. $\Delta v_{10},\, \Delta v_{50},\, \Delta v_{90}$ are 10th, 50th and 90th percentiles of the velocity separations.}
  \begin{tabular}{||ccccccccc||}
    \hline
    &$\rm log\, M_{\star}$ & $\rm \langle log\, M_{\star}\rangle$ & $\rm \sigma_{log\, M_{\star}}$ & $\langle\Delta v\rangle$ & $\sigma$ & $ \Delta v_{\rm 10}$ & $ \Delta v_{\rm 50}$  & $ \Delta v_{\rm 90}$ \\
    & $\rm[M_{\odot}]$ & $\rm[M_{\odot}]$ & $\rm[M_{\odot}]$ & [\kms] & [\kms]& [\kms] & [\kms]&[\kms]\\
    \hline
    ELGs: & Full Sample & 10.4 & 0.33 & -55 & 543& -768 & 28 & 325\\
    LRGs: & Full Sample & 11.4 & 0.33 & 51 & 529& -462 & 33 & 584\\
    \hline
     & [9, 10] & 9.8 & 0.2 & 151 & 341 & -68 & 52 & 513\\
     ELGs: & [10, 10.5] & 10.3 & 0.14 & -47 & 531 & -760 & 39 & 360\\
    & [10.5, 11] & 10.7 & 0.15 & -120& 576 & -1100 & -17 & 278\\ \\
    & [10, 11] & 10.5 & 0.24 & 122 & 394 & -100 & 85 & 286\\
    LRGs: & [11, 11.5] & 11.3 & 0.12 & 28 & 547 & -455 &-34 & 607\\
    & [11.5, 12] & 11.7 & 0.12 & 66 & 528 & -498 & 125 & 578\\
    
    \hline
  \end{tabular}
  \label{tab:dv_sm}
\end{table*}

\subsection{The Relative Kinematics of Galaxies and Absorbers}

With robust spectroscopic information for both the absorbing gas clouds and parent galaxies we can study the line-of-sight velocities to constrain the relative motion between the two. For this purpose we estimate the line-of-sight velocity separation, $\Delta v=c \Delta z/(1+z)$, where $c$ is the speed of light and $z$ is the galaxy redshift.

In the left panel of Figure~\ref{fig:dv_vs_dist} we show the distribution of $\Delta v$ for Mg\,\textsc{ii} absorber-galaxy pairs, separating ELGs (blue) from LRGs (red). Here we show absorbers with $\rm{EW}_{\rm 2796}>0.4$ \AA\, within $10\, \rm kpc< D_{\rm proj}<400$ kpc from the galaxy. The velocity distribution of Mg\,\textsc{ii} absorbers around ELGs (shown in blue solid line) is well characterized by a single gaussian with mean $\langle\Delta v\rangle\sim -11(\pm8)$ \kms and dispersion, $\sigma_{v}\sim135$ \kms (shown in the dashed blue line). The velocity distribution for LRGs (shown in the red solid line) in the central region is best-fit by a single gaussian of mean $\langle\Delta v\rangle \sim -9(\pm10)$ \kms and dispersion, $\sigma_{v} \sim 200$ \kms (shown in the dashed red line). The mean velocity difference is in agreement with previous studies such as \citet{huang21}, even though the samples and methods are significantly different. This further suggests that our analysis yields consistent trends. The typical dark matter halo velocity dispersion ($\Delta v_{200}$) for ELGs is $\sim 140$ \kms\, and $\sim 350$ \kms\, for LRGs \citep{elahi18}. This implies $\rm \sigma_{ELG}\sim \sigma_{ELG, \, halo}$ while $\rm \sigma_{LRG}\sim 0.6\,\sigma_{LRG, \, halo}$. While the cool CGM is consistent with virial motion around star-forming galaxies, it is sub-virial for massive quiescent systems.

Visual inspection shows that a single gaussian is not a good representation of the tails of these two $\Delta v$ distributions. A multi-component gaussian has been shown to better characterize the velocity distribution around LRGs \citep{huang16, chen17}. We therefore fit a double gaussian (shown with the dashed black line) to the LRG distribution. The best-fit means are consistent with zero, while the two velocity dispersions represent a narrow component with $\sigma_{\rm narrow}\sim 180$ \kms, describing the majority of pairs ($\approx 85$ percent, central distribution), together with a broad component with $\sigma_{\rm broad}\sim 1100$ \kms\, describing the tails. We discuss possible interpretations in Section \ref{discussion}.

In the right panel of Figure~\ref{fig:dv_vs_dist} we show the relative velocity separation of galaxy-absorber pairs as a function of projected distance from the galaxy. The median values of the distribution in each radial bin are shown in colored circles for ELGs (blue) and LRGs (red). The vertical bars show the 5th to 95th percentiles of the distribution in each radial bin. The corresponding dashed color lines show the decreasing escape velocity for a dark matter halo mass, estimated using the average masses of the galaxies. The shaded gray region shows the increasing Hubble flow at the average redshift of the galaxy samples. 

The median values are always within $\sim 100$ \kms and well below the escape velocity of the dark matter haloes. There is a small positive offset for ELGs, possibly due to gas moving away from ELGs towards the observer powered by strong galactic outflows, while gas behind the galaxy is detected less frequently due to obscuration effects. At small distances, $D_{\rm proj}<50$ kpc, the distributions are narrow and the vast majority of galaxy-absorber pairs have low relative velocities. At larger distances $D_{\rm proj}>200$ kpc the tails of the distributions imply a small fraction of absorbing clouds with velocities higher than the escape velocity of dark matter halo for both ELGs and LRGs.

To study the dependence of gas kinematics on the stellar mass of galaxies we divide the galaxy samples into stellar mass bins and estimate the mean $\langle\Delta v\rangle$ and  standard deviation $\sigma$, as well as the 10th, 50th, and 90th percentiles of the velocity distributions within $10 \, \rm kpc < D_{\rm proj}<400\, \rm kpc$. The values are compiled in Table ~\ref{tab:dv_sm}. For both ELGs and LRGs, we see an increasing trend of gas velocity dispersion $\sigma$ with stellar mass. For ELGs, this can be attributed to the positive correlation between SFR and stellar mass, and consequently stronger galactic outflows. However, for LRGs, this is more likely related to a trend of gas accretion rate with halo mass, either from the IGM or from satellite galaxies and their interaction with pre-existing halo gas. 

\begin{figure}
    \centering
    \includegraphics[width=0.9\columnwidth]{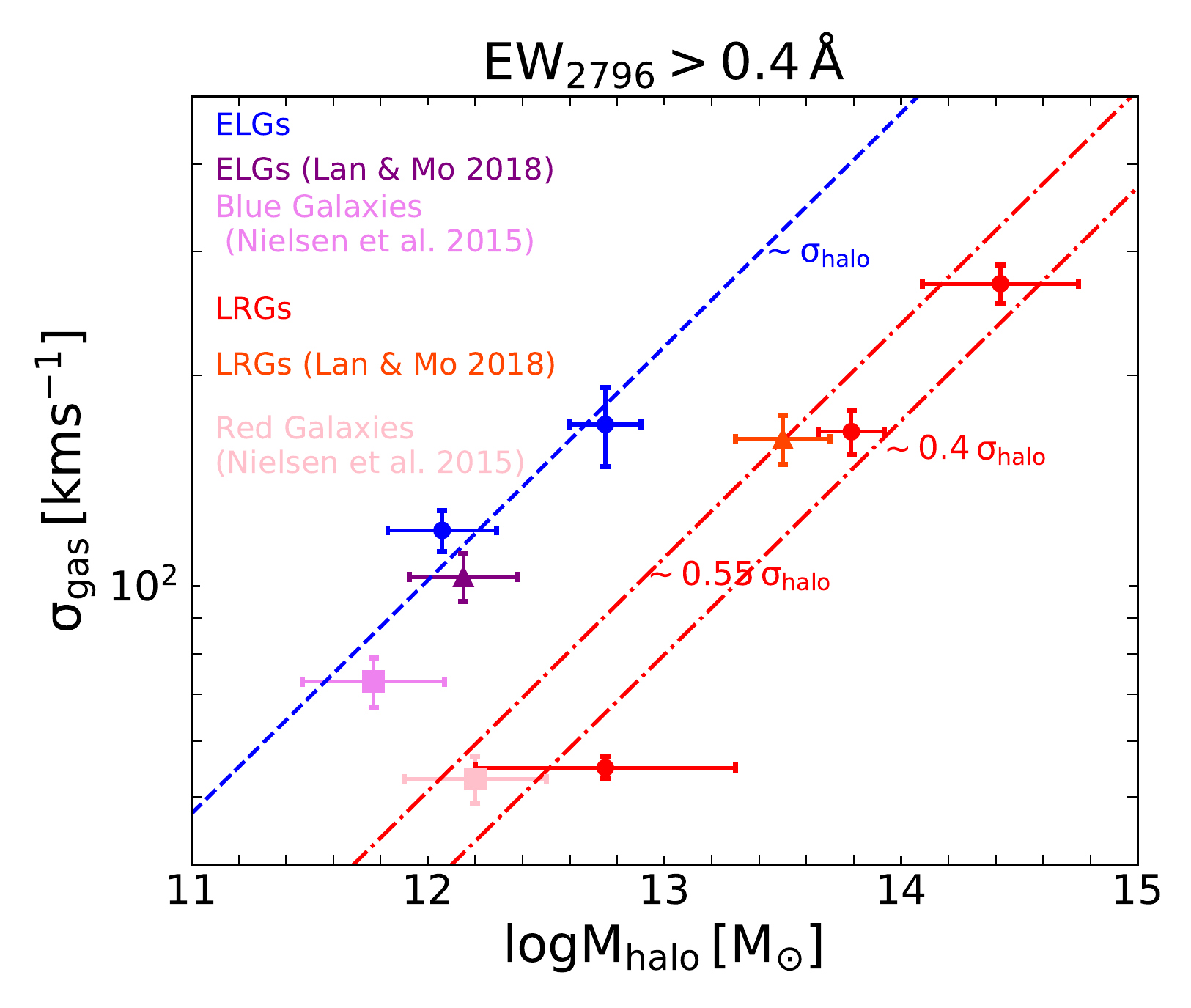}
    \caption{The line-of-sight velocity dispersion, $\sigma_{\rm gas}$, of galaxy-absorber pairs, within $\rm 10\, kpc< D_{\rm proj}<400\, kpc$, as a function of halo mass and galaxy types. The dark blue and red dots represent ELGs and LRGs, respectively. The dashed blue line shows the expected velocity dispersion as a function of halo mass of ELGs. The two dash-dotted red lines show the $0.5$ and $0.4$ times the dark matter velocity dispersion of LRGs. We also show the comparison with previous measurements from \citet{nielsen15} in violet and pink squares, and \citet{lan18} in purple and orange-red triangles. The motion of gas around LRGs is suppressed relative to the expected velocity dispersion given their dark matter halo masses.}
    \label{fig:vel_vs_hm}
\end{figure}

Finally, we directly assess how our measured dispersion values depend on halo mass. The result is shown in Figure~\ref{fig:vel_vs_hm}. For simplicity the velocity dispersion is estimated by fitting a single gaussian in each halo mass range for both ELGs and LRGs. We clearly see that the gas motion around LRGs is suppressed ($\rm 0.4-0.55\, \sigma_{halo}$, as indicated by the dot-dashed red lines) over a large range of halo masses ($\rm M_{halo}\sim 10^{13}-10^{14.5}\, M_{\odot}$). This is in contrast to ELGs (blue points), which exhibit dispersions similar to $\sigma_{\rm halo}$, as indicated by the dashed blue line. The trend for ELGs is also visible up to $\rm M_{halo}\sim 10^{13}\, M_{\odot}$. We contrast against measurements from previous studies \citep{nielsen15,lan18}, which we here extend to a larger mass range -- this comparison is discussed in the next section.

\section[]{Discussion: The Nature of Cold Gas around Galaxies at 0.5 < z < 1}\label{discussion}

\subsection{Comparison with Previous Studies}

\subsubsection{Mg\,\textsc{ii} Covering Fraction}

\begin{figure}
    \centering
    \includegraphics[width=0.9\columnwidth]{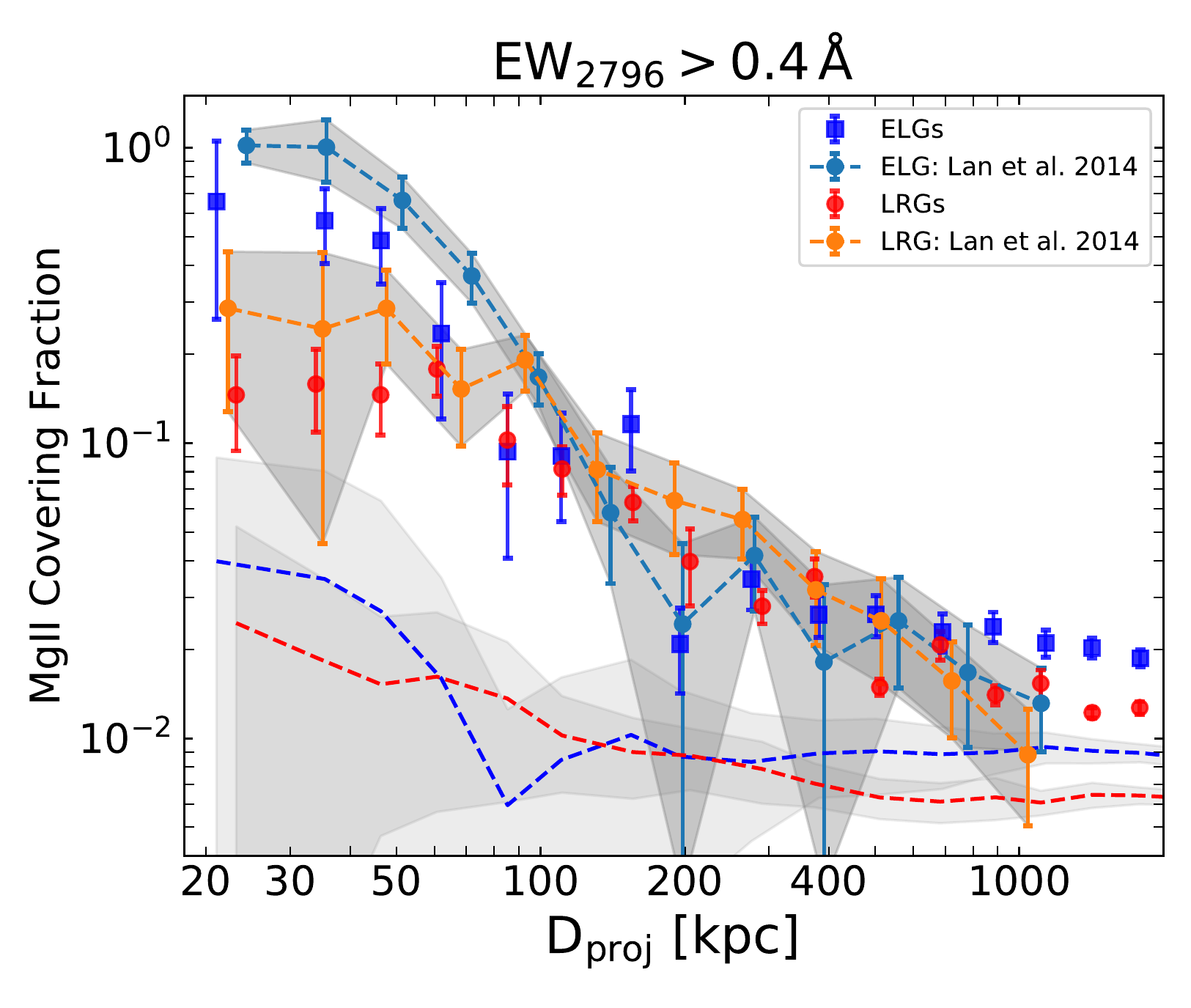}
    \includegraphics[width=0.9\columnwidth]{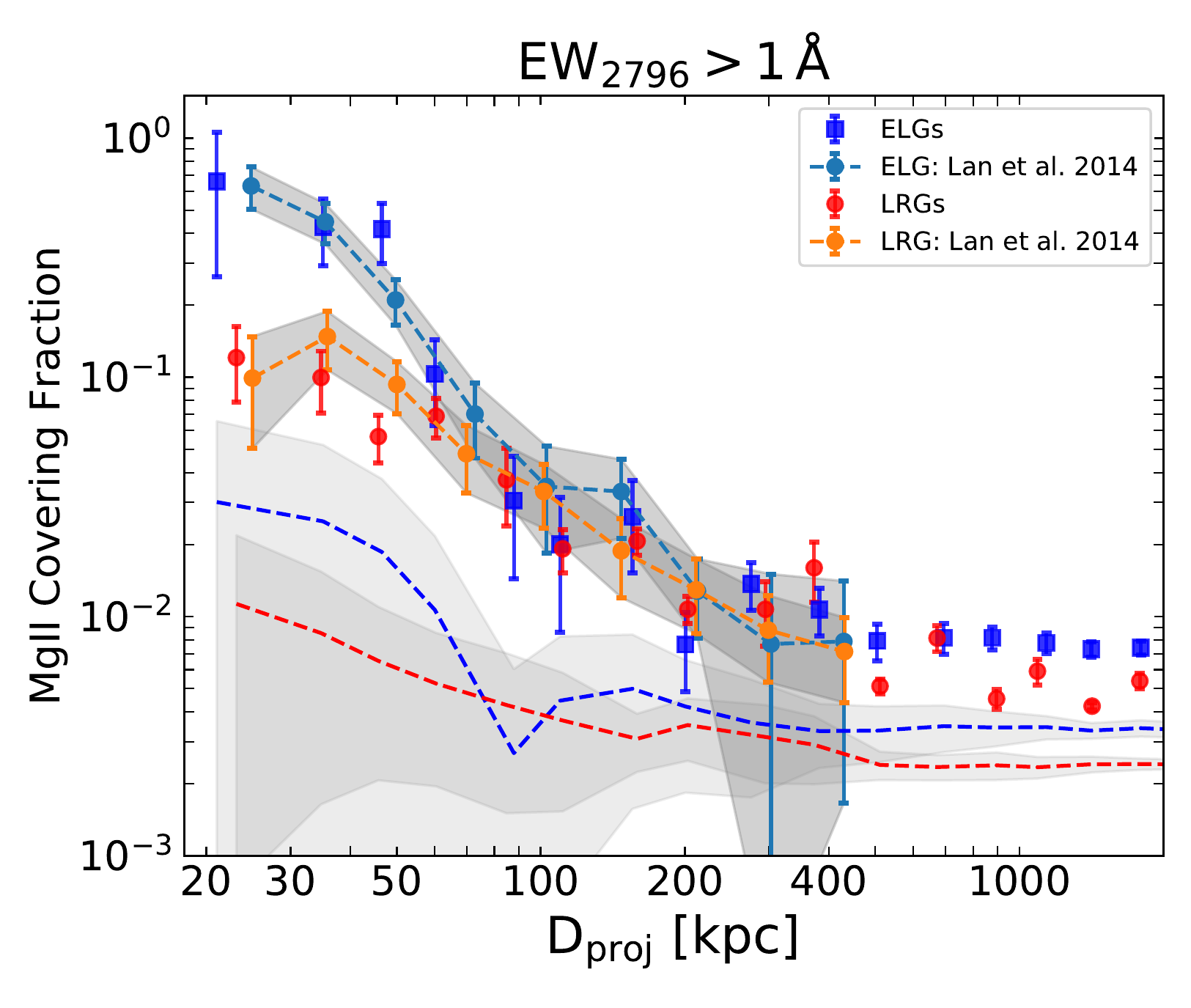}
    \caption{Comparison of the differential Mg\,\textsc{ii} covering fraction with values obtained in \citet{lan14} and \citet{lan18} around ELGs and LRGs. \textbf{Top:} Comparison for Mg\,\textsc{ii} absorbers with $\rm{EW}_{\rm 2796}>0.4$ \AA.\,\textbf{Bottom:} Comparison for Mg\,\textsc{ii} absorbers with $\rm{EW}_{\rm 2796}>1$ \AA. Our results agree well with these literature values, recovering the same qualitative and relative trends with respect to galaxy type, despite significant methodological and analysis differences.}
    \label{fig:lan_fc_vals}
\end{figure}

We now make a comparison between our new results and previous studies. First, we quantitatively compare our covering fractions with values obtained in \citet{lan14} for $\rm{EW}_{\rm 2796}>1$ \AA\, and \citet{lan18} for $\rm{EW}_{\rm 2796}>0.4$ \AA, shown in Figure~\ref{fig:lan_fc_vals}. Our analysis extends over larger spatial scales due to the increased statistics offered in DR16, but within the same scales probed, the results agree well within the error bars. We observe that for $\rm{EW}_{\rm 2796}>0.4$ \AA,\, $f_{c}$ slightly lower than \citet{lan18} below $D_{\rm proj} \sim 50$ kpc. However, a similar covering fraction dependence with galaxy type is present. The minor  discrepancy is likely due to the non-trivial differences in methodologies and analysis choices.

For $\rm{EW}_{\rm 2796}>0.4$ \AA\, \citet{lan18} estimated covering fraction using a stacking analysis. They assumed that the average Mg\,\textsc{ii} absorption is dominated by absorbers with $\rm{EW}_{\rm 2796}>0.4$ \AA\, and can be expressed as the product of covering fraction and the average value of $\rm{EW}_{\rm 2796}$ estimated from the intrinsic distribution of individual absorbers toward random quasar sightlines. \citet{lan14} estimated the excess in the number of galaxies found within a given impact parameter w.r.t reference quasars as opposed to random positions. Then by counting galaxies around the selected quasars they estimated the denominator in the covering fraction expression. The error bars were estimated with Poisson statistics. 

The uncertainties on \citet{lan18} values are smaller than ours even though the absorber sample is smaller. This is due to the small errors on the average Mg\,\textsc{ii} strength in stacked spectra. For $\rm{EW}_{\rm 2796}>1$ \AA\,absorbers we agree well with \citet{lan14} at $D_{\rm proj} < 100$ kpc even though our methods and resulting absorber catalogues are substantially different. For instance, \citet{lan14} used the \citet{zhu13a} Mg\,\textsc{ii} catalogue (based on SDSS DR7 quasars) which has different completeness characteristics than our catalogue. We also note that the \citet{lan18} results show less clear evidence for a sudden decline in covering fraction at a projected radius of 50 kpc.

\subsubsection{Trends with Galaxy Properties}
 
A  detailed quantitative comparison of the stellar mass dependence  of the Mg\,\textsc{ii} covering fraction with results from previous studies 
lies beyond the scope of this paper. The study of \citet{lan20} of  the stellar mass dependence of
covering fraction indicates that it is important to study these trends in relatively narrow redshift
intervals, because there are relatively strong evolutionary trends in covering fraction
with redshift at fixed mass. When this is accomplished, \citet{lan20} find that after normalizing the projected distance by the virial radii the covering fraction
of absorbers with rest equivalent widths ($\rm EW_{2796}$) greater than 1 \AA\ shows
very little dependence on stellar mass for star-forming galaxies, but decreases with stellar mass for
the passive galaxy population. The weak dependence on mass that we find for our ELG sample
is in agreement with these findings.

For LRGs we see a strong anti-correlation of covering fraction with stellar mass after normalizing projected distance by the virial radii, such that $f_c$ is lower for more massive galaxies (bottom panel of Figure~\ref{fig:mass_rvir_fc_mgii}). These results
are also in agreement with \citet{lan20} for strong absorbers ($\rm{EW}_{\rm 2796}>1$ \AA), who found that the covering fraction (on projected scale) varies weakly with stellar mass as $f_{\rm c} \propto M_{\star}^{0.3}$, and varies from $2$ percent from low mass to 3 percent for high mass LRGs at $D_{\rm proj}\gtrsim 100$ kpc. For similar stellar masses, we find that the covering fraction varies from 3 percent to 7 percent. Note that our LRGs are more massive than DESI passive galaxies.

\subsubsection{The Kinematics of Galaxy-Absorber Pairs}

By connecting galaxy-absorber pairs in velocity space we estimated the ensemble velocity distributions of gas clouds associated with galaxies. Using single Gaussian profiles, we have characterized the velocity dispersion for absorbers (Figure~\ref{fig:vel_vs_hm}). Using the line width of the Mg\,\textsc{ii} doublet in stacked spectra, similar velocity dispersion values were obtained in \citet{lan18}. That work also reported a large velocity dispersion comparable to the dark matter halo velocity dispersion for ELGs and sub-virial gas motion ($\sim 0.5\sigma_{\rm halo}$) around LRGs. The sub-virial velocity of cold gas was also observed by \citet{huang16} for both passive LRGs and [O\,\textsc{ii}] emitting LRGs \citep[see also][]{chen17}. Furthermore, the velocity dispersion scaling is in rough agreement with \citet{nielsen15, nielsen16} who characterized the galaxy-absorber relative velocity around blue and red galaxies using high-resolution spectra of quasars. The existence of broad components in the velocity distributions of Mg\,\textsc{ii} absorbers  around
massive red galaxies is
in agreement with the results of  \citet{kauffmann17}, who proposed that the large velocity separation Mg\,\textsc{ii} absorbers trace gas that has been pushed out of the dark matter haloes by multiple episodes of AGN-driven mechanical feedback acting over long time-scales. 

\subsubsection{The main new results in this paper}
Thanks to the larger quasar and galaxy samples that are included in our study, we are able to characterize the scale dependence of Mg\,\textsc{ii} absorber covering fraction and rest equivalent width  with greater accuracy.
We find that the  Mg\,\textsc{ii} covering fraction for
both LRGs and ELGs  declines very rapidly
at a projected radius of 50 kpc. Our results are consistent with previous findings, but the greater S/N of our measurements allows us to better characterize the sharpness of the break, indicative of a rather sudden transition between the regime where physical properties of the CGM are regulated by galactic outflows and the regime where the CGM is in thermal balance set by the gravitational field of the dark matter halo.

On scales of a few hundred kpc, the average total 
Mg\,\textsc{ii} EW per absorber dips below the characteristic field value. This is seen
both for LRGs and ELGs, but is a stronger effect for the latter. We speculate that this dip is related to gas inflow processes onto dark matter haloes.

\subsection{Clues to the origin of Mg \textsc{ii} absorbers}

The velocity separation between galaxy-absorber pairs provides clues to the kinematics of absorbing gas around galaxies. The gas around galaxies can acquire motion due to several physical processes. The parent dark matter halo can accrete pristine gas from the intergalactic medium which can be enriched by metals ejected via galactic outflows, stellar winds, or supernovae in star-forming galaxies. The inflow velocity of gas accreting from the IGM can be smaller or similar to the velocities expected from gravitational free-fall. In addition, cold gas can be in virial equilibrium with orbiting halo gas e.g. rotating with the halo orbital velocity. Alternatively, the metals forming inside stars can be thrown out of the galaxies by powerful supernovae with velocities up to $v\sim 500-1000$ \kms \citep{dalla08, sharma13}. The powerful outflows from supermassive black holes can likewise eject gas out to $\sim$ tens of kpc with velocities exceeding $\sim 3000$ \kms \citep{circosta18, perrotta19, nelson19}.

The high velocity dispersion around ELGs suggests that the origin of cold gas traced by Mg\,\textsc{ii} absorbers around ELGs is likely due to powerful galactic outflows. Such powerful outflows around star-forming galaxies have been observed \citep{steidel10, bordoloi14, rubin14, zhu15}. This conclusion is also supported by the particularly strong enhancement of the Mg\,\textsc{ii} covering fraction within $D_{\rm proj}<50$ kpc. Using higher fidelity imaging, \citet{lan18} found an enhancement of Mg\,\textsc{ii} absorption along the minor axes of ELGs, which also supports the outflow scenario. In contrast, accreted gas from the halo or satellite galaxies is predicted to align along the major axes of galaxies \citep{peroux20}. Finally, we also see a strong positive correlation between the SFR of ELGs and covering fractions (differential values) (Figure~\ref{fig:fc_sfr}), evidence that star driven outflows play a pivotal role in metal enrichment at $\lesssim 100$ kpc or $D_{\rm proj}\lesssim r_{\rm vir}$ in star-forming galaxies.

On the other hand, the suppression of gas motion around LRGs indicates that the absorbing gas associated with them is gravitationally bound and unlikely to have originated from galactic outflows. This is further supported by the low star formation activity in LRGs and negligible contribution of stellar activity to the abundance of metals around them \citep{afruni19}. A similar suppression of gas velocity has been found for $\rm Ly\, \alpha$, and O\,\textsc{vi} gas around COS-halos galaxies \citep{tumlinson11, tumlinson13}. The interaction between cold and hot gas inside the halo also plays an important role in deciding the fate of cold gas around LRGs. As pointed in \citet{huang16}, the drag exerted by the hot CGM can slow down the gas motion around LRGs and, cool clumps of gas can fall towards the central galaxy due to orbital decay. However, the evaporation time is significantly smaller than the infall time and most cool clumps would therefore evaporate before reaching the galaxy, or small distances \citep{zahedy19}. The observed existence of cold gas around LRGs implies that there is some form of balance between heating and cooling inside the halo such that the cold gas is routinely formed and destroyed in the halo \citep{sharma12,voit18}. The origin, and survival, of such clouds of cool gas remains an unsolved theoretical question \citep{schneider18,nelson20,kishore21}. 

Furthermore, our analysis also shows that a non-negligible fraction ($\approx 15$ percent) of LRGs have high-velocity gas clouds associated with them. As it is unlikely that galactic outflows powered by stellar activity could account for this population, one possible explanation could be AGN-driven outflows. The origin of cool gas around LRGs is undoubtedly therefore diverse and results from a combination of processes including accretion from the IGM, ram-stripping of the gas from satellite galaxies, mass loss from massive halo stars, and metal-enrichment by AGN driven outflows \citep{bordoloi11,huang16,lan18}.

Combining our results with previous measurements (Figure~\ref{fig:vel_vs_hm}), we see that the different behaviour of the gas motion around star-forming versus quiescent galaxies supports the picture that the origin of cold gas is fundamentally different for star-forming versus passive galaxies. 


\section{Summary of Conclusions}\label{conclude}

In this work we have developed a fully automated quasar continuum estimator and absorption line detection pipeline. We have run the pipeline on the SDSS DR16 quasar sample and compiled the largest Mg\,\textsc{ii} metal absorber catalogue to date. Our main findings are: 

\begin{itemize}
    \item Our Mg\,\textsc{ii} catalogue contains 160,000 Mg\,\textsc{ii} absorbers ($ 0.36<z<2.3$) based on the SDSS DR16 final quasar sample. The median redshift and Mg\,\textsc{ii} absorption strength of the catalogue are $z\sim 1.14$ and $EW_{\rm 2796}\sim 1.3$ \AA,\, respectively. Fe\,\textsc{ii} lines are also commonly detected and have smaller rest equivalent widths than Mg\,\textsc{ii} as they have weaker oscillator strengths.
    
    \item Stacking quasar spectra in the rest-frame of Mg\,\textsc{ii} absorbers, we detect the presence of other weak metal lines including Si\,\textsc{i} , C\,\textsc{iv}, Al\,\textsc{ii}, Al\,\textsc{ii}, Mg\,\textsc{i} and Fe\,\textsc{ii}. We study their properties as a function of Mg\,\textsc{ii} line strength. 
    
    \item The measured doublet ratio of Mg\,\textsc{ii} absorbers shows that most lie within the theoretical limit of the Mg\,\textsc{ii} doublet ratio, i.e. 1 (saturated case) and 2 (unsaturated case). Strong absorbers are almost always saturated.
    
    \item To investigate the completeness of our detection pipeline we simulate $\sim$ 33 million fake absorbers. Completeness is a strong function of both rest equivalent width and redshift of absorbers. Our method is naturally less complete for weak absorbers ($EW_{\rm 2796}<0.5$ \AA) and reasonably complete for strong absorbers. Our detection pipeline performs excellent on quasars with relatively high S/N. On the other hand purity is also very high ($\sim 95\%$) for our catalogue.
\end{itemize}

In the second half of the paper we connect Mg\,\textsc{ii} absorbers to the latest SDSS DR16 catalogue of $\sim$ 1.3 million galaxies, divided into star-forming, emission line galaxies (ELGs) and quiescent, luminous red galaxies (LRGs). We use these samples to study the incidence and properties of metal absorption in the circumgalactic medium and nearby intergalactic medium. We investigate the absorber-galaxy correlation based on several properties of galaxies to understand the physics which most affect the properties of gas around galaxies. Our main results are:

\begin{itemize}
    \item The mean surface density of Mg\,\textsc{ii} absorbers is larger than the expected random background for both ELGs and LRGs. At large distance, it reaches the expectation for random sightlines at a  distance ($D_{\rm proj}\sim 15$ Mpc) for both ELGs and LRGs. The mean surface density is also a strong function of galaxy type within $D_{\rm proj}<50$ kpc, with ELGs showing a strong enhancement in the inner halo. 
    
    \item The covering fraction of Mg\,\textsc{ii} absorbers varies strongly with galaxy type. ELGs have 2 - 4 times higher covering fractions than LRGs within $D_{\rm proj}<50$ kpc, regardless of $\rm{EW}_{\rm 2796}$ strength. The covering fraction decreases with projected distance and decreases strongly with increasing rest equivalent width (EW). For $D_{\rm proj}<400$ kpc, the covering fraction is larger than the expected random background and converges to this value only at $D_{\rm proj}\gtrsim 10$ Mpc. A similar trend is visible for the cumulative covering fraction, which is clearly enhanced in star-forming versus passive galaxies up to $D_{\rm proj}\lesssim 200$ kpc. 
    
    \item The average rest equivalent width of Mg\,\textsc{ii} absorption per absorber shows a clear enhancement close to galaxies, $D_{\rm proj}<50$ kpc for both ELGs and LRGs. LRGs always have systematically higher values than ELGs. There is a weak evolution of EW with galaxy redshift, such that higher $z$ galaxies have slightly higher EWs at all $D_{\rm proj}\gtrsim 1$ Mpc. In all redshift bins, the Mg\,\textsc{ii} average rest equivalent width declines to the field value beyond $\sim$ 150 kpc. 
    
    \item When normalizing projected distances by the virial radii of galaxies we find that the covering fraction varies strongly with galaxy type. There is a weak stellar mass trend for ELGs, but a strong stellar mass trend for LRGs, whereby the covering fraction decreases with increasing stellar mass, consistent with previous studies. The trend is visible even at $D_{\rm proj}\sim 2r_{\rm vir}$.
    
    \item Splitting ELGs by star-formation activity we find that Mg\,\textsc{ii} covering fraction is positively correlated with SFR, supporting the idea of a galactic outflow origin for the cold gas around ELGs.
    
    \item We trace the kinematics of galaxy-absorber pairs with the distribution of line-of-sight relative velocity $\Delta v$. Around ELGs, this is well-characterized by a single gaussian profile, while LRGs require a second, broad component with $\sigma_{v}\sim1100$ \kms\, to capture high-velocity tails, likely indicative of AGN-driven outflows. The velocity dispersion is higher around LRGs than ELGs. The majority of the absorbers close to galaxies ($D_{\rm proj}<100$) kpc have low velocities and are likely gravitationally bound. 
    
    \item The velocity dispersion of cool CGM gas increases with stellar (and halo) mass for both samples. The gas motions around ELGs are similar to the expected dark matter halo velocity dispersion, while gas motions around LRGs are suppressed and closer to $\sim 0.5 \sigma_{\rm DM,halo}$. The different properties and, moreover, trends of Mg\,\textsc{ii} absorption with galactic properties between ELGs and LRGs implies that cool circumgalactic gas around star-forming versus quiescent galaxies has a fundamentally different physical origin.
\end{itemize}

In this study we have benefited from the cross-correlation of a large Mg\,\textsc{ii} absorber catalogue with the similarly expansive spectroscopic galaxy samples of SDSS. However, in understanding the galaxy-CGM connection we were limited in the physical properties of galaxies available. The ground-based wavelength limitations also restricted the available redshift range to connect galaxies and absorbers. Finally, available SDSS imaging does not generally allow fine-grained galactic morphologies to be inferred. However, such a morphology based analysis can be performed with the recent data from the Dark Energy Survey Instrument (DESI) Legacy Imaging Survey \citep{dey19}. 

In the future, other large imaging surveys such as the Large Synoptic Survey Telescope (LSST) at the Rubin Observatory will provide enormous datasets of galaxies, up to higher redshifts, and with high quality imaging. Together with upcoming large spectroscopic galaxy surveys such as PFS on the Subaru telescope \citep{tamura16}, statistical analyses of the circumgalactic medium will be an ever more powerful tool to understand the formation and evolution of galaxies across cosmic time.

\section*{Data Availability}

Data directly related to this publication and its figures is available on request from the corresponding author. This work is based on data which is publicly available in its entirety as part of DR16 of the Sloan Digital Sky Survey (SDSS; \url{www.sdss.org}).


\section*{Acknowledgements}

We thank the anonymous referee for a very constructive and insightful report that has helped us improve the quality of the paper. We also thank Ting-Wen Lan, Hong Guo and Guangtun Ben Zhu for discussions and assistance with various aspects of this paper. AA would also like to thank his colleagues Mohammadreza Ayromlou, Periklis Okalidis and Ilkham Galiullin for insightful discussions. We make use of Python libraries such as \textsc{Numpy}, \textsc{Scipy} and \textsc{Matplotlib} for data analysis and visualization.  

Funding for the Sloan Digital Sky Survey IV has been provided by the Alfred P. Sloan Foundation, the U.S. Department of Energy Office of Science, and the Participating Institutions. SDSS acknowledges support and resources from the Center for High-Performance Computing at the University of Utah. The SDSS web site is \url{www.sdss.org}.
SDSS is managed by the Astrophysical Research Consortium for the Participating Institutions of the SDSS Collaboration including the Brazilian Participation Group, the Carnegie Institution for Science, Carnegie Mellon University, Center for Astrophysics | Harvard \& Smithsonian (CfA), the Chilean Participation Group, the French Participation Group, Instituto de Astrof{\'i}sica de Canarias, The Johns Hopkins University, Kavli Institute for the Physics and Mathematics of the Universe (IPMU) / University of Tokyo, the Korean Participation Group, Lawrence Berkeley National Laboratory, Leibniz Institut f\"{u}r Astrophysik Potsdam (AIP), Max-Planck-Institut f\"{u}r Astronomie (MPIA Heidelberg), Max-Planck-Institut f\"{u}r Astrophysik (MPA Garching), Max-Planck-Institut f\"{u}r Extraterrestrische Physik (MPE), National Astronomical Observatories of China, New Mexico State University, New York University, University of Notre Dame, Observat{\'o}rio Nacional \/ MCTI, The Ohio State University, Pennsylvania State University, Shanghai Astronomical Observatory, United Kingdom Participation Group, Universidad Nacional Aut{\'o}noma de M{\'e}xico, University of Arizona, University of Colorado Boulder, University of Oxford, University of Portsmouth, University of Utah, University of Virginia, University of Washington, University of Wisconsin, Vanderbilt University, and Yale University.


\bibliographystyle{mnras}
\bibliography{refs}
\appendix


\section{Receiver Operating Characteristic (ROC) Analysis for the parameters}\label{appendix:roc}

To select an optimized value for the $\alpha$ parameter (see Section \ref{pipeline}), we make use of the available DR12 Mg \textsc{ii} catalogue compiled by \citet{zhu13a} to test our detection pipeline. The reference catalogue has $76,148$ Mg \textsc{ii} absorbers detected in $47,065$ quasars. After applying all the masks as described in the section \ref{pipeline}, we end up with $39,219$ Mg \textsc{ii} absorbers in $25,716$ quasars. Then we run our complete detection pipeline on this set of quasars for several values of $\alpha$ and estimate the True Positive Rate (TPR) and False Positive Rate (FPR) to analyse the ROC curve to choose an optimized value for $\alpha$. To estimate the TPR and FPR we only use Mg \textsc{ii} absorbers with $\rm EW_{2796}>0.1$ \AA\, ($\rm N = 38,313$). We define TPR and FPR as follows:

\subsection{Adaptive SNR approach: Mg \textsc{ii} doublet criteria}

In this approach (using Rule 1, 2, and 3 described in section \ref{pipeline}) we define TPR as the ratio of absorbers that passed our Mg \textsc{ii} criteria to the total number of Zhu absorbers. FPR is defined as the ratio of all legitimate absorbers that passed our Mg \textsc{ii} criteria, but did not match with Zhu absorbers, to the maximum of total Zhu absorbers and all absorbers detected with our pipeline.

\begin{figure}
    \centering
  \includegraphics[width=0.9\columnwidth]{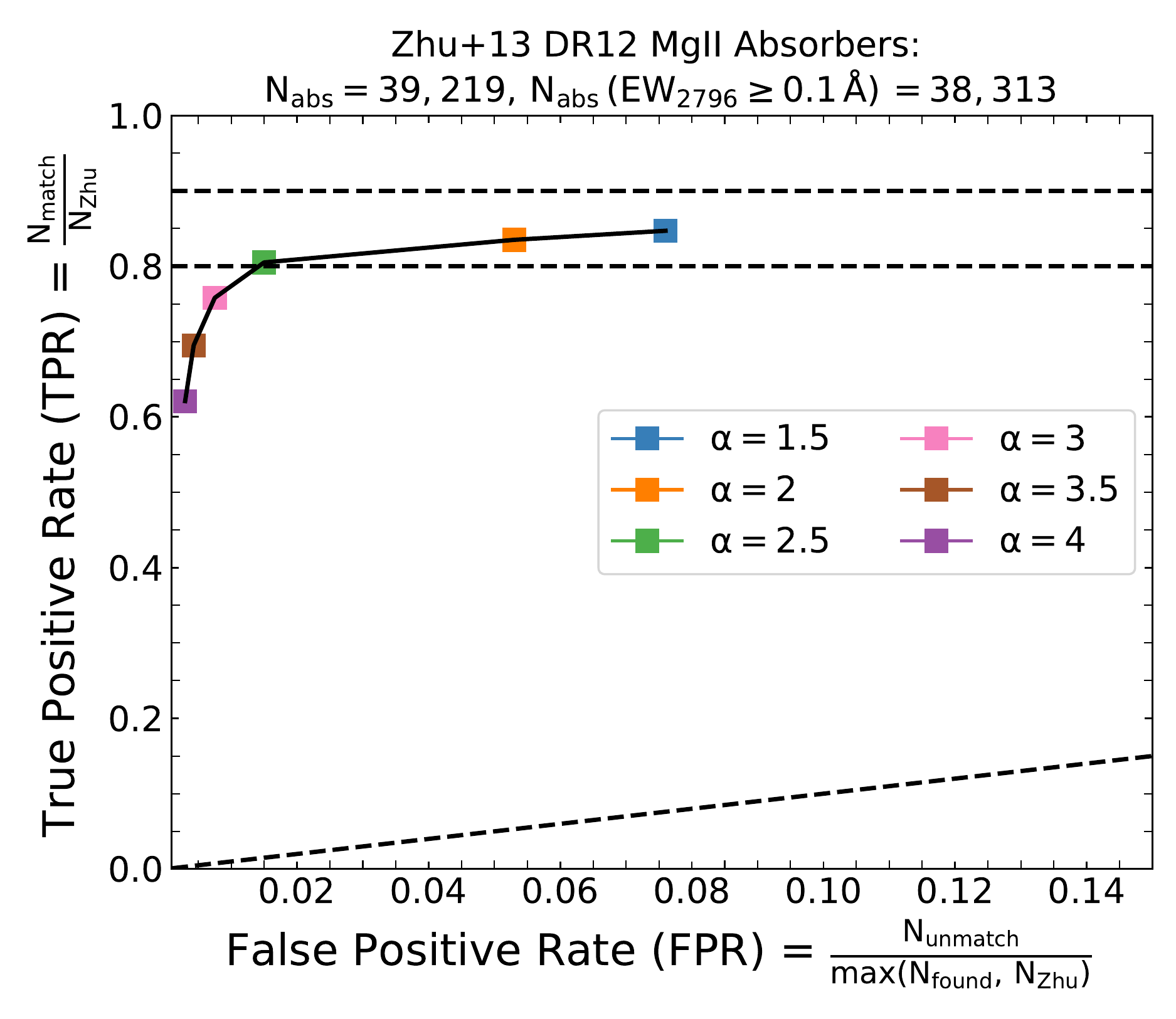}
  \caption{ROC curve: True Positive Rate (TPR) as a function of False Positive Rate (FPR) for different $\alpha$ values using selection based on SNR condition. Higher $\alpha$ corresponds to lower TPR and FPR as expected. Overall the FPR is significantly lower in the adaptive S/N approach. Note that, x-axis is zoomed for clarity. The dashed horizontal lines show 0.8 and 0.9 on TPR axis and the dashed inclined line shows one-to-one line.}
  \label{fig:roc}
\end{figure}

\begin{figure}
    \centering
  \includegraphics[width=0.85\columnwidth]{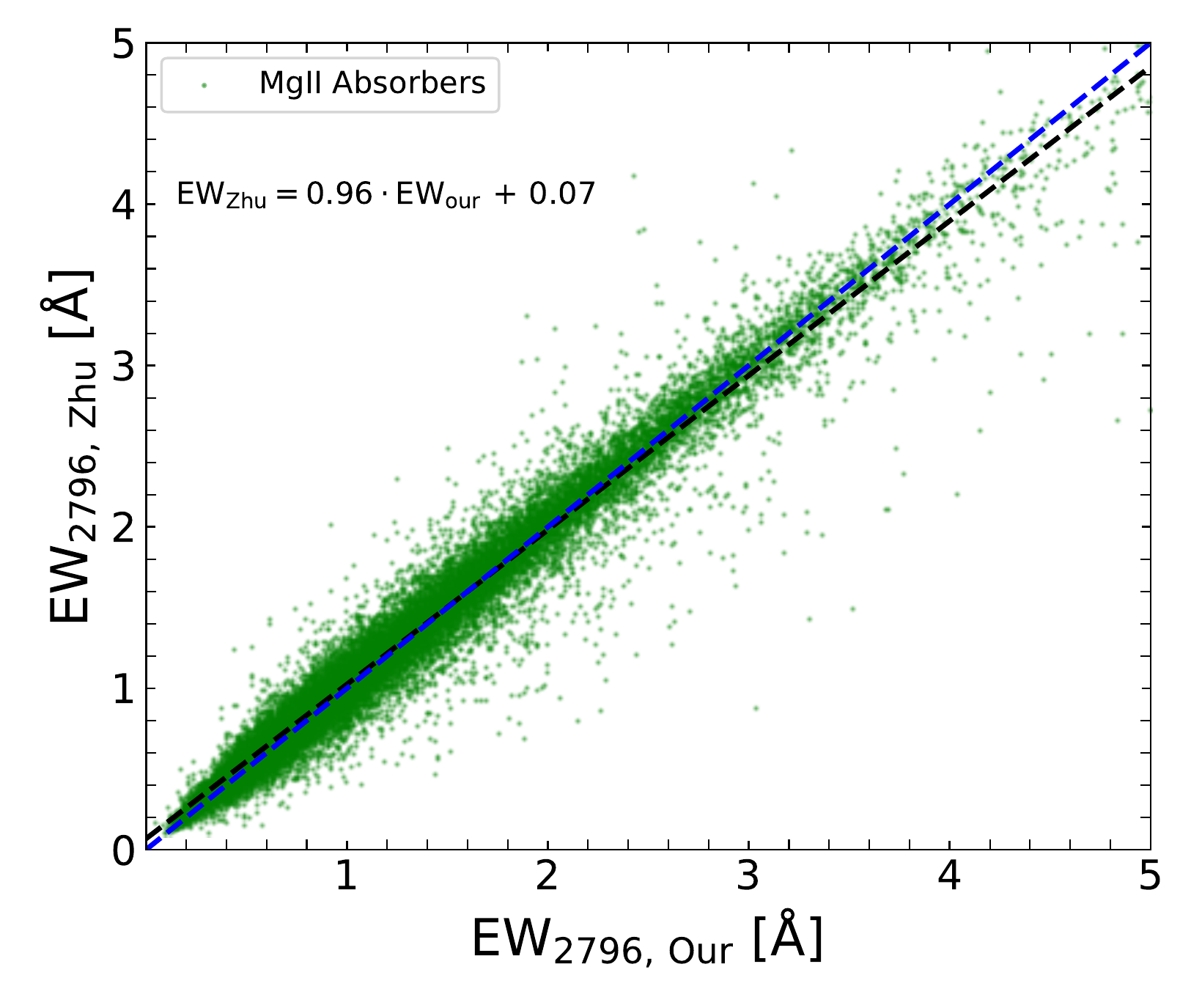}\\
  \includegraphics[width=0.9\columnwidth]{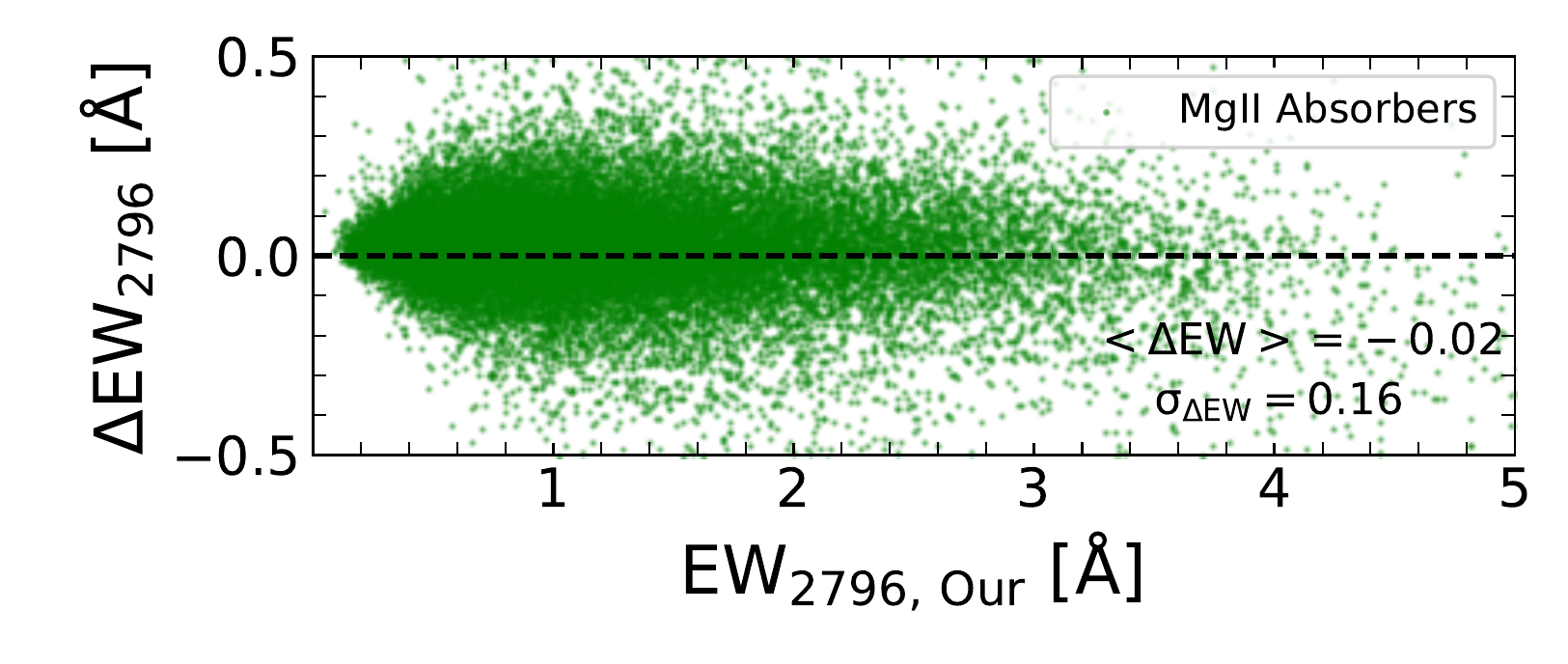}
  \caption{Comparison with \citet{zhu13a} DR12 Mg \textsc{ii} catalogue, for $\alpha=2.5$. \textbf{Top:} $\rm EW_{2796,\, Zhu}$ vs. $\rm EW_{2796, \, our}$, showing good agreement, though our EWs are slightly higher (as visible in best-fit dashed line shown in black). \textbf{Bottom:} Difference between these two values as a function of $\rm EW_{2796,\, our}$. The majority of the absorbers have $\rm |\Delta EW|\leq0.2$ \AA\, and the typical error between our values versus Zhu is $\sim0.16$ \AA.}
  \label{fig:compare_zhu}
\end{figure}

We take the maximum of these two quantities to account for the case when our pipeline finds several absorbers that are not matched with the \citet{zhu13a} catalogue. Figure~\ref{fig:roc} show the ROC curve estimated for different $\alpha$ values in adaptive S/N approach. We clearly see that applying the adaptive S/N condition significantly reduces the false cases. For example at $\alpha=2.5$ the corresponding TPR and FPR are $\sim 0.8$ and $\sim 0.02$, respectively. Therefore, we take advantage of the S/N of the spectra to reduce the incidence of false cases. As expected for larger values of $\alpha$, the TPR and FPR are both low while, increasing for smaller $\alpha$ values. 

Finally, for our current catalogue, we conservatively choose $\alpha=2.5$ because our ROC analysis is based on quasars (from the DR12 absorber catalogue only) with possibly high S/N compared to the entire DR16 quasar set used in the current study. By choosing a slightly higher $\alpha$ value we select relatively strong potential absorbers at the convolution step thus reducing the pipeline runtime significantly. However, we miss a good fraction of absorbers as shown in the completeness analysis (see section \ref{completeness}, Figure~\ref{fig:completeness}). On the other hand, this guarantees a very high purity for our catalogue as shown in Figure~\ref{fig:purity}.

\section{Rest equivalent widths}\label{appendix:eqv_width}

\begin{figure}
    \centering
  \includegraphics[width=0.85\columnwidth]{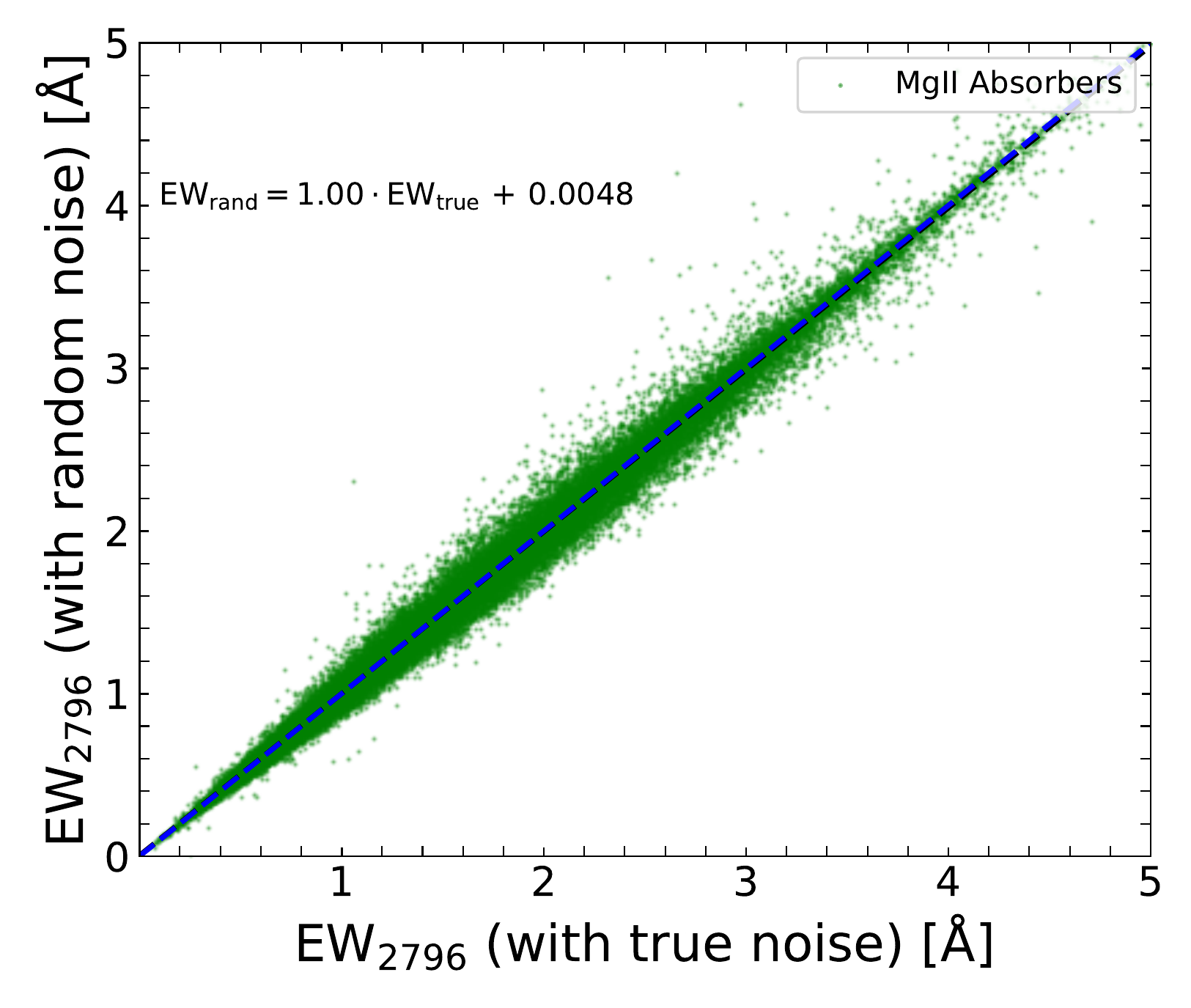}\\
  \includegraphics[width=0.9\columnwidth]{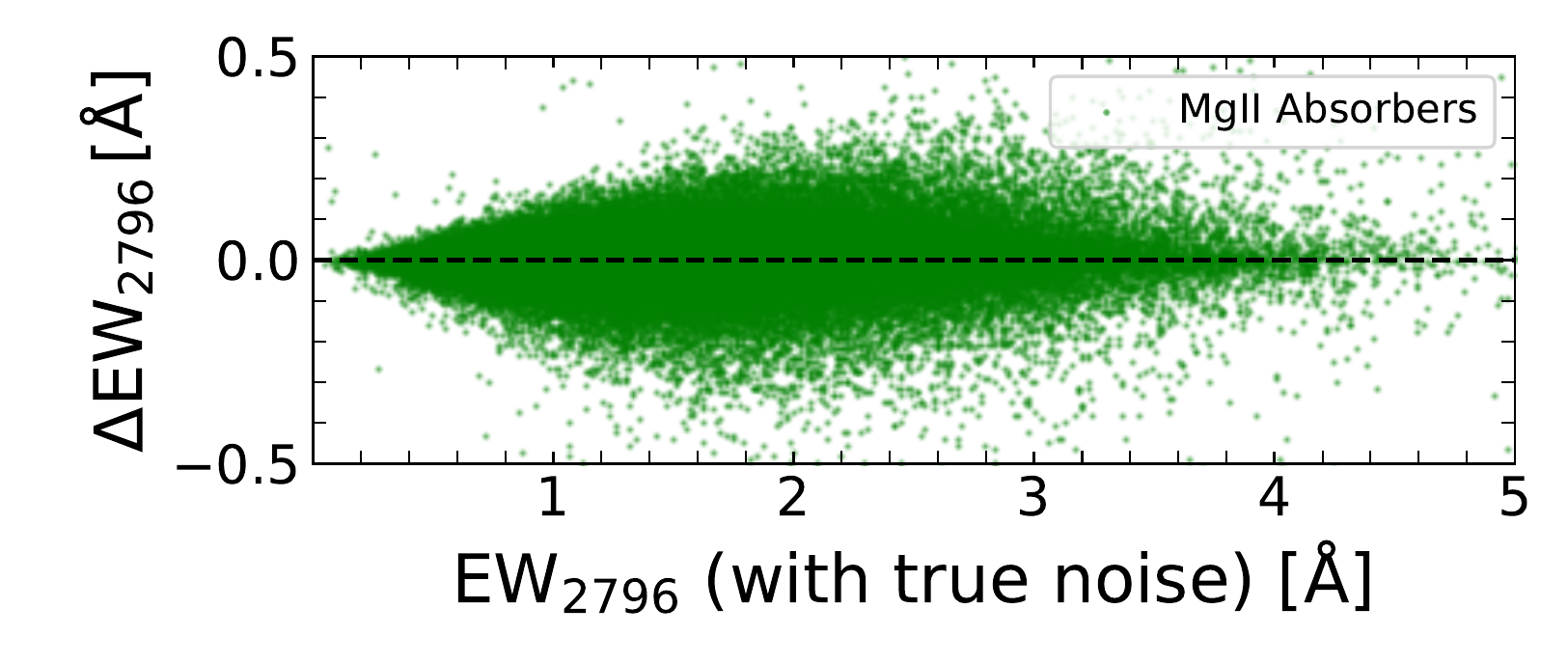}\\
  \includegraphics[width=0.85\columnwidth]{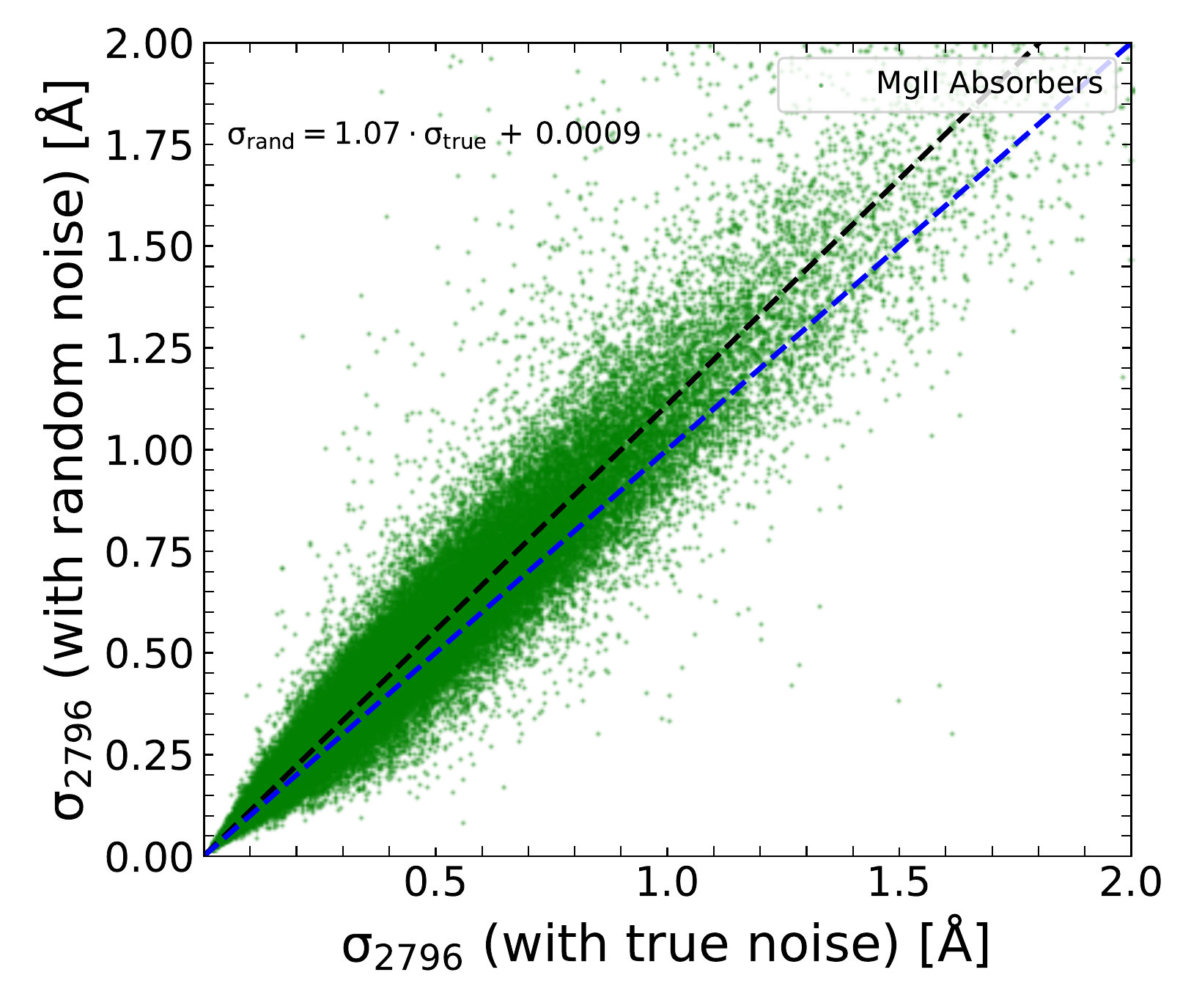}\\
  \includegraphics[width=0.9\columnwidth]{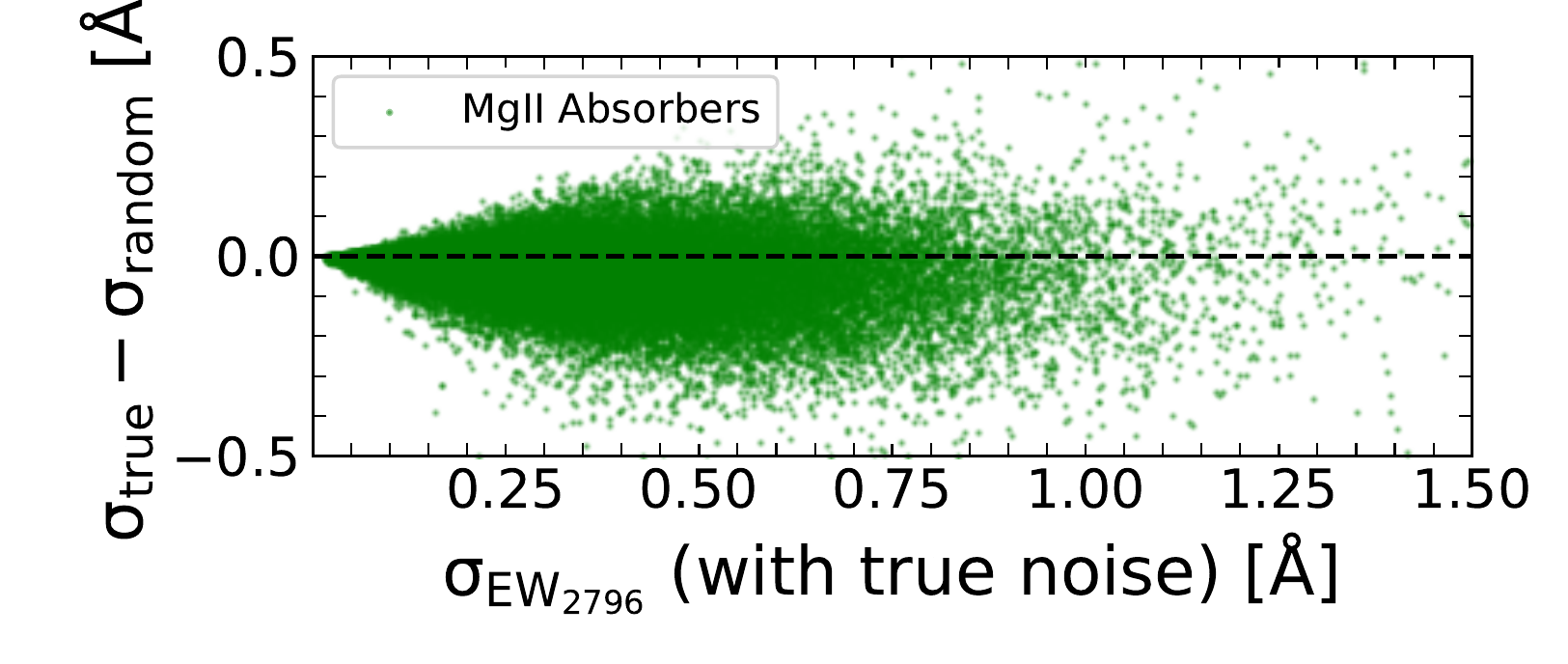}
  \caption{Comparison of rest EW and error measurements with random and true noise. \textbf{Top:} Measured rest $\rm EW_{2796}$ (by adding random noise) as a function of rest $\rm EW_{2796}$ (by adding true noise). The corresponding smaller panel shows the difference between two as a function of rest $\rm EW_{2796}$ (by adding true noise). The agreement is very good and this further supports that measurements are robust. \textbf{Bottom:} Corresponding EW errors measured using these approaches. They match very well though error from random noise approach is slightly higher (as visible in best-fit dashed line shown in black).}
  \label{fig:error_plot}
\end{figure}

As described in section \ref{ew_boot} we measured the rest equivalent widths and the uncertainties of Mg \textsc{ii} doublet by adding \textit{true noise from the spectra} as well as purely \textit{random noise} to the absorption feature. We find that rest EWs and errors measured with these two approaches agree well. This further supports the fact that a very high fraction of our absorbers are genuine. The comparison of these two approaches is shown in Figure ~\ref{fig:error_plot}. We find that the difference is similar to the typical uncertainty in the measurement. The uncertainties measured from adding random noise are slightly ($\sim 5\%$) higher than the errors measured by adding true noise from the spectra. However, the overall match is very good, and we use measurements based on true spectral noise for our study.

\section{Purity analysis of the Pipeline}\label{purity_analysis}

In order to quantify the quality of our catalogue we also estimate the purity of simulated absorbers detected in our pipeline. As described in Section \ref{completeness}, we estimate the purity in each $EW\, - \, z$ bin as the ratio of detected absorbers to the total absorbers that pass our Mg \textsc{ii} criteria.

\begin{figure}
    \centering
  \includegraphics[width=0.95\columnwidth]{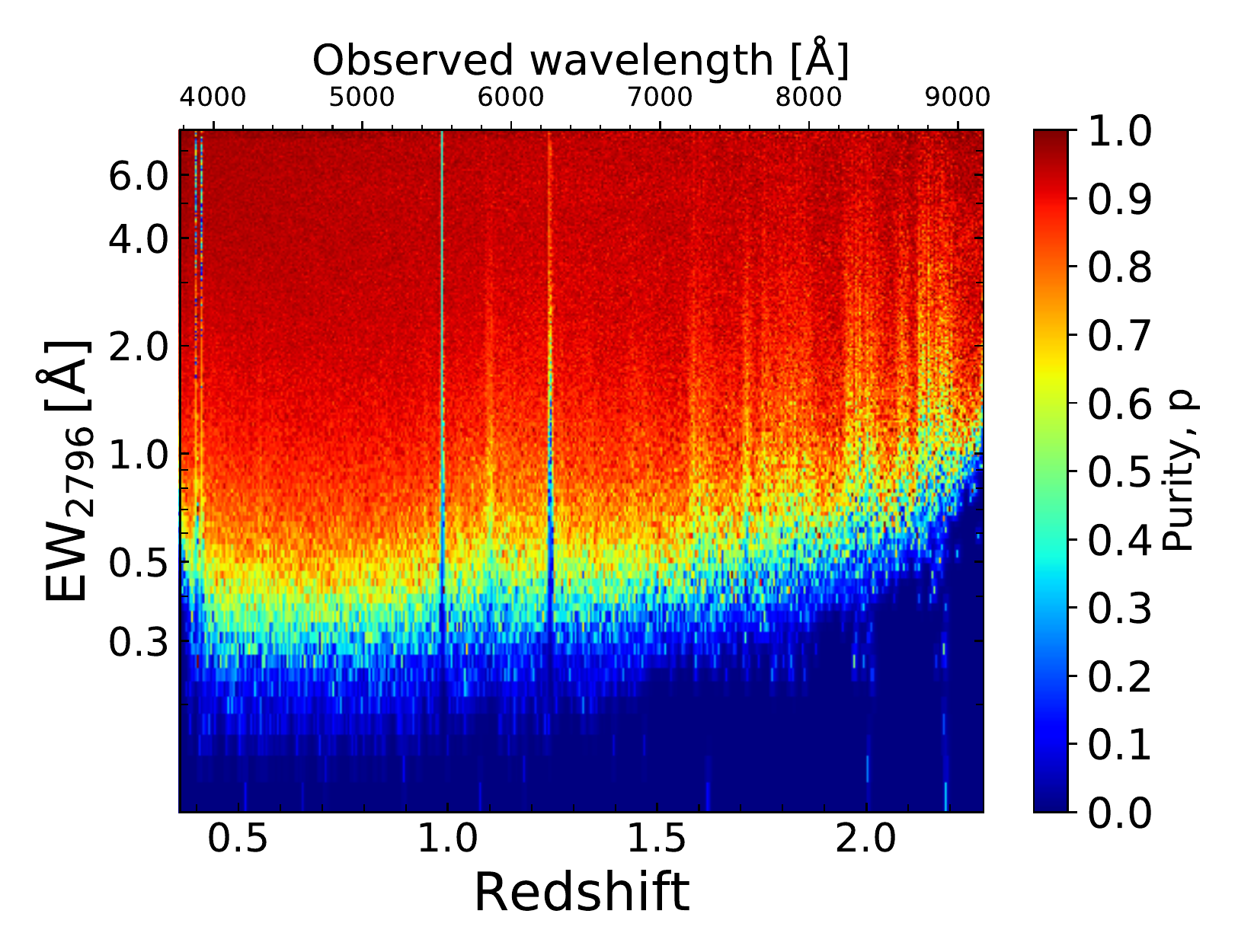}
  \caption{Purity function $p({EW}_{\rm 2796},z)$ for QSOs with $\rm S/N_{QSO}>2$. 2D distribution as a function of rest equivalent width and redshift. The purity is very high for detected absorbers. For very weak absorbers ($EW_{\rm 2796}<0.3\,$\AA) the purity is lower because the completeness is similarly low, and the pipeline is more likely to identify false cases due to noisy features.}
  \label{fig:purity}
\end{figure}

We show the 2D purity distribution of simulated absorbers as a function of rest EW and redshift of the absorber in Figure ~\ref{fig:purity}. We clearly see that the purity of detected absorbers is very high in almost every bin for $EW_{\rm 2796}>0.4$ \AA\, systems. For very weak absorbers ($EW_{\rm 2796}<0.3\,$\AA) the purity is low because the corresponding completeness is also low, and possibly the pipeline finds false cases due to noisy features. The overall purity of our detection pipeline is $\sim 95\%$. Note that in Figure~\ref{fig:purity} we only show the purity for absorbers detected in QSOs having $\rm S/N_{QSO}>2$.

\end{document}